\documentclass[10pt,twocolumn,preprintnumbers,amsmath,amssymb,nofootinbib,superscriptaddress]{revtex4-1}

\usepackage{epsfig}
\usepackage{url}
\usepackage[normalem]{ulem}
\usepackage{latexsym}
\usepackage{epsfig}
\usepackage{amsmath}
\usepackage{amssymb}
\usepackage{wasysym}
\usepackage{graphicx}
\usepackage{verbatim}
\usepackage{enumerate,mdwlist}
\usepackage[titletoc]{appendix}
\usepackage{amsfonts}
\usepackage{pgfplots}
\usepackage[export]{adjustbox}
\usepackage{bbold}
\usepackage[normalem]{ulem}
\usepackage{float}

\usepackage[linktocpage=true]{hyperref}
\hypersetup{
colorlinks=true,
citecolor=ultramarine,
linkcolor=cadmiumgreen,
urlcolor=indigo(dye),
pdfauthor={},
pdftitle={},
pdfsubject={}
}

\usepackage[all]{xy} 
\usepackage{amsfonts}

\newcommand{\ba}{\begin{eqnarray}}
\newcommand{\ea}{\end{eqnarray}}
\newcommand{\be}{\begin{equation}}
\newcommand{\ee}{\end{equation}}

\newcommand{\gwphaseurl}{https://github.com/ezquiaga/phazap}

\definecolor{grey}{rgb}{0.4,0.4,0.4}
\definecolor{dullmagenta}{rgb}{0.4,0,0.4}
\definecolor{darkblue}{rgb}{0,0,0.4}
\definecolor{midblue}{rgb}{0,0,0.5}
\definecolor{midred}{rgb}{0.5,0,0}
\definecolor{orange}{rgb}{1,0.5,0}
\definecolor{lightbrown}{rgb}{0.75,0.5,0.25}
\definecolor{tan}{cmyk}{0.14,0.42,0.56,0}
\definecolor{djunglegreen}{cmyk}{0.99,0,0.52,0}
\definecolor{lightgreen}{rgb}{0,1,0}
\definecolor{olivegreen}{cmyk}{0.64,0,0.95,0.40}
\definecolor{midgreen}{rgb}{0.0,0.675,0.0}
\definecolor{darkgreen}{rgb}{0,0.5,0}
\definecolor{ultramarine}{rgb}{0.07, 0.04, 0.56}
\definecolor{cadmiumgreen}{rgb}{0.0, 0.42, 0.24}
\definecolor{indigo(dye)}{rgb}{0.0, 0.25, 0.42}

\begin{document}

\title{Diffraction around caustics in gravitational wave lensing}

\author{Jose Mar\'ia Ezquiaga}
\email{jose.ezquiaga@nbi.ku.dk}
\author{Rico K.~L.~Lo}
\email{kalok.lo@nbi.ku.dk}
\author{Luka Vujeva}
\email{luka.vujeva@nbi.ku.dk}
\affiliation{Center of Gravity, Niels Bohr Institute, Blegdamsvej 17, 2100 Copenhagen, Denmark}

\begin{abstract}
Gravitational lensing magnification is maximal around caustics. At these source locations, an incoming wave from a point source would formally experience an infinite amplification in the high-frequency or geometric optics limit. This divergence reflects the break-down of the mathematical formalism, which is regularized by either the finite size of the source or its wavelength. We explore diffraction around caustics and their implications for the distortion of waveforms from point sources, focusing on three types of caustics: point singularities, folds, and cusps. We derive analytical results for the amplitude and phase of the diffracted waves, and compare those against the stationary phase approximation. We then study the observational signatures and detectability of these distortions on gravitational waves. We find that the lensing distortions could be detectable, but that the stationary phase approximation is still a good description of the system even close to the caustic, when the repeated gravitational wave chirps interfere with each other. We also quantify the possibility of distinguishing lensed signals from different caustics by performing Bayesian parameter estimation on simulated signals. Our results demonstrate that the universal distortions due to diffraction around caustics could be used to single out a gravitational wave event as lensed. 
\end{abstract}

\date{\today}

\maketitle

%-----
%SECTION: INTRODUCTION
%-----
\section{Introduction}

Through gravitational lensing, the Universe offers natural and universal ``gravitational telescopes'' to unveil objects that are otherwise too faint and distant to be observable. 
As a consequence, and irrespective of the messenger's nature, gravitational lensing is intimately related to the study of the furthest sources in the cosmos, unveiling early Universe physics, and enabling the discovery of new classes of objects. 
In certain configurations, gravitational lensing of electromagnetic waves can be extraordinarily efficient, magnifying signals by many orders of magnitude~\cite{Diego:2018fzr}. 
Lensing has, for example, enabled detecting an individual star in a galaxy more than 12 billion light years away that has been magnified by $\sim 10^4-10^5$ times~\cite{Welch_2022}, and to observe the furthest galaxies so far at redshifts of $z \approx 13$~\citep{2023ApJ...957L..34W}. 

The maximal gravitational lensing magnifications occur near caustics~\cite{Ohanian:1983}. 
For this reason, caustics are of great observational interest and have been studied in considerable detail~\cite{Schneider:1992}. 
Extreme magnifications are expected, for instance, when stars cross the caustic of a galaxy cluster~\cite{Miralda-Escude:1991}. 
In general, high-magnification regions serve to amplify small scale structures and expose the presence of ``microlenses''~\cite{Chang_Refsdal:1979,Chang_Refsdal:1984,Kayser_Refsdal_Stabell:1986}. 
From a mathematical perspective, caustics are also very interesting, since they predict universal behaviors for the lensing observables~\cite{Blandford:1986zz}. 
Moreover, caustics signal out the failure of the geometric optics approximation, which would imply infinite magnifications at those points.  
Typically, in the electromagnetic spectrum, these infinities are regularized by the physical size of the source, given that their wavelength is much smaller. 
For instance, there is a large hierarchy of scales between the kiloparsec size of a galaxy and the optical emission from its center. 
There are exceptions to this general rule, such as pulsar scintillation by the interstellar medium~\cite{Narayan1992_pulsar_scintillation}. 
Fast radio bursts could also be diffracted by small compact objects ranging from 0.1 to 100 Earth masses~\cite{Jow:2020rcy}.  

For gravitational waves (GWs) from compact binary coalescences, on the contrary, it is always the wave effects that render the magnification finite. 
The wavelength of these systems is associated to their orbital motion, which decreases towards merger at most down to their Schwarzschild radius~\cite{misner2017gravitation}. 
If the sources are merging black holes or neutron stars, then 
high magnifications are limited to lenses more massive than stars, as otherwise, the wave is only weakly lensed~\cite{Bontz_Haugan:1981}. 
In order to include the finite wavelength effects, it is necessary to solve the gravitational lensing problem in the wave optics regime. 
Wave optics effects are appealing because they can distort the waveforms~\cite{Deguchi:1986zz,NakamuraDeguchi}, which may change the notion of their arrival time in a multi-messenger event~\cite{Ezquiaga:2020spg}. 
They could also break some of the degeneracies in time-delay cosmography~\cite{Cremonese:2021puh,Chen:2024xal}. 
Lensing waveform modifications are particularly relevant for GWs because, different to other transients, they are both coherently emitted and detected. 
Moreover, GWs effectively do not interact with the medium along their propagation, so, within general relativity, they can only be modified by lensing~\cite{Misner:1974qy}. 
Because the emitted waveforms of compact binaries merging in vacuum are well understood from first principles, ``just'' solving Einstein's field equations, one can search for these signatures of lensing in the waveform distortions. 

Caustics are well understood and classified from catastrophe theory~\cite{Berry_Upstill}. 
There is only a finite number of stable catastrophes. 
In two dimensional mappings such as gravitational lensing, there are only two: the \emph{fold} and the \emph{cusp} caustics. 
There have already been great efforts in solving the diffraction integral around caustics, many of them several decades ago, cf. Ref.~\cite{Berry_Upstill} for a seminal 1980's review and Ref.~\cite{Feldbrugge:2019fjs} for a modern revision motivated by radio astronomy.  
However, because these studies were driven by optical experiments or astrophysical electromagnetic sources, they focused on quantifying changes to the intensity of the signal for different source positions, e.g.~\cite{Ohanian:1983}, or frequency ranges, e.g.~\cite{Ulmer:1994ij}. 
For GWs this is not enough, as one needs to account for modifications in both the amplitude and phase of a signal that evolves in frequency. 
Modern codes already incorporate all this information by solving the problem numerically, e.g.~\cite{Villarrubia-Rojo:2024xcj}, but no detailed study has focused on the wave optics features around caustics. 
At any rate, when accounting for these effects, it was found that caustics can imprint ``smoking-gun'' observational evidence of lensing on GW signals~\cite{Lo:2024wqm}. 

Signatures of gravitational lensing of GWs are being searched with data from the LIGO \cite{LIGOScientific:2014pky}, Virgo \cite{VIRGO:2014yos} and KAGRA \cite{KAGRA:2020tym} detectors, although there is no evidence yet \cite{Hannuksela:2019kle,LIGOScientific:2021izm,Janquart:2023mvf,LIGOScientific:2023bwz}. 
These searches look for two classes of observables: repeated signals of the same event, and distorted waveforms. 
Repeated signals are expected when a GW is strongly lensed by a galaxy or cluster of galaxies, leading to typical time delays between the signals ranging from hours to months~\cite{Oguri:2019fix,Xu:2021bfn,Vujeva:2025kko}. 
Distorted waveforms are typically expected when a GW encounters a small, compact lens~\cite{Takahashi:2003ix}. 
They can also appear for some of the strongly lensed repeated signals~\cite{Dai:2017huk,Ezquiaga:2020gdt}, although to a lesser degree. 
The diffraction signatures of lensing around caustics presented in this work could be added to current search efforts for wave optics features, which focus on simple isolated lens models (point masses and singular isothermal spheres)~\cite{Wright:2021cbn}, or phenomenological templates of multiple-images overlapping~\cite{Liu:2023ikc}.

The paper is structured as follows. 
We begin in \S\ref{sec:caustic_intro} with an introduction to the concept of a caustic and the mathematical framework to study gravitational lensing. 
Then, from \S\ref{sec:point_caustic} to \ref{sec:cusp_caustic}, we move to study the particular cases of point, fold and cusp caustics. 
In all the cases we study both the limiting solutions when multiple images form, and the general solution in wave optics. 
We then move in \S\ref{sec:implications} to investigate the implications for the detectability of diffraction in gravitational lensing of gravitational waves. 
We investigate the detectability by analyzing the effect of lensing on the match of a simulated signal with an unlensed template, the signal-to-noise time series at the detectors, and the Bayesian inference of the signal's properties. 
We finish the paper with the conclusions in \S\ref{sec:conclusions}. 

%-----
%SECTION: FORMALISM
%-----
\section{Diffraction around caustics}
\label{sec:caustic_intro}

Gravitational lensing involves solving for the wave propagation of a small perturbation over a curved background. 
For the following work, we will restrict ourselves to the weak gravity regime in which the perturbed metric is controlled by the Newtonian potential of the mass distribution responsible of producing the gravitational lensing of the wave. 
We use this section to review the basics of this formalism, cf. \cite{Schneider:1992} for more details, and set the notation and conventions for the rest of the paper. 
For static lenses, it is convenient to Fourier transform the problem so that instead of solving the (real) time domain signal $S_L(t)$, one focuses on its (complex) frequency domain counterpart $\tilde S_L(\omega)$. 
Each of them are related by
\begin{equation}
    S_L(t)=\int \frac{d\omega}{2\pi}\tilde S_L(\omega)e^{-i\omega t}\,.
\end{equation}
The frequency-domain lensed signal is connected to the unlensed one, $\tilde{S}(\omega)$, by the (complex-valued) amplification function $F(\omega)$:
\begin{equation}
    \tilde S_L(\omega) = F(\omega)\tilde S(\omega)\,.
\end{equation}
In that case, the problem then reduces to solving the diffraction integral, which in the thin-lens approximation takes the form:
\begin{equation}
    F(\omega,\vec{\theta}_s)=\frac{\tau_D\omega}{2\pi i}\int \mathrm{d}^2\theta e^{i\omega t_d(\vec\theta,\vec\theta_\mathrm{S})}\,,
\end{equation}
where $\vec\theta_\mathrm{S}$ is the unlensed source angular position and the integral is over the lens plane. Here $t_d$ is the time delay from the source to the observer at each angular position $\vec\theta$ for a given source position $\vec\theta_\mathrm{S}$. %The time delay adds together the geometric and Shapiro time delays. 
In the thin lens approximation the time delay is given by
\begin{equation}
    t_d\approx \tau_D\left[\frac{1}{2}|\vec\theta-\vec\theta_\mathrm{S}|^2 - \Psi(\vec\theta)\right]
\end{equation}
where each term corresponds to the geometric and Shapiro delays, respectively, with $\Psi = -t_\Phi/\tau_D= 2/(\tau_Dc^3)\int \Phi \mathrm{d}s$ and $c$ is the speed of light. 
We also define for convenience the time associated to the effective distance between the source--lens--observer: 
\begin{equation}
    \tau_D\equiv \frac{D_LD_S}{cD_{LS}}\,,
\end{equation}
where $D_L$, $D_{S}$ and $D_{LS}$ are the angular diameter distances to the lens, the source, and between the lens and source, respectively. 
Therefore, the diffraction integral effectively acts as a path integral, summing over all possible time delays across the lens. 
It will turn out to be useful to rewrite sometimes this integral in terms of the dimensionless frequency and time delay
\begin{equation}
    w \equiv \tau_D \theta_*^2 \omega\,,\qquad T_d \equiv \frac{t_d}{\tau_D \theta_*^2}\,,
\end{equation}
where we have introduced a reference angular scale $\theta_*$. 
This angular scale is lens model dependent. 
For a point mass characterized by a lens mass $M_L$, the typical reference scale is the Einstein angle, which corresponds to
\begin{equation} \label{eq:Einstein_angle_point_mass}
    \theta_E=\sqrt{\frac{4GM_L}{c^2}\frac{(1+z_L)D_{LS}}{D_LD_S}}\,,
\end{equation}
where $z_L$ is the redshift of the lens, and $G$ is the Newton's gravitational constant.
For a singular isothermal sphere described by a velocity dispersion $\sigma$, the commonly defined Einstein angle is
\begin{equation} \label{eq:Einstein_angle_sis}
    \theta_E = 4\pi \left(\frac{\sigma}{c}\right)^2\frac{D_{LS}}{D_S}\,.
\end{equation}
The definition above allows us to define the dimensionless source and image positions
\begin{equation}
    \vec y = \vec{\theta}_\mathrm{S}/\theta_*\,,\qquad \vec x = \vec{\theta}/\theta_*\,,
\end{equation}
to arrive at 
\begin{equation} \label{eq:F_dimensionless}
    F(w,\vec y)=\frac{w}{2\pi i}\int \mathrm{d}^2x e^{iw T_d(\vec x,\vec y)}\,,
\end{equation}
where all quantities are dimensionless. 
The dimensionless time delay surface is also referred as the Fermat potential $\phi$:
\begin{equation} \label{eq:fermat_potential}
    T_d(\vec x,\vec y) =\phi(\vec x,\vec y)= \frac{1}{2}|\vec x -\vec y|^2-\psi(\vec x)\,,
\end{equation}
where the lensing potential needs to be re-scaled $\psi \equiv \Psi /\theta_*^2$. 
By Fermat's principle, images will form at the extrema of the time delay surface~\cite{Blandford:1986zz}. 
Note that the diffraction integral is properly normalized so that when there is no lensing, $\Psi\to0$, then $F\to1$. 
Moreover, because the original lensed signal in time domain is a real function $S_L(t)$, the amplification function satisfies $F(-w)=F^*(w)$. 
Therefore, without loss of generality, we will assume that $w>0$ in the rest of the paper.

The diffraction integral can be solved in different limits. The most relevant one being the \emph{stationary phase approximation}, in which the time delay between the stationary phase points is much larger than the inverse of the frequency, i.e. $w\cdot T_d\gg1$. 
In this limit, the stationary points represent distinct images at locations $\vec x_j$:
\begin{equation} \label{eq:stationary_points}
    \left.\frac{\partial T_d}{\partial \vec x}\right\vert_{\vec x=\vec x_j}=0\,.
\end{equation}
By Taylor expanding the time delay around the stationary points up to quadratic order,
\begin{equation}
    T_d(\vec x) = T_d(\vec x_j) + \frac{1}{2}\frac{\partial^2 T_d}{\partial x_a\partial x_b}(x_a-x_{j,a})(x_b-x_{j,b})\,,
\end{equation}
and solving a two dimensional Gaussian integral, one finds that (see e.g. \cite{Ezquiaga:2020gdt})
\begin{equation}
    F(w)\approx\sum_j \sqrt{|\mu(\vec x_j)|}\exp\left(iw T_d(\vec x_j) - i n_j \pi/2\right)\,.
\end{equation}
Therefore, each image has a different arrival time $T_j$, magnification $\mu_j$, and phase shift $n_j$. 
The magnification comes from the determinant of the Hessian matrix of the time delays
\begin{equation}
    \mu_j= 1/\mathrm{det}\left(T_{ab}(\vec x_j)\right)\,,\qquad T_{ab} \equiv \frac{\partial^2 T_d}{\partial x_a\partial x_b}\,. 
\end{equation}
The phase shift takes integer values of 0, 1, or 2, depending of the type of extrema of the lensing potential (minimum, saddle point or maximum). 
This phase can also be related to the sign of the eigenvalues of $T_{ab}$, which in turn determines the parity of the images as given by the sign of the magnification. When both have the same sign, the image is said to have positive parity. 
When both eigenvalues are positive, $\mathrm{det}(T_{ab})>0$ and $\mathrm{Tr}(T_{ab})>0$, the image is of type I. 
When the eigenvalues have opposite signs, $\mathrm{det}(T_{ab})<0$, the image has negative parity and is of type II. 
When both eigenvalues are negative, $\mathrm{det}(T_{ab})>0$ and $\mathrm{Tr}(T_{ab})<0$, the image is of type III. 

The stationary phase approximation results coincide with those of the \emph{geometric optics} limit, in which the initial wave propagation equation in curved spacetime is expanded for short wavelengths compared to the curvature of the background metric, $\lambda\ll L_B$. 
Geometric optics predicts that each image will propagate along distinct rays that follow null geodesics. 
Conservation of phase-space volume (Liouville's theorem) in curved space-time along a ray bundle determines the magnification of the images~\cite{Misner:1974qy,Schneider:1992}. 
Because of this reason, many authors choose to refer to geometric optics when multiple images are produced in gravitational lensing, to contrast with the \emph{wave optics} regime in which different rays can interfere and diffract with each other, which in the context above translates into solving the diffraction integral without approximations.   

The Hessian matrix can be related to the Jacobian of the mapping between the source position $\vec y$ and the image positions $\vec x_j$. 
When this Jacobian is singular, the magnification diverges. 
The images positions where the transformation becomes degenerate, 
\begin{equation}
    \mathrm{det}(T_{ab})=0\,
\end{equation}
define the \emph{critical curves}. 
These critical curves map back to the \emph{caustics} in the source plane.
The divergence of the magnification highlights that the stationary phase approximation \emph{breaks down} at these critical points. 
In general, the critical curves only appear when the observer is located at some distance from the lens. This defines the \emph{focal length}~\cite{Bontz_Haugan:1981}.

The properties of the images can be constrained from the global topology of the time delay surface~\cite{Blandford:1986zz}. 
If the total mass of a single, transparent thin-lens is finite, and the deflection angle $\vec{\alpha}$ is bounded, then the lens is \emph{regular} and the ``odd-number theorem'' establishes that there is an odd total number of images, $n_{I}+n_{II}+n_{III}=2n+1$~\cite{Burke:1981}. 
As a corollary, the number of even parity images exceed that of odd parity by one, $n_I+n_{II}=n+1$. 
In that case, crossing a critical curve always adds or subtracts 2 images. 
Moreover, under the same assumptions, the image arriving first must come from a global minimum, type I, and it is at least equally magnified, if not more, than the original signal~\cite{Schneider:1984}. 
These theorems are not respected by \emph{singular} lenses, which exhibit a singularity in the lensing potential. 
Examples of singular lenses are black holes and cosmic strings, which differ by the number and parity properties of their images~\cite{Blandford:1986zz}. 
Far from their horizon, in the weak field limit, black holes are well described by point lenses. 
Such model is characterized by having an infinite density at the origin that translates into a critical point. 
Therefore, the focal point is already located in the lens plane making the focal length zero. 
The singularity infinitely de-magnifies the ``odd image''. 
As a consequence, point lenses always produce two images. 
For a non-singular lens, the focal length has to be computed for its particular density profile. 

In the following, we are going to study lensing around caustics. 
We will focus on the \emph{local} images produced around the critical curves, their interference and any diffraction effect. 
To that end, we will expand the time delay surface around the critical curve. 
This local expansion is only valid so long the frequency of the wave times the time delay with the other images is large, $w\cdot T(x_{\rm near-critical-curve})-T(x_{\rm other})\gg 1$. 
In other words, when lensing is \emph{not} in the wave optics regime for the \emph{global} Fermat potential.\footnote{When considering entire lensing scenarios in the wave optics regime, the Uniform Approximation approach can serve to expand around a fold caustic \cite{Serra:2025kbw}.} 
When considering the entire lensing potential under the stationary phase approximation, 
more images are indeed possible, arising from the extrema of the global time delay surface far from the caustic. 
In certain setups, the above theorems can be useful in predicting the existence of additional global images in regular lenses. 

Although the gravitational lensing phenomena described in this section is independent of the type of source, the relevant observables vary. 
For most electromagnetic sources, the only measurable quantities are the images' intensities, $\mathcal{I}\propto |S_L(t)|^2$, which may be individually resolved or blended, and may expand a wide range of angular positions if the source is extended.  
As a consequence, the magnification $|\mu(\vec x_j)|$ is the pertinent number to quantify how much the intensity, and thus the detectability, will increase or decrease due to lensing. 
For GWs, on the other hand, the time domain signal $S_L(t)$ is coherently detected. 
Therefore, both the amplitude and phase of the amplification function $F(\omega)$ are relevant to characterize the detectability. 
In the regime of multiple images, the signal-to-noise ratio of a GW will scale with $\sqrt{|\mu(\vec x_j)|}$. 
The image types can be derived from the phase difference of the signals. 
In the regime of wave optics, the distortions of the full signal need to be accounted to asses the impact on the detectability.  

%-----
%SECTION: POINT
%-----
\section{Point caustic}
\label{sec:point_caustic}

We begin by considering the family of lens systems with axial symmetry. This encompasses all lensing potentials that only depend on radial distance to the center of the lens in the image plane, i.e.
\begin{equation}
    T_d = \frac{1}{2}|\vec x -\vec y|^2-\psi(|\vec x|)\,.
\end{equation}
Because of the symmetry of this system, it will be advantageous to work in polar coordinates: $(x_1,x_2)\to(x,\alpha)$. Moreover, we will always have the freedom to rotate the coordinate system so that the source position $(y_1,y_2)$ is located at $\alpha=0$, being described by a single (impact) parameter $y$. With these choices, the time delay simplifies to 
\begin{equation}
    T_d = \frac{1}{2}\left[(x\cos\alpha-y)^2+x^2\sin^2\alpha\right] - \psi(x)\,.
\end{equation}
In the following we proceed to investigate the properties of this class of lenses for source positions close to its center. 
First we study this in the stationary phase approximation where distinct images are produced. 
Then, we solve the diffraction integral. 
It is important to note that point caustics are in general unstable as they arise solely by the axi-symmetry of the lens. 
Any perturbation in the lensing potential that breaks this symmetry will erase the caustic. 

The wave optics features of axi-symmetric lenses have been studied with great detail~\cite{NakamuraDeguchi,Takahashi:2003ix,Bulashenko:2021fes}, and efficient algorithms have been developed to study them~\cite{Tambalo:2022plm}. 
Extensions of axi-symmetric lenses have also been considered in wave optics; for example adding external shear~\cite{Ulmer:1994ij} or considering a binary lens~\cite{Feldbrugge:2020ycp}. 
Recently, these examples have also been considered in the context of GWs~\cite{Mishra:2021xzz,Yeung:2021chy,Meena:2025gry}. 

\begin{figure*}
    \centering
    \includegraphics[width=\linewidth]{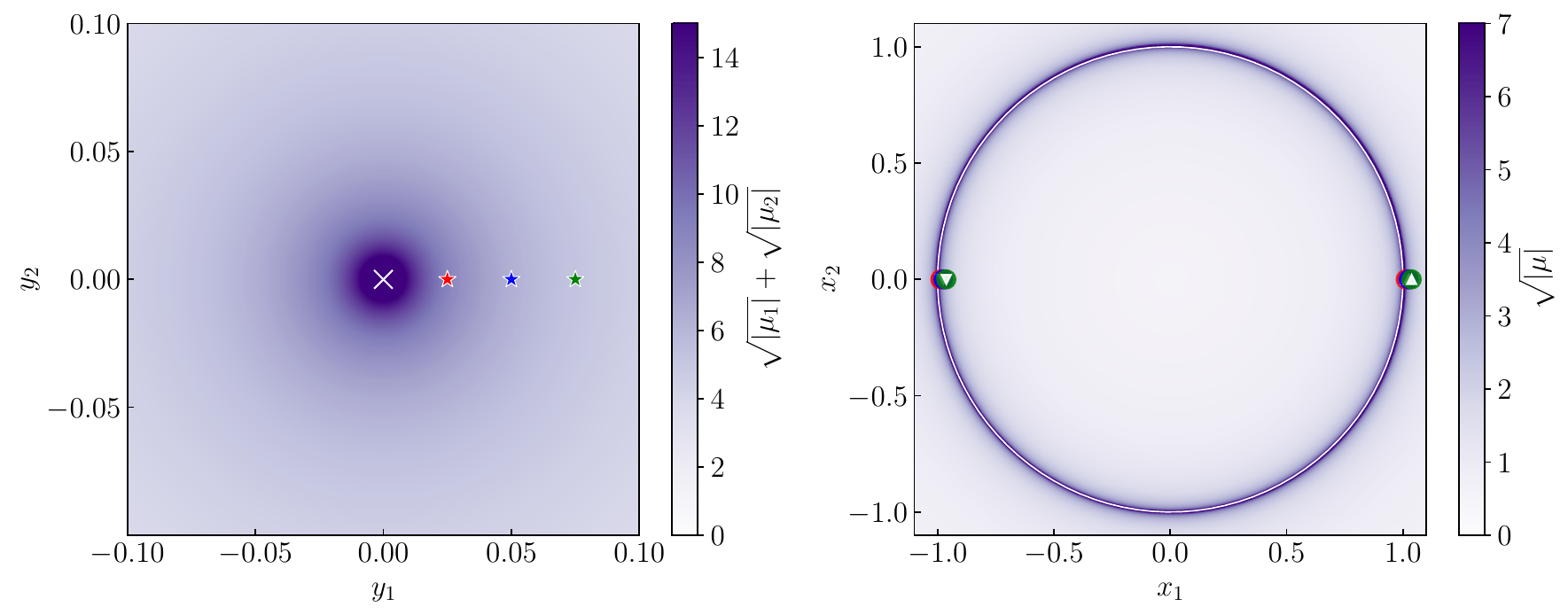}
    \vspace{-20pt}
    \caption{Source (left) and image (right) planes for an axi-symmetric lens close to the central caustic (white cross). The lens is characterized by a critical curve at $x_c=1$ (white circle) and a lensing potential with $\psi(x_c)=0$ and $\psi''(x_c)=-1$. 
    Each different source position (indicated by a different colored star) will produce two images at opposite sides of the center of the lens (indicated by a circle of the same color as its source position), and with opposite parity. 
    The triangle within each circle ($\bigtriangleup/\bigtriangledown$) denotes the positive/negative parity of the image. 
    The color map on the right displays the square root of the magnification $\mu$ at each possible image position. 
    In the source plane, we show the sum of the square root of the magnifications of the images generated at each source position (both are equally amplified).}
    \label{fig:source_image_planes_point_caustic}
\end{figure*}

%-----------
%SPA
%-----------
\subsection{Stationary phase approximation}

As described before, the stationary points of the time delay surface are defined by its extrema. For the axially-symmetric case, this is
\begin{align}
    & \frac{\partial T_d}{\partial \alpha}= xy\sin\alpha =0\,, \\
    & \frac{\partial T_d}{\partial x} = x -y\cos\alpha -\psi'=0\,, 
\end{align}
where primes indicate derivatives with respect to $x$: $\psi'\equiv\partial \Psi/\partial x$. 
For impact parameters $y>0$, the first equation implies that $\alpha=0,\pi$.  
Their radial position is determined by solving the lens equation $x=\psi'-y$. 
In other words, for each solution of the (radial) lens equation $x^{(i)}$, there are two possible image positions aligned with the source, and located equidistantly at opposite sides of the center of the lens. 
In general, these two images will arrive at different times as determined by 
\begin{equation}
    T_{\pm}=\frac{1}{2}(\pm x^{(i)}-y)^2-\psi(x^{(i)})\,.
\end{equation}
The image on the other side of the lens ($-$) will always arrive later.

In the case in which the source, lens, and observer are aligned ($y=0$), then any angular position with critical radial distance
\begin{equation} \label{eq:critical_curve_axisym}
    x_c=\psi'(x_c)
\end{equation}
is a solution. 
Therefore, a point source degenerates into a circle of images or ``Einstein ring''. 
This ring arrives at the observer at the time
\begin{equation}
    T_c = \frac{1}{2}x_c^2-\psi(x_c)\,.
\end{equation}

The magnifications of these images are determined by the inverse Jacobian of the lens mapping
\begin{equation}
\begin{split}
    \mu(x,y) &=1/\mathrm{det}(T_{ab}) = 
    \left|\begin{pmatrix}
        xy\cos\alpha & y\sin\alpha \\
        y\sin\alpha & 1-\psi''(x) 
    \end{pmatrix}\right|^{-1} \\
    & = \frac{1}{y\left(x\cos\alpha (1-\psi''(x))-y\sin^2\alpha\right)}\,.
\end{split}
\end{equation}
From here, it is clear that there is a singularity at the center, when $y\to0$. The point $y=0$ will define a caustic point that maps into a critical curve defined by the circle in (\ref{eq:critical_curve_axisym}). 
Axi-symmetric lenses could exhibit other critical points depending on the functional form of $\psi(x)$, but we will not consider those here. 

To gain further insights into the image properties, we can compute the parity of the images, which is determined by the sign of the magnification. 
The images on opposite sides of the lens, $\alpha=0,\pi$, also have opposite parity 
\begin{equation} \label{eq:mu_pm_point}
    \mu_{\pm}=\pm \frac{1}{y(\psi'-y)(1-\psi'')}.
\end{equation}
The trace of the Hessian at the image position is given by
\begin{equation}
    \mathrm{Tr}(T_{ab}(\vec x_\pm))=1 \pm y(\psi'-y)-\psi''\,.
\end{equation}

%-----------
%SPA expansion
%-----------
\subsection{Expansion around the caustic}

As we have just seen, for axi-symmetric lenses, there is a caustic point at the center of the lens that maps onto a critical curve that forms a circle. As the source position approaches the center, the images of opposite parities on opposite sides of the center approach the critical curve. 
We wish to solve the diffraction integral around this caustic in the next section. 
For that, we first expand the time delay around the critical curve $x_c$ and caustic $y_c=0$ up to quadratic order:
\begin{equation} \label{eq:Td_exp_point}
    \begin{split}
        T_d 
        &= T_{d,c}+\frac{1}{2}y^2+\frac{1}{2}(1-\psi'')x_c^2 \\
        &+\frac{1}{2}(1-\psi'')x^2-x(y\cos\alpha + (1-\psi'')x_c)\,, 
    \end{split}
\end{equation}
where we have defined the time delay at the critical curve $T_{d,c} \equiv x_c^2/2-\psi(x_c)$. 
Note that the first line only contains terms independent of the image positions, while the second line depends on $x$. 
We can now repeat the procedure in the last section and find the properties of the images in the limit of source positions close to the caustic. We find that the two stationary points $x_\pm$ are at 
\begin{equation}
    x_\pm= x_c\pm \frac{1}{(1-\psi'')}y\,,\qquad \cos\alpha_\pm=\pm1\,.
\end{equation}
The time delay between the two images is 
\begin{equation}
    \Delta T = T_{-} - T_{+} = 2x_cy\,.
\end{equation}
Since $\Delta T>0$, the positive parity image always arrives first. 
Therefore, at leading order in $y$ in Eq. (\ref{eq:mu_pm_point}), the magnifications scale inversely with the time delay: 
\begin{equation}
    \mu_\pm = \pm \frac{2}{(1-\psi'')\Delta T}\,.
\end{equation}
The trace of the Hessian matrix will then be
\begin{equation}
    \mathrm{Tr}(T_{ab}(\vec x_\pm))=1-\psi''\pm\frac{1}{2}\Delta T + \frac{y^2}{1-\psi''}\,.
\end{equation}
As a consequence, if $1-\psi''>0$, the the positive parity image will be type I. 
If $1-\psi''<0$, it will be type III. 
This defines all the necessary information to construct the amplification function
\begin{equation}
\label{eq:F_two_images_point}
F(w,y) = \pm\sqrt{\mu_+}e^{i w T_+}\left( 1 + \sqrt{\mu_\mathrm{rel}} e^{i(w \Delta T \mp \pi/2)}\right)\,,
\end{equation}
where at leading order, the relative magnification is $\mu_\mathrm{rel}\approx1$. 
The $+/-$ solutions correspond to the positive parity image being type I/III. 
This results implies that, modulo the overall magnification and phase, the amplification function in the stationary phase approximation for a point singularity only depends on the time delay between the images $\Delta T$. 
This quantity can then be related to the two parameters defining the lensing potential: $\psi''$ and $x_c$. 

The results of the stationary phase approximation for the point caustic are summarized in Fig. \ref{fig:source_image_planes_point_caustic}. 
For an example potential with $\psi(x_c)=0$ and $\psi''(x_c)=-1$, we plot the caustic and critical curve in the source and image plane respectively. 
Then, we choose different source positions and present their corresponding images at each side of the critical curve in opposite angular positions. 
The parity of each image is also indicated with an upwards or downwards arrow. 
In addition, we include a magnification map in the image plane. 
The magnification is narrowly peaking around the Einstein ring, with significant de-magnification ($\mu<1$) at the center.  
To quantify the amplitude of the amplification function $|F|$ in the stationary phase approximation, which is the relevant quantity for GW detectability, we also plot the sum of the square root of the magnifications of the images generated at each possible source position. 
The result is smoothly and monotonically increasing in magnification towards the center where the magnification is maximal. 
Each image contributes equally. 

%-----------
%WO
%-----------
\subsection{Wave optics}

\begin{figure}
    \centering
    \includegraphics[width=\columnwidth]{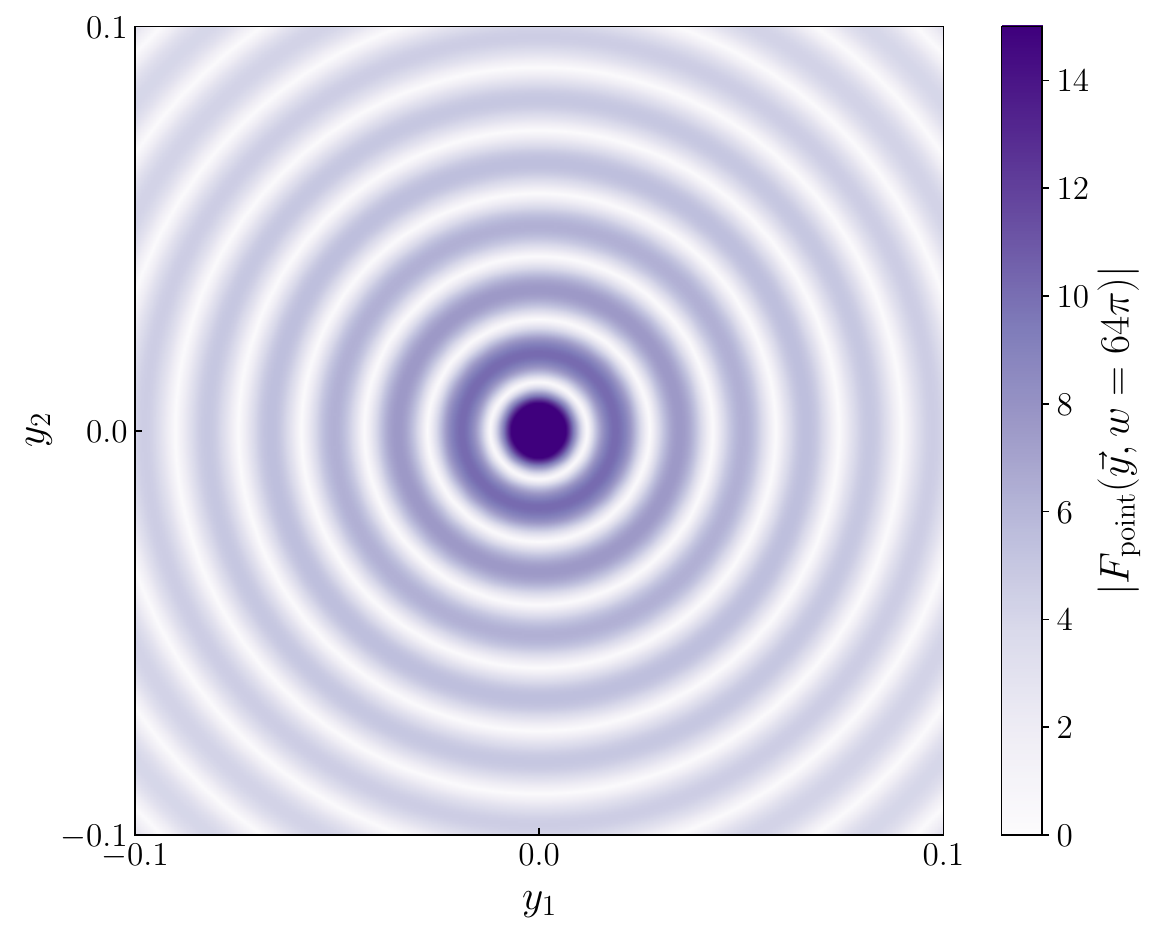}
    \vspace{-20pt}
    \caption{Amplification function $F$ of a point caustic as a function of the dimensionless source positions $(y_1,y_2)$ for a fixed dimensionless frequency $w$ of $64\pi$. 
    The caustic is located at $y=\sqrt{y_1^2+y_2^2}=0$.}
    \label{fig:F_point_y1y2}
\end{figure}

We now proceed to solve the diffraction integral (\ref{eq:F_dimensionless}). 
In polar coordinates, and for the time delay around the caustic (\ref{eq:Td_exp_point}), the amplification function becomes
\begin{equation}
    F=\frac{we^{iwT_d^{(0)}}}{2\pi i}\int_0^{\infty}dx xe^{iwa(x^2-2x_cx)}\int_0^{2\pi}d\alpha e^{-iwyx\cos\alpha}\,,
\end{equation}
where $T_d^{(0)}$ is the part of the time delay independent of $x$, and for shortness, we have defined 
$a\equiv (1-\psi'')/2$. 
The angular part of the system can be solved analytically in terms of a zeroth-order Bessel function $J_0(x)$, 
whose integral definition is given by~\cite{abramowitz+stegun}
\begin{equation}
    J_n(z)=\frac{1}{2\pi i^n}\int_0^{2\pi}e^{iz\cos\phi}e^{in\phi}d\phi\,,
\end{equation}
to arrive at
\begin{equation}
    F=\frac{we^{iwT_d^{(0)}}}{i}\int_0^{\infty}dx xJ_0(wyx)e^{iwa(x^2-2x_cx)}\,.
\end{equation}
To solve the remaining radial integral, we can change variables to be around the critical curve, $z\equiv x-x_c$. 
In these coordinates, the integral becomes
\begin{equation}
    F=\frac{we^{iw(T_d^{(0)}-ax_c^2)}}{i}\int_{-x_c}^{\infty}dz (z+x_c)J_0(wy(z+x_c))e^{iwaz^2}\,.
\end{equation}
If we concentrate on regions very close to the critical curve, $z\ll 1$, then the integral can be approximated to 
\begin{equation} \label{eq:F_point_caustic}
\begin{split}
    F&\approx\frac{we^{iw(T_d^{(0)}-ax_c^2)}}{i}\int_{-\infty}^{\infty}dz (z+x_c)J_0(wy(x_c))e^{iwaz^2} \\
    &= e^{iw(T_{d,c}+\frac{1}{2}y^2)}i^{-1/2}\sqrt{\frac{2\pi w}{(1-\psi'')}}x_cJ_0(wyx_c)\,,
\end{split}
\end{equation}
where in the last equality, we have taken out the term proportional to $z$ that vanished by symmetry, and solved the remaining Gaussian integral.\footnote{Specifically, we have used $\int_{-\infty}^\infty e^{-a(x+b)^2}dx=\sqrt{\pi/a}.$} 
In the final result, we have reverted to the original variables to arrive at our final result. 
The amplitude of the amplification function $|F|$ as a function of the source positions is plotted in Fig. \ref{fig:F_point_y1y2} for a fixed dimensionless frequency. 
Here, it is clear that the symmetry of the system and the maximum magnification occur at $y=0$. 
Moreover, the interference pattern is also apparent. 
The separation of the fringes is determined by the reference frequency. 
We can compare this plot with its stationary phase approximation analog, the map of the sum of the square root of the magnifications for each source position, displayed in the left panel of Fig. \ref{fig:source_image_planes_point_caustic}. 
The contrast is clear, as the map is monotonically increasing towards the center. 

\begin{figure}
    \centering
    \includegraphics[width=\columnwidth]{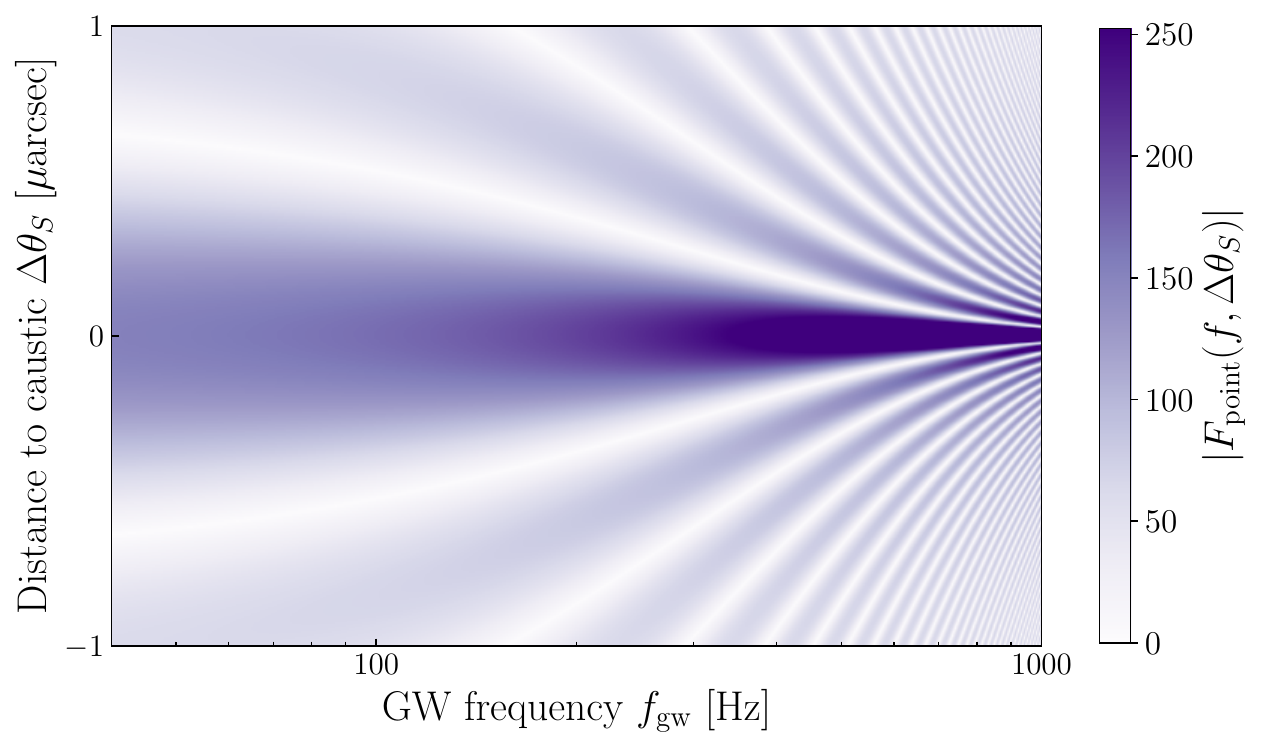}
    \vspace{-20pt}
    \caption{Amplification factor around a point caustic as a function of the distance to the caustic $\Delta\theta_\mathrm{S}$, and the GW frequency $f_\mathrm{gw}$ for a lens at $z_L=0.5$ with mass $M_L=10^6M_\odot$. 
    The diffraction pattern is symmetric around $\Delta \theta_\mathrm{S}=0$.}
    \label{fig:F_point_fy}
\end{figure}

Next, we introduce physical units. For that, we consider a lens whose characteristic angular scale is only characterized by the mass of the lens as in Eq. (\ref{eq:Einstein_angle_point_mass}). 
This corresponds to the Einstein radius of a point mass. 
In this case, the product of the squared angular scale and the time delay distance simplify to $\tau_D\theta_*^2=4GM_L(1+z_L)/c^3$, which is nothing but the dilated Schwarzschild diameter crossing time. 
Focusing on a lens of $10^6M_\odot$, which could correspond to a supermassive black hole at the center of a galaxy, we plot the corresponding diffraction pattern for the relevant frequency range of current ground-based detectors in Fig. \ref{fig:F_point_fy}. 
The pattern is symmetric around the caustic at $\Delta\theta_\mathrm{S}=0$, where $\Delta \theta_\mathrm{S}$ is the distance to the caustic. 

In order to compute the maximum magnification around a point caustic, one can note that $J_0(z)$ has a maximum of 1 at $z=0$, i.e. the center of the lens. 
Therefore, taking the modulus square of $F$ in (\ref{eq:F_point_caustic}), we obtain
\begin{equation}
    \left(\frac{\mu_\mathrm{max}}{10^4}\right)\simeq \frac{x_c^2}{(1-\psi'')}\left(\frac{f}{100\mathrm{Hz}}\right)\left(\frac{(1+z_L)M_L}{10^{6}M_\odot}\right)\,.
\end{equation}
The maximum magnification is limited by the point at which wave optics effects are dominant. This occurs at $\omega\Delta t_d\sim1$. For this case, we find then
\begin{equation}
    \omega\Delta t_d \simeq \frac{1}{(1-\psi'')}\left(\frac{10^4}{|\mu|}\right)\left(\frac{f}{100\mathrm{Hz}}\right)\left(\frac{(1+z_L)M_L}{10^{6}M_\odot}\right)\,.
\end{equation}
So, for typical sources of current ground-based detectors, the signal's strain $h$ could not be amplified by more than $\sim10^2$ ($h_L\propto \sqrt{|\mu|}h$) by a lens of $\sim10^6M_\odot$.

%-----
%SECTION: GENERAL CAUSTICS
%-----
\section{General caustics}
\label{sec:general_caustic}

After considering the special case of axi-symmetric lenses and the appearance of a caustic at its center, we wish to consider caustics in more general setups. 
For that purpose, we consider the leading order corrections around a generic caustic $\{\vec y_c\}$ and critical curve $\{\vec x_c\}$. 
In this process, we choose a convenient coordinate system, and explicitly indicate the components of the source and image positions, i.e. $\vec y = (y_1,y_2)$ and $\vec x = (x_1,x_2)$. 
In particular, we can choose coordinates in which the caustic and critical curve are located at the center of their respective planes, $\vec y_c = 0$ and $\vec x_c = 0$. 
Because the expansion of the time delay will involve higher order derivatives with respect to the image positions, we will indicate them as $g_{i_1\cdots i_{n}}\equiv\partial^n g /\partial x_{i_1}\cdots\partial x_{i_n}$, where $g$ is a generic function. 
For derivatives with respect to the time delay $T_d$, we avoid the subindex $d$ for simplicity, writing the partial derivatives as $T_{i_1\cdots i_{n}}$. 
For example, the critical curve is defined as the image positions $T_1=T_2=0$ where
\begin{equation} \label{eq:detTab}
    D\equiv\mathrm{det}(T_{ab}) = T_{11}T_{22} - T_{12}^2=0\,.
\end{equation}  
The properties of the caustic are determined by the rank of the Hessian matrix $T_{ab}$, which can be either 0 or 1. 
Moreover, the derivatives of the determinant of the Hessian around the critical curve will also be important, in particular the first derivative $D_a$ and second derivative $D_{ab}$. 
For lens mappings of rank 1, 
we can further choose the axes to be parallel so that $T_{ab}$ will be diagonal with a single non-zero eigenvalue. 
Thus in these coordinates, we can set $T_{12}=0$ and $T_{22}=0$. 

There are many classes of critical curves that satisfy the defining condition (\ref{eq:detTab}). 
However, there are only two which are stable in a two dimensional mapping from the source to the image plane, as determined by Whitney's theorem~\cite{Whitney1992}. 
The stable critical curves are defined by the Hessian matrices of rank 1 in which their normal vector $\vec n = \nabla D$ does not vanish. Using the coordinates above, this implies $n_i = -T_{11}T_{22i}\neq0$. 
This is satisfied whenever at least one of $T_{122}$ (which is equal to $T_{221}$) or $T_{222}$ are non-zero. 
We can rotate 90 degrees from the normal vector to obtain the tangent vector: $\vec \chi = T_{11}(-T_{222},T_{122})$. 
By projecting the tangent vector into the Hessian matrix, one finds
\begin{equation}
    T_{ab}\chi^b = - T_{11}^2T_{222}\,.
\end{equation}
Therefore, two qualitatively different behaviors will occur depending on whether $T_{222}$ vanished or not. 
When $T_{222}\neq 0$, this defines a \emph{fold} caustic. 
The opposite defines a \emph{cusp} caustic. 
In passing, let us note that in the special case of axi-symmetric lenses considered in \S\ref{sec:point_caustic}, only fold caustics are possible in addition to the already discussed (unstable) point caustic at its center. 

Folds and cusps have been studied in great detail in the stationary phase approximation. 
The expressions derived below are in agreement with the classical results summarized in Ref.~\cite{Schneider:1992}. 
More recent calculations have also been performed using different methods, e.g. in Refs.~\cite{Gaudi:2001fp,Gaudi:2002hu,Congdon:2008pm}. 

%-----
%SECTION: FOLDS
%-----
\section{Fold caustics}
\label{sec:fold_caustic}

From the above description, a fold caustic is defined by a lens mapping in which the determinant of the Hessian matrix vanishes ($D=0$), its normal vector does not vanish ($\vec n \neq 0$), and the projection of the tangent vector into the Hessian does not vanish either ($T_{222}\neq0$). 
With the notation and coordinate choices in the previous section, the time delay around a fold caustic will be of the form
\begin{equation}
    \begin{split}
        T_d& = T_c + \frac{1}{2}(y_1^2+y_2^2) - x_1y_1 - x_2y_2  + \frac{1}{2}T_{11}x_1^2 \\ 
        +& \frac{1}{6}T_{111}x_1^3 + \frac{1}{2}T_{112}x_1^2x_2 + \frac{1}{2}T_{122}x_1x_2^2 +\frac{1}{6}T_{222}x_2^3\,,
    \end{split}
\end{equation}
up to cubic order. 
In this expression, the derivatives of the time delay surface could take any value, except for $T_{11}$ and $T_{222}$, which we must ensure are non-zero. 
By choosing $T_{122}=0$, we fix the tangent to the critical curve to be parallel to the $x_1$ axis. 
By choosing $T_{111}=T_{112}=0$, we are defining the critical curve as a straight line along the $x_1$ axis. 
Note that this is a restrictive assumption, but one that will become increasingly a better approximation as the source gets closer to the caustic. 
For this reason, we take the following time delay expression as a starting ansatz:
\begin{equation}\label{eq:Td_fold}
    T_d = T_c + \frac{T_{11}}{2}x_1^2-x_1y_1-x_2y_2+\frac{T_{222}}{6}x_2^3\,.
\end{equation}
Note that because of the choices above, the properties of the solutions should be independent of the coordinate $y_1$, and we can parametrize all observables (time delays, magnifications, etc.) in terms of the perpendicular distance to the caustic $y_2$. 

Before we proceed to solve the problem in the stationary phase approximation and wave optics limit, it is important to emphasize that the solutions will only be valid so long as our initial ansatz of zooming in around the caustic is a good approximation. 
The region of validity of this ansatz will vary from lens system to lens system. 
However, close enough to a caustic, the solutions will be valid for any lens, defining a \emph{universal} behavior in the high-magnification limit~\cite{Blandford:1986zz}. 

\subsection{Stationary phase approximation}

\begin{figure*}
    \centering
    \includegraphics[width=\linewidth]{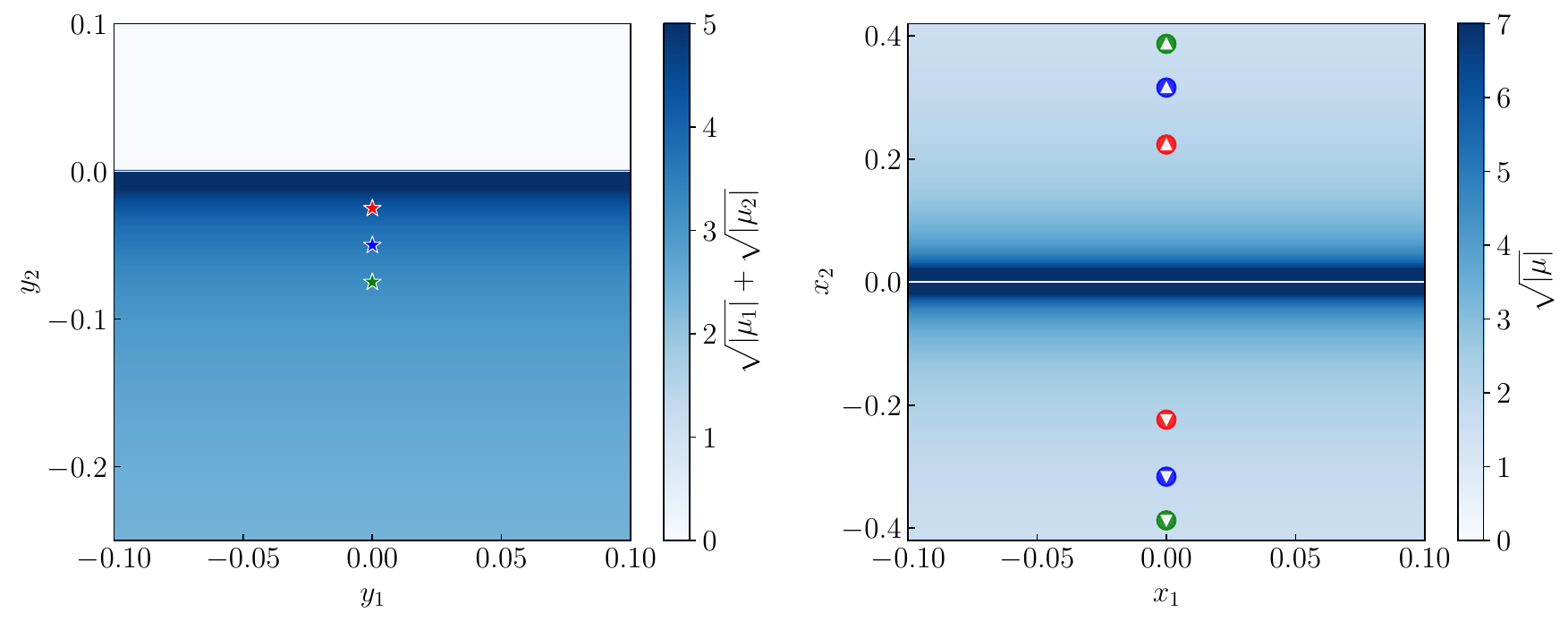}
    \vspace{-20pt}
    \caption{Source (left) and image (right) planes for a fold caustic (dark line). The lens is characterized by  $T_{11}(x_c)=1$ and $T_{222}(x_c)=-1$. 
    Each different source position (indicated by different colored stars) will produce two images (indicated by a circle of the same color as the source position) at opposite sides of the critical curve (white line). 
    The triangle within each circle ($\bigtriangleup/\bigtriangledown$) denotes the positive/negative parity of the image.
    The color map on the right displays the square root of the magnification $\mu$ at each possible image position. 
    In the source plane, we show the sum of the square root of the magnification of the images generated at each source position. Two equally amplified images are only generated when $y_2<0$.}
    \label{fig:source_image_planes_fold_caustic}
\end{figure*}

We begin by studying the properties of the stationary points of (\ref{eq:Td_fold}). 
The image positions can be obtained by solving the lens equation, which defines a quadratic equation with two solutions:
\begin{align}
    x_1 &= \frac{y_1}{T_{11}}\,, \\
    x_2 &= \pm \sqrt{\frac{2y_2}{T_{222}}}\,.
\end{align}
In the following, without loss of generality, we assume $T_{222}<0$. 
Therefore, the two images only exist for $y_2<0$, and they form on each side of the critical curve. When $y_2>0$ no images are formed. 
Note that the region with higher image multiplicity always corresponds to the area ``inside'' the caustic, or closer to the center of the lens. 
With the image positions at hand, it is easy to compute the magnifications of the images
\begin{equation}
    \mu_\pm=\pm\frac{1}{T_{11}\sqrt{2T_{222}y_2}}\,.
\end{equation}
This expression is coordinate system invariant and only depends on the perpendicular distance to the caustic $y_2$ as anticipated. 
The absolute value of the magnification is the same for both images, and their parities are opposite. 
If $T_{11}>0$, the $\vec x_+$ image will have positive parity. The trace of the Hessian matrix for these images is given by
\begin{equation}
    \mathrm{Tr}\left(T_{ab}(\vec x_\pm)\right)=T_{11}\pm \sqrt{2T_{222}y_2}\,.
\end{equation}
Therefore, if $T_{11}>0$, the positive parity image $\vec x_+$ will be type I. 
If $T_{11}<0$, the positive parity image $\vec x_-$ will be type III. 
Finally, substituting the image position into the time delay we can obtain
\begin{equation}
    T_{\pm}= T_c \mp\frac{2\sqrt{2}}{3}\frac{y_2^{3/2}}{|T_{222}|^{1/2}}-\frac{y_1^2}{2T_{11}}\,.
\end{equation}
This means that the time delay between the images only depends on $y_2$:
\begin{equation}
    \Delta T =T_--T_+= \frac{4\sqrt{2}}{3}\frac{|y_2|^{3/2}}{|T_{222}|^{1/2}}\,.
\end{equation}
We can therefore relate the magnification to the time delay by:
\begin{equation}
    |\mu|=\frac{2^{1/3}}{3^{1/3}|T_{11}||T_{222}|^{2/3}}\frac{1}{\Delta T^{1/3}}\,.
\end{equation}
The final ingredient to obtain the amplification function in the stationary phase approximation is to note that because the images form at each side of the caustic, they have opposite parities. 
Putting everything together, we find
\begin{equation} \label{eq:F_two_images}
    F=\pm\sqrt{|\mu|}e^{iwT_0}\left(1+e^{i(w\Delta T \mp\pi/2)}\right)\,,
\end{equation}
where we have defined the global time delay $T_0\equiv T_c-y_1^2/2T_{11}$. 
The $+/-$ correspond to the positive parity image being type I/III. 
This expression is formally similar to the point caustic. 
However, the scaling of the magnification with the time delay is now $1/\Delta T ^{1/3}$ compared to $1/\Delta T$. 
When marginalizing over the overall amplitude and phase, the fold caustic is defined by a single parameter $\Delta T$. 
This observable can then be related to the derivatives of the lensing potential defining the fold: $T_{11}$ and $T_{222}$. 
For the case of a type III--II configuration, the theorems described at the end of \S\ref{sec:caustic_intro} imply that the global time delay should contain at least one additional type I image arriving first. 
In general, the ``odd-number theorem'' implies that if a fold is realized on a regular lens, an additional image should be present. 

The stationary phase approximation phenomenology for a fold caustic is summarized in Fig. \ref{fig:source_image_planes_fold_caustic}. 
We choose three different source positions and present their corresponding images produced at both sides of the critical curve. 
The source positions are within the caustic, $y_2>0$, as otherwise no images would form. 
The images are equidistant to the critical curve and have opposite parities (indicated by the upper/lower arrow). 
They also have equal magnifications. 
To show this last point more explicitly,
we plot the square root of the absolute value of the magnification of the images in the image plane. 
The magnification only changes in the axis perpendicular to the critical curve. 
To translate this magnification map into the source plane, we plot the sum of the square root of the magnifications of the two images generated at each source position. 
Above the caustic, no images are formed. 
Below the caustic, the magnification map grows continuously towards its maximum at $y_2=0$. 
This color map quantifies the amplitude of the two images contributing to the amplification function in the stationary phase approximation, cf. (\ref{eq:F_two_images}).

\subsection{Wave optics}

The diffraction integral of the fold caustic can be solved analytically starting from the time delay ansatz (\ref{eq:Td_fold}). 
We first decompose the integral along the different image plane axes: 
\begin{equation}
    \begin{split}
        F=\frac{we^{iwT_c}}{2\pi i}&\int_{-\infty}^\infty dx_2 e^{iw\left(\frac{T_{222}}{6}x_2^3-y_2x_2\right)} \\
        &\int_{-\infty}^\infty dx_1 e^{iw\left(\frac{T_{11}}{2}x_1^2-y_1x_1\right)}\,.
    \end{split}
\end{equation}
The integral in $x_1$ can be solved using a Gaussian integral.\footnote{Specifically, we use $\int_{-\infty}^\infty e^{-(ax^2+bx+c)}=\sqrt{\pi/a}e^{\frac{b^2}{4a}-c}$.} 
We obtain
\begin{equation}
    \begin{split}
        F=\frac{w^{1/2}e^{iwT_0}}{2^{1/2}\pi^{1/2} i^{5/2}T_{11}^{1/2}}&\int_{-\infty}^\infty dx_2 e^{iw\left(\frac{T_{222}}{6}x_2^3-y_2x_2\right)}\,,
    \end{split}
\end{equation}
where, again, we have introduced the global time delay of the signal $T_0$ as in the case of the stationary phase approximation. 
With a simple change of variables, the integral in $x_2$ can be solved analytically in terms of an Airy function, 
whose integral form definition is~\cite{abramowitz+stegun}
\begin{equation}
    \mathrm{Ai}(z)= \frac{1}{2\pi}\int_{-\infty}^\infty e^{i(t^3/3+zt)}dt\,,
\end{equation}
to arrive at the final expression
\begin{equation} \label{eq:F_fold}
    F = \frac{2^{5/6}\pi^{1/2}w^{1/6}e^{iwT_0}}{i^{5/2}T_{11}^{1/2}|T_{222}|^{1/3}}\mathrm{Ai}\left[\frac{2^{1/3}y_2w^{2/3}}{|T_{222}|^{1/3}}\right]\,,
\end{equation} 
where we are continuing with our assumption that $T_{222}<0$. 
The Airy function has well known asymptotic limits.  
At large positive values the amplitude decays exponentially, while at large negative values it oscillates sinusoidally.\footnote{Following 10.4.59-60 in Ref.~\cite{abramowitz+stegun}, at large $|z|$, the leading term of the expansion at positive values is $\mathrm{Ai}(z)\sim z^{-1/4}e^{-\frac{2}{3}z^{3/2}}$, and at negative values is $\mathrm{Ai}(-z)\sim z^{-1/4}\sin\left(\frac{2}{3}z^{3/2}+\frac{\pi}{4}\right)$.} 
The spacing of the fringes can then be derived from the period, $\Delta w= 3\pi\, 2^{-3/2}y_2^{-3/2}/|T_{222}|^{-1/2}$, as pointed out in the appendix of Ref.~\cite{Ulmer:1994ij}.

\begin{figure}
    \centering
    \includegraphics[width=\columnwidth]{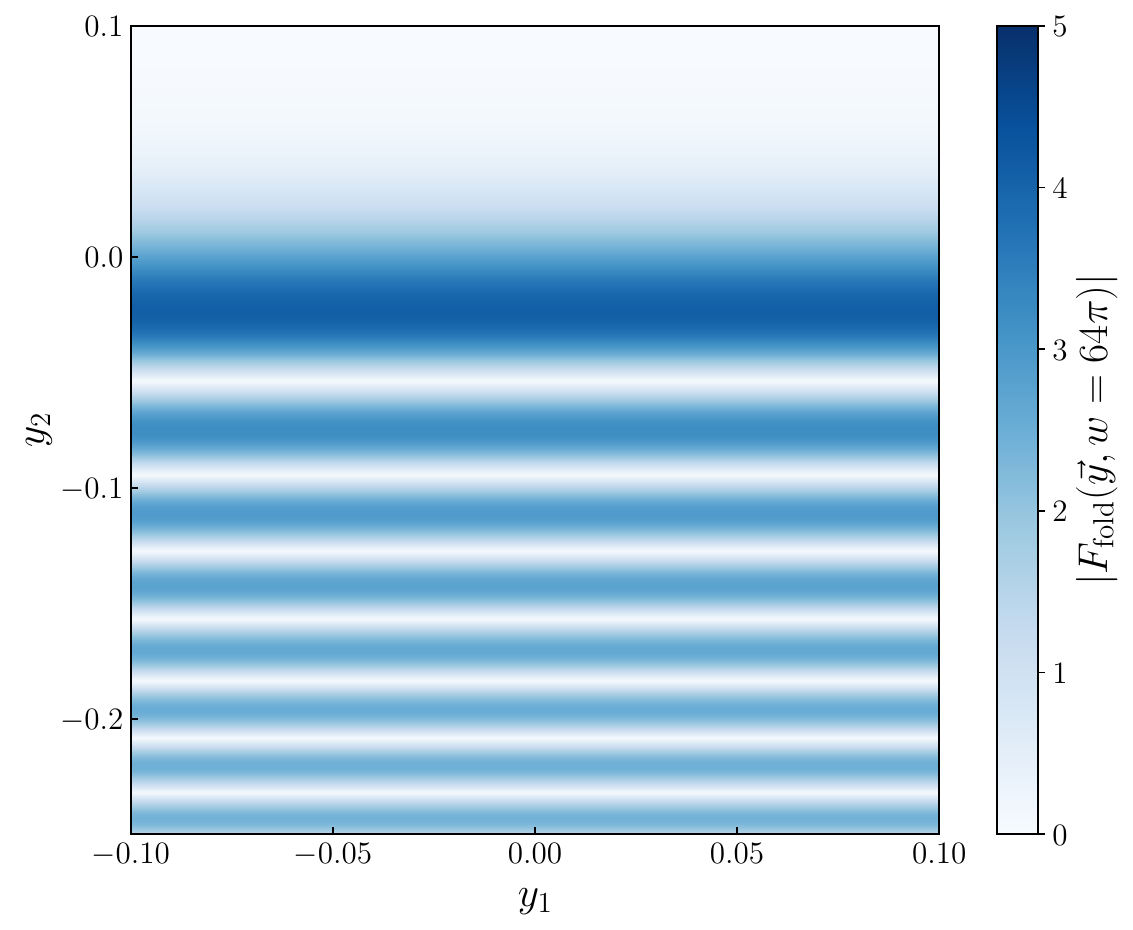}
    \vspace{-20pt}
    \caption{Amplification function $F$ of a fold caustic as a function of the dimensionless source positions $(y_1,y_2)$ for a fixed dimensionless frequency $w$ of $64\pi$. 
    The caustic is located along the $y_2=0$ axis.}
    \label{fig:F_fold_y1y2}
\end{figure}

The amplitude of the amplification function as a function of the source positions for a fixed dimensionless frequency is presented in Fig. \ref{fig:F_fold_y1y2}. 
This is the wave optics counterpart of Fig. \ref{fig:source_image_planes_fold_caustic}. 
On one side of the caustic, one can see that $|F|$ is highly oscillatory (representing two local interfering images), and on the other, $|F|\to0$ (representing a limiting solution with no local images). 
The size of the fringes depends on the (dimensionless) frequency of the wave, which for this example, we have fixed to $64\pi$. 
Interestingly, the transition from the 2-image to the 0-image regions is smooth in wave optics. For sources close to and above the caustic, there is still a region with amplification. 
Moreover, it is also evident from this plot that the maximum magnification does not occur at the caustic, $y_2=0$, but rather a bit inside. 
This contrasts with the source-plane magnification map in the left panel of Fig. \ref{fig:source_image_planes_fold_caustic}.

\begin{figure}
    \centering
    \includegraphics[width=\columnwidth]{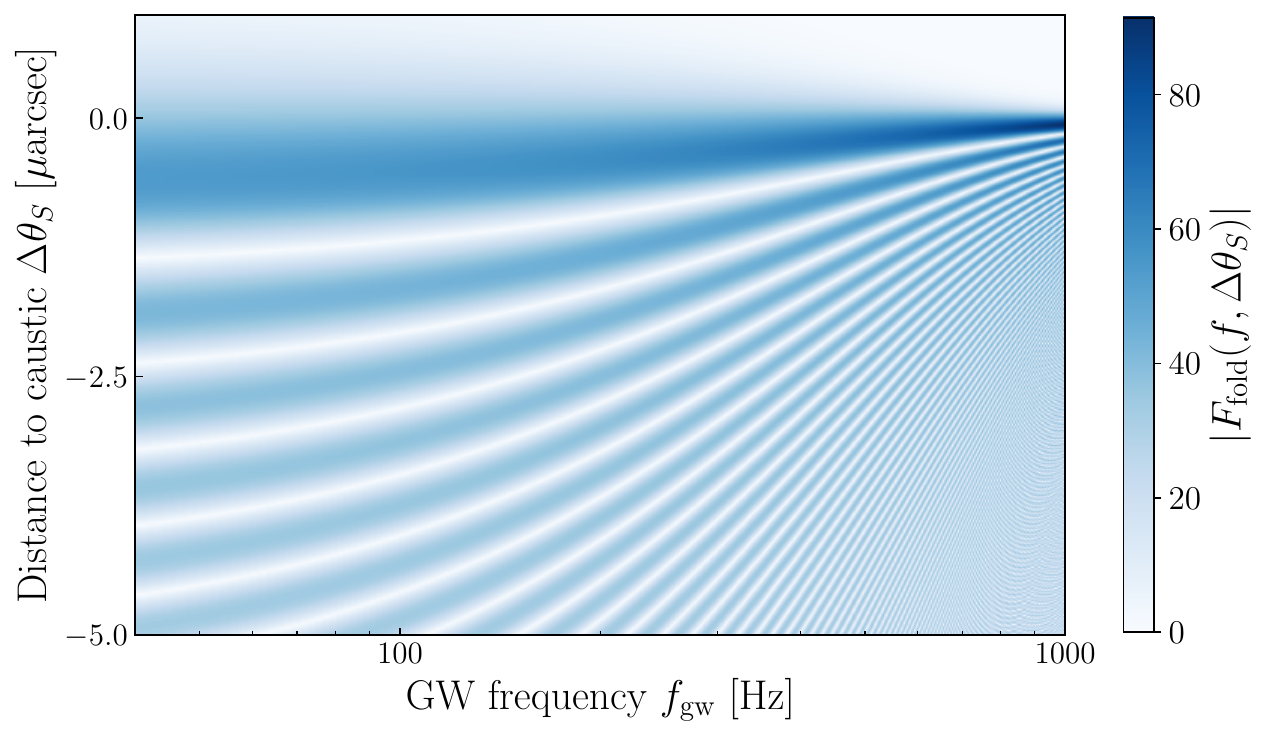}
    \vspace{-20pt}
    \caption{Amplification factor around a fold caustic as a function of the distance to the caustic $\Delta\theta_\mathrm{S}$ and the GW frequency $f_\mathrm{gw}$ for a lens at $z_L=0.5$ with velocity dispersion $\sigma=200$km/s and a source at $z_S=2$. 
    Negative $\Delta\theta_\mathrm{S}$ means getting closer to the center of the lens or, in other words, represents the side of the caustic leading to higher multiplicity. 
    }
    \label{fig:F_fold_fy2}
\end{figure}

We also examine the case of a fold caustic in a typical galaxy-scale lens with velocity dispersion $\sigma=200$km/s in Fig. \ref{fig:F_fold_fy2}. 
In this case, we plot the amplitude of the amplification function as a function of the GW frequency (in Hz), and the perpendicular distance of the source to the caustic with physical units, i.e. $\Delta \theta_\mathrm{S}=y_2*\theta_*$, where $\theta_*$ is the relevant angular scale of the lens. 
In this example, we take the simplifying assumption that $\theta_*$ follows the behavior of the singular isothermal sphere model, cf. (\ref{eq:Einstein_angle_sis}). 
We particularize for GW frequencies relevant for ground-based detectors ranging from 10 to 1000 Hz. 
We observe that when the source is micro arc-seconds away from the caustic, the absolute value of $F$ can reach $\sim80$. 
For a typical galaxy lens, at those distances from the caustic, the leading terms of the expansion of the time delay surface considered here, cf. (\ref{eq:Td_fold}), are enough to describe the fold~\cite{Lo:2024wqm}. 

The maximum magnification, $\mu_\mathrm{max}=\mathrm{max}(|F|^2)$, can be computed taking into account that the maximum of the Airy function is at $z\simeq 0.54$, thus leading to
\begin{equation} \label{eq:maximum_magnification_fold}
\begin{split}
    \left(\frac{\mu_\mathrm{max}}{10^3}\right) \simeq &\frac{3(1+z_L)^{1/3}}{|T_{222}|^{2/3}|T_{11}|}\left(\frac{f}{100\mathrm{Hz}}\right)^{1/3} \\
    &\left(\frac{c\tau_{\mathrm{D}}}{1\mathrm{Gpc}}\right)^{1/3}\left(\frac{\sigma}{200\mathrm{km\,s}^{-1}}\right)^{4/3}\,,
\end{split}
\end{equation}
where $c\tau_{\mathrm{D}}$ is the typical distance between the source, lens and observer, assuming that $D_{LS}\approx D_L$.  
The maximum extent to which a signal can be magnified is in turn related to whether or not diffraction is taking place. 
Wave optics is relevant when $\omega\Delta t_d\sim1$, which in terms of the parameters of the problem translates into:
\begin{equation}
\begin{split}
    \omega\Delta t_d \simeq &\frac{(1+z_L)}{|T_{222}|^2|T_{11}|^3}\left(\frac{10^3}{|\mu|}\right)^3 \left(\frac{f}{100\mathrm{Hz}}\right) \\
    &\left(\frac{c\tau_{\mathrm{D}}}{1\mathrm{Gpc}}\right)\left(\frac{\sigma}{200\mathrm{km\,s}^{-1}}\right)^4\,.
\end{split}
\end{equation}
This implies that for the type of compact binary coalescences that we observe with current detectors, we should not expect the signal's strain to be amplified by more than $\sqrt{1000}\sim30$ if the lens is a typical galaxy.

%-----
%SECTION: CUSPS
%-----
\section{Cusp caustics}
\label{sec:cusp_caustic}

As discussed in \S\ref{sec:general_caustic}, in addition to the fold, the other stable caustics in a two dimensional mapping are cusps. 
These catastrophes are characterized (in the coordinate system used before) by $T_{11}\neq0$ and $T_{222}=0$. 
This implies that we need to include quartic derivatives in the $x_2$ direction, i.e. $T_{2222}\neq0$. 
The time delay expanded around the cusp should include
\begin{equation}\label{eq:Td_cusp}
\begin{split}
    T_d &= T_c + \frac{T_{11}}{2}x_1^2-\vec x\vec y
    +\frac{1}{2}T_{122}x_1x_2^2+\frac{1}{24}T_{2222}x_2^4\,,
\end{split}
\end{equation}
where we have neglected the term proportional to $T_{112}$ because it is negligible compared to the center when approaching to the cusp at $\vec x\to0$. 

\begin{figure*}
    \centering
    \includegraphics[width=\linewidth]{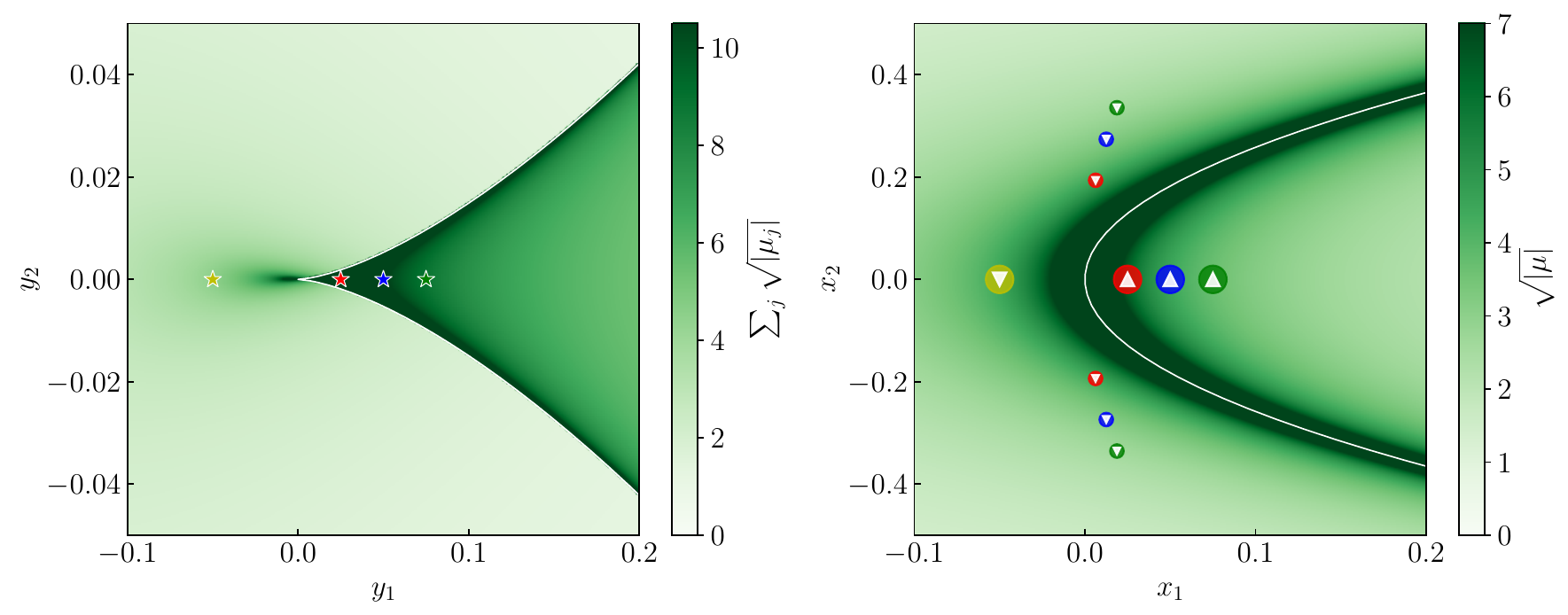}
    \vspace{-20pt}
    \caption{Source (left) and image (right) planes for a cusp caustic (white line). The lens is characterized by  $T_{11}(x_c)=1$, $T_{122}(x_c)=1$, and $T_{2222}(x_c)=-1$. 
    Each different source position (indicated by a different colored star) will produce either one or three images (indicated by circles of the same color as the source position) that can be inside or outside of the critical curve (white line). 
    The triangle within each circle ($\bigtriangleup/\bigtriangledown$) denotes the positive/negative parity of the image. 
    The relative size of the circles scales with the relative magnification. 
    The color map displays the square root of the magnification $\mu$ at each possible image position. 
    In the source plane, we show the sum of the square root of the magnification of the images generated at each source position. Within the caustic, there are three images, and only one outside.}
    \label{fig:source_image_planes_cusp_caustic}
\end{figure*}

\subsection{Stationary phase approximation}

Because of the quartic order of the time delay, the image positions are obtained solving the system of equations
\begin{align}
    &\frac{1}{2}\left( \frac{T_{2222}}{3}-\frac{T_{122}^2}{T_{11}}\right)x_2^3+\frac{T_{122}y_1}{T_{11}}x_2-y_2=0\,, \\
    & x_1 = -\frac{T_{122}}{2T_{11}}x_2^2+\frac{y_1}{T_{11}}\,,
\end{align}
which are of cubic order for $x_2$. Therefore, a cusp could in general have regions with 3 images, characterized by 3 real roots, and 1 image, characterized by 1 real root and two complex roots. 
The properties of the roots of a cubic equation are determined by whether the discriminant $\Delta$ of the equation\footnote{For a cubic equation of the form $a x^3 + cy+d=0$, the discriminant simplifies to $\Delta = -4ac^3-27a^2d^2$.}
is positive (3 real roots) or negative (1 real root). 
Therefore, $\Delta=0$ defines the source positions in which the image multiplicity changes, i.e. the caustic. 
It will be convenient to define the parameter
\begin{equation} \label{eq:gamma}
    \gamma\equiv \frac{T_{122}^2}{T_{11}} - \frac{T_{2222}}{3}. 
\end{equation}
With this notation, the caustic is defined by 
\begin{equation} \label{eq:cusp_caustic}
    y_1^3=\frac{27\gamma}{8}\frac{T_{11}^3}{T_{122}^3} y_2^{2}\,.
\end{equation}
Without loss of generality, we will choose the convention that the caustic is always contained in the $y_1>0$ region. 
Therefore, if $\gamma >0$, then $T_{122}/T_{11}>0$, and vice versa. 
As a consequence, there will always be only one real root for $y_1<0$. For regions between caustics, $y_2<\left(8T_{122}^3y_1^3/(27\gamma T_{11}^3)\right)^{1/2}$, there will be three images. 
We plot a representative example of a cusp caustic in the left panel of Fig. \ref{fig:source_image_planes_cusp_caustic}. 

On the other hand, the critical curves are more easily obtained from the divergence of the magnification
\begin{equation}
    \mu = \frac{1}{\left(T_{11}T_{2222}/2-T_{122}^2\right)x_2^2+T_{11}T_{122}x_1}\,,
\end{equation}
which defines the parabola 
\begin{equation} \label{eq:critical_curve_cusp}
    x_1 = \left(\frac{T_{122}}{T_{11}}-\frac{T_{2222}}{2T_{122}}\right)x_2^2\,.
\end{equation}
There are two different qualitative behaviors for the critical curve. 
When the term in parentheses in (\ref{eq:critical_curve_cusp}) is negative, $2T_{122}^2>T_{11}T_{2222}$, the critical curve will be contained in the $x_1>0$ region. 
Conversely, when $2T_{122}^2<T_{11}T_{2222}$, it will be located in the $x_1<0$ region. 
Moreover, when $2T_{122}^2=T_{11}T_{2222}$, the critical curve becomes a line at $x_1=0$.  
The sign of the curvature of the parabola will also have implications for the parity of the images. 
If $\gamma>0$, when $2T_{122}^2>T_{11}T_{2222}$, two negative parity images will appear outside of the critical curve, and one positive parity image will be created inside, cf. Fig.~\ref{fig:source_image_planes_cusp_caustic}. When $\gamma>0$ and $2T_{122}^2<T_{11}T_{2222}$, one positive parity image appears outside and two negative parity images are created inside. 
If $\gamma<0$, the same behavior will occur for opposite values of the sign of the parabola. 

It will be interesting to later consider the case in which the source positions are located along the symmetry axis of the cusp caustic, i.e. $y_2=0$. 
In that case, there will be always an image along the $x_1$-axis:
\begin{equation}
    x_1^{(0)}=y_1/T_{11}\,,\qquad x_2^{(0)}=0\,.
\end{equation} 
The other two possible images exist when $\gamma>0$ and $y_1>0$ or $\gamma<0$ and $y_1>0$. 
They are located at
\begin{equation}
    x_{1}^{(1,2)}=\left(T_{11} - \frac{T_{122}^2}{\gamma}\right)\frac{y_1}{T_{11}^2}\,, 
    \quad x_{2}^{(1,2)}=\pm\sqrt{\frac{2T_{122} y_1}{\gamma T_{11}}}\,.
\end{equation}
The image in the $x_1$-axis always arrives first for $\gamma>0$ with a time delay of 
\begin{equation}
    T_d(\vec x^{(0)})=T_c-\frac{1}{2}\frac{y_1^2}{T_{11}}\,.
\end{equation}
The other two images, when present, arrive at the same time
\begin{equation}
    \Delta T = T_d(\vec x^{(1,2)})-T_d(\vec x^{(0)})= \frac{1}{2}\frac{T_{122}^2y_1^2}{T_{11}^2\gamma}\,.
\end{equation}
In this symmetric case, their magnifications are given by:
\begin{equation} \label{eq:magnification_cusp_symmetric}
    \mu^{(0)}=-2\mu^{(1,2)}=\frac{1}{T_{122}y_1}=\frac{T_{11}}{\sqrt{2\gamma\Delta T}}\,.
\end{equation}
Therefore, the image at $x_2=0$ has opposite parity to the other pair of images that have the same parity and absolute magnification. 
For the image with $x_2=0$, 
the trace of the Hessian matrix $T_{ab}$ is
\begin{equation}
    \mathrm{Tr}\left(T_{ab}^{(0)}\right)=T_{11}+\frac{T_{122}}{T_{11}}y_1\,,
\end{equation}
while the other pair of images have
\begin{equation}
    \mathrm{Tr}\left(T_{ab}^{(1,2)}\right)=T_{11}-2\frac{T_{122}}{T_{11}}y_1+2\frac{T_{122}^3}{T_{11}^2\gamma}y_1\,.
\end{equation}
Therefore, for $\gamma>0$ and $y_1>0$, if $T_{11},T_{122}>0$, then $\vec x^{(0)}$ will be a type I image part of a triplet with a pair of type II images. 
For $y_1<0$, $\vec x^{(0)}$ will be a single type II image. 
Otherwise, if $T_{11},T_{122}<0$, for $y_1>0$, $\vec x^{(0)}$ will be a negative parity image in a triplet with two type III images. 
For $y_1<0$, $\vec x^{(0)}$ will be a single type III image. 
If $\gamma<0$ and $y_1>0$, $\vec x^{(0)}$ will be a type III image in a triplet (with two type II images) if $T_{11}<0$, otherwise it will be a type II image in a triplet (with two type I images). 
If $\gamma<0$ and $y_1<0$, it will be a single type II image if $T_{11}<0$, otherwise it will be a single type I image. 
For this setup, as one approaches the cusp, $y_1\to0$, all images arrive at the same time with a divergent magnification. 
Interestingly, even for source positions in the one-image region, the single image can be highly magnified. 

With all this information about the image properties, we can construct the amplification function in the three-image region. 
Focusing on source positions with $y_2=0$, in our convention, the multiple image region corresponds to $y_1>0$. When there is one positive parity image and two negative parity images, the amplification function can be written as
\begin{equation} \label{eq:F_three_image_one_positive}
    F = \pm\sqrt{|\mu_0|}e^{iwT_0}\left(1+\frac{2}{\sqrt{2}}e^{i(w\Delta T\mp\pi/2)}\right)\,,
\end{equation}
where the $\pm$ corresponds to the cases where the positive parity image is type I and III. 
When there are two positive parity images, the amplification function becomes 
\begin{equation} \label{eq:F_three_image_one_negative}
    F = \sqrt{|\mu_0|}e^{i(wT_0-\pi/2)}\left(1+\frac{2}{\sqrt{2}}e^{i(w\Delta T\pm\pi/2)}\right)\,,
\end{equation}
where the $\pm$ correspond to the pair of positive parity images being type I and III. 
Note that because in the above solutions we are restricting ourselves to source positions with $y_2=0$, the additional pair of images arrives at the same time. 
Therefore, the amplification function behaves effectively as a two-image model. 
When the source positions depart from $y_2=0$, this symmetry is broken, cf. Fig.~\ref{fig:xaust_angles}. 
In fact, as one approaches the caustic from the interior ($y_1>0$), the solution tends to that of the pair of images of a fold, with an extra image. 
For the cases with a type III--II--II or II--I--I configuration, the theorems described at the end of \S\ref{sec:caustic_intro} imply that the global time delay should contain at least two additional images: one type I image arriving first and an extra image to make the total number odd. 
If the cusp has two odd-parity images, then the two additional global images should have even parity. 
In the other case, the additional images should have opposite parities. 

We summarize the behavior of the images generated near a cusp in Fig. \ref{fig:source_image_planes_cusp_caustic}. 
We consider three different source positions within the cusp along its symmetry axis. 
Each of those produces three images: two equidistantly above and below the critical curve, and the other in the symmetry axis. 
The two images outside the critical curve have opposite parities to the image inside, as indicated by the direction of the arrow at the image positions. 
The size of the markers indicates the relative magnifications between the images. 
The interior image arrives first, and is brighter. 
As an example, we also show a source position outside of the cusp, which produces a single image. 
The image plane is colored according to the square root of the magnification of the images. 
Magnifications are maximal along the critical curve. 
To complement this information, we plot the sum of the square root of the magnifications of the images generated at each source position.  
This serves as a proxy for the absolute value of the amplification function $|F|$ in (\ref{eq:F_three_image_one_positive}). 
As mentioned before, it is interesting that in the single-image region just outside the cusp it is possible to have large magnifications. 
This is something that does not occur for the point and fold caustics, where high magnification occurs for a pair of images. 

It is important to emphasize that when approaching the cusp through different trajectories, the properties of the images will be different than those along its symmetry axis, cf. Fig.~\ref{fig:xaust_angles}. 
In fact, depending on the locations, the system could resemble more a fold with an extra image, see App. \ref{app:caustic_angles} for further details. 
Interestingly, in all cases, it is satisfied that the image within the cusp, $\vec{x}^{(0)}$, has an absolute magnification equal to the absolute value of the sum of the images outside, $\vec{x}^{(1,2)}$, which have the same parity~\cite{Schneider_cusp_1992}, i.e.
\begin{equation} \label{eq:cusp_relative_magnifications}
    |\mu_0|=|\mu_1|+|\mu_2|\,.
\end{equation}
Note that this relation is satisfied in the symmetric case of Eq. (\ref{eq:magnification_cusp_symmetric}). 
Having this relation is advantageous because we can then write the amplification function as
\begin{equation}
    F\propto 1 + \sqrt{|\mu_\mathrm{rel}|}e^{i(w\Delta T_{10}\mp\pi/2)}+ \sqrt{1-|\mu_\mathrm{rel}|}e^{i(w\Delta T_{20}\mp\pi/2)}\,,
\end{equation}
where $\mu_\mathrm{rel}=\mu^{(1)}/\mu^{(0)}$. 
Therefore, there are three variables $\mu_\mathrm{rel}$, $\Delta T_{10}$ and $\Delta T_{20}$ that can be related to the derivatives of the potential defining the cusp: $T_{11}$, $T_{122}$ and $T_{2222}$. 

\subsection{Wave optics}

\begin{figure}
    \centering
    \includegraphics[width=\columnwidth]{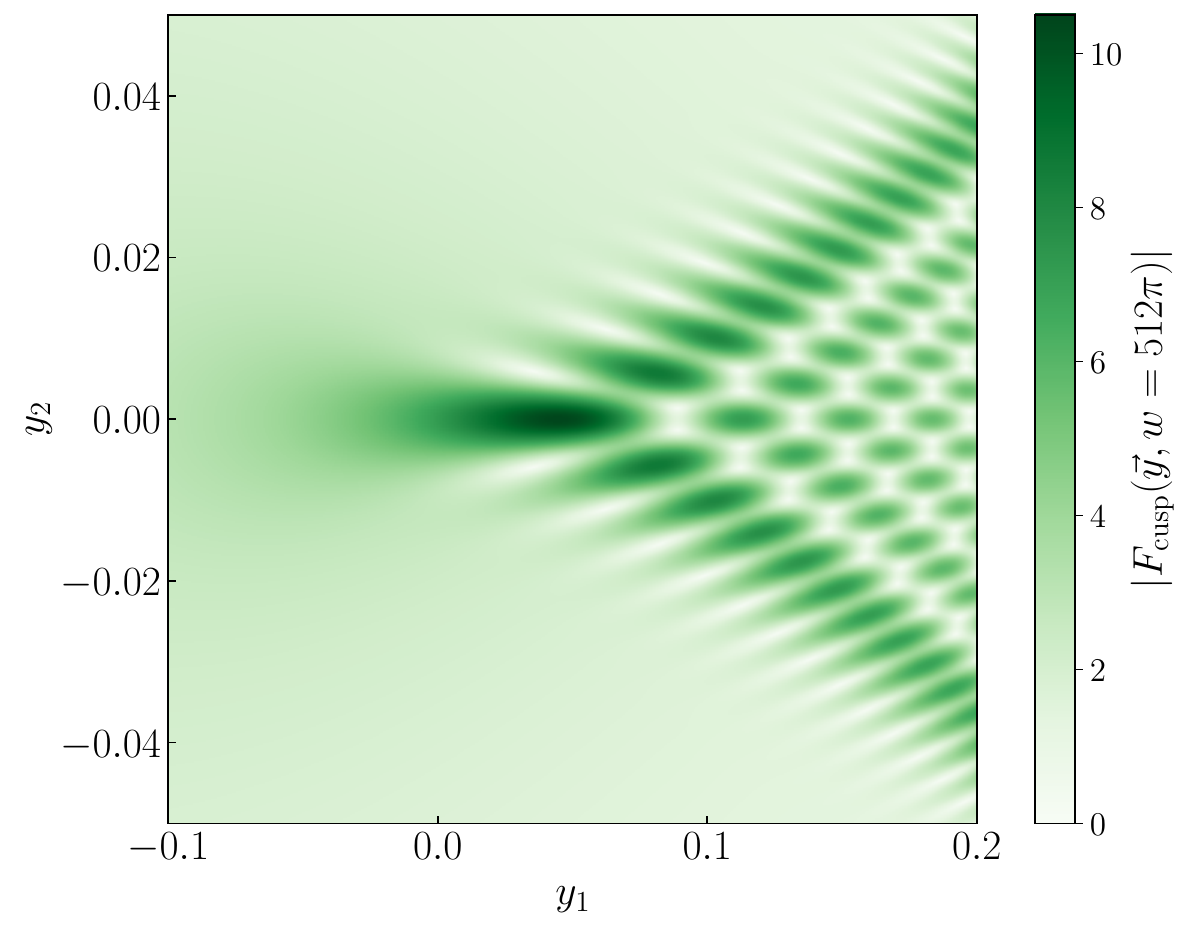}
    \vspace{-20pt}
    \caption{Amplification function $F$ of a cusp caustic as a function of the dimensionless source positions $(y_1,y_2)$ for a fixed dimensionless frequency $w$ of $512\pi$. 
    The cusp has the same configuration of Fig. \ref{fig:source_image_planes_cusp_caustic}.  
    }
    \label{fig:F_cusp_y1y2}
\end{figure}

Similarly to the fold caustic, the diffraction integral of the cusp can be decomposed into the integrals along the different image plane axes: 
\begin{equation}
    \begin{split}
        F=\frac{we^{iwT_c}}{2\pi i}&\int_{-\infty}^\infty dx_2 e^{iw\left(\frac{T_{2222}}{24}x_2^4-y_2x_2\right)} \\
        &\int_{-\infty}^\infty dx_1 e^{iw\left(\frac{T_{11}}{2}x_1^2+(\frac{T_{122}}{2}x_2^2-y_1)x_1\right)}\,.
    \end{split}
\end{equation}
The integral in $x_1$ can be solved yet again using a Gaussian integral. 
We obtain
\begin{equation} \label{eq:cusp_F_derivation}
    F=\sqrt{\frac{w}{2\pi i T_{11}}}e^{iwT_0}\int_{-\infty}^\infty dx_2 e^{iw\left(-\frac{\gamma}{8}x_2^4 +\frac{T_{122}y_1}{2T_{11}}x_2^2-y_2x_2\right)}\,,
\end{equation}
where we have introduced the parameter $\gamma$ from (\ref{eq:gamma}). 
Interestingly, $\gamma$ is related to the definition of the cusp, 
and its sign determines the parity of the images. 
On the other hand, the time delay $T_0\equiv T_c - y_1^2 /2T_{11}$. 
Note that $T_0$ corresponds to the arrival time of the first image in geometric optics when $y_2=0$. 
The last integral in the amplification function $F$ can be solved in terms of a Pearcey integral, 
\begin{equation} \label{eq:pearcey_integral}
    P(a,b) = \int_{-\infty}^\infty e^{i(t^4+at^2+bt)}dt\,,
\end{equation}
named after the scholar who first studied the structure of electromagnetic fields in the neighborhood of a cusp of a caustic~\cite{Pearcey01051946}. 
Note, however, that the sign of the quartic term in (\ref{eq:cusp_F_derivation}) is the opposite to the one of the Pearcey integral in (\ref{eq:pearcey_integral}). 
This is not a problem if one solves the diffraction for the negative frequencies first, and then computes the positive frequency solution by complex conjugation, $F(w)=F^*(-w)$ as discussed in the introduction below Eq. (\ref{eq:fermat_potential}).   
Taking this approach, we arrive at the solution 
\begin{equation} \label{eq:F_cusp}
    F(w)=\frac{2^{1/4}\nu^{1/4}e^{iwT_0}}{i^{1/2}\pi^{1/2} T_{11}^{1/2}}P^*\left(\tilde y_1,\tilde y_2\right)\,,
\end{equation}
where 
\begin{align} \label{eq:tilde_y1}
    &\tilde{y}_1\equiv -2^{1/2}\nu^{1/2}\frac{T_{122}}{T_{11}}y_1\,, \\
    &\tilde y_2\equiv 2^{3/4}\nu^{1/4}w^{1/2}y_2\,,
\end{align}
and we have rescaled the dimensionless frequency as $\nu\equiv w/\gamma$ assuming $\gamma>0$. 
The absolute value of the amplification factor $|F|$ as a function of the source positions is presented in Fig. \ref{fig:F_cusp_y1y2}, where we solve the Pearcey integral numerically.\footnote{Our numerical routine solving the Pearcey integral can be found at \cite{githubpearcey}.} 
One can clearly see the interference patterns of the 3 images within the caustic, and the possibility of having high magnifications in the one-image region, for example at $y_1<0$ and $y_2\approx0$. 
The interference pattern forms a distorted hexagonal lattice very different from the regular fringes occurring near a fold caustic, cf. Fig. \ref{fig:F_fold_y1y2}. 
This lattice is characterized by the destructive interference of the three images that dims the amplification function to vanish. 
Such points are produced by the wavefront dislocations~\cite{Nye_Berry:1974}. 
Moreover, we also see that the maximum magnification occurs within the cusp. 
This rich diffraction and interference pattern can be compared with the magnification map plotted in the left panel of Fig. \ref{fig:source_image_planes_cusp_caustic}. 
The single-image high-magnification region is diffused in a broader region by diffraction, and the destructive interference of the images occurs even over the side caustics. 

If one restricts to source positions along the $y_1$ axis, the Pearcey integral simplifies since (\ref{eq:cusp_F_derivation}) reduces to
\begin{equation} 
    F=\frac{2^{1/4}\nu^{1/4}e^{iwT_0}}{\pi^{1/2} T_{11}^{1/2}}\int_{-\infty}^\infty d\tilde x_2 e^{-\left(\tilde x_2^4 -\tilde y_1\tilde x_2^2\right)}\,,
\end{equation}
where we have already included the definition of $\tilde y_1$ in (\ref{eq:tilde_y1}). 
When $\tilde y_1>0$, the solution can be written as 
\begin{equation}\label{eq:F_cusp_y1}
\begin{split}
    F(w)=& \frac{(-1)^{1/8}\pi^{1/2}|\tilde y_1|^{1/2}\nu^{1/4}e^{i(wT_0+\tilde y_1^2/8)}}{2^{5/4} T_{11}^{1/2}} \\
    & \cdot\left[(-1)^{-1/4}J_{-1/4}^{(1)}\left(\tilde y_1^2/8\right)+J_{1/4}^{(1)}\left(\tilde y_1^2/8\right)\right]\,,
\end{split}
\end{equation}
where $J_n(z)$ is a Bessel function of the first-kind. 
When $\tilde y_1 <0$, it follows
\begin{equation}\label{eq:F_cusp_y1}
    F(w)=\frac{\sqrt{(-1-i)\pi|\tilde y_1|}\nu^{1/4}e^{i(wT_0+\tilde y_1^2/8)}}{4 T_{11}^{1/2}}H_{-1/4}^{(2)}\left(\tilde y_1^2/8\right)\,,
\end{equation}
where $H_n^{(2)}(z)$ is the Hankel function of the second-kind, related to the Bessel function of first- and second-kind by $H_n^{(1)}(z)=J_n(z)-iY_n(z)$. 

\begin{figure}
    \centering
    \includegraphics[width=\columnwidth]{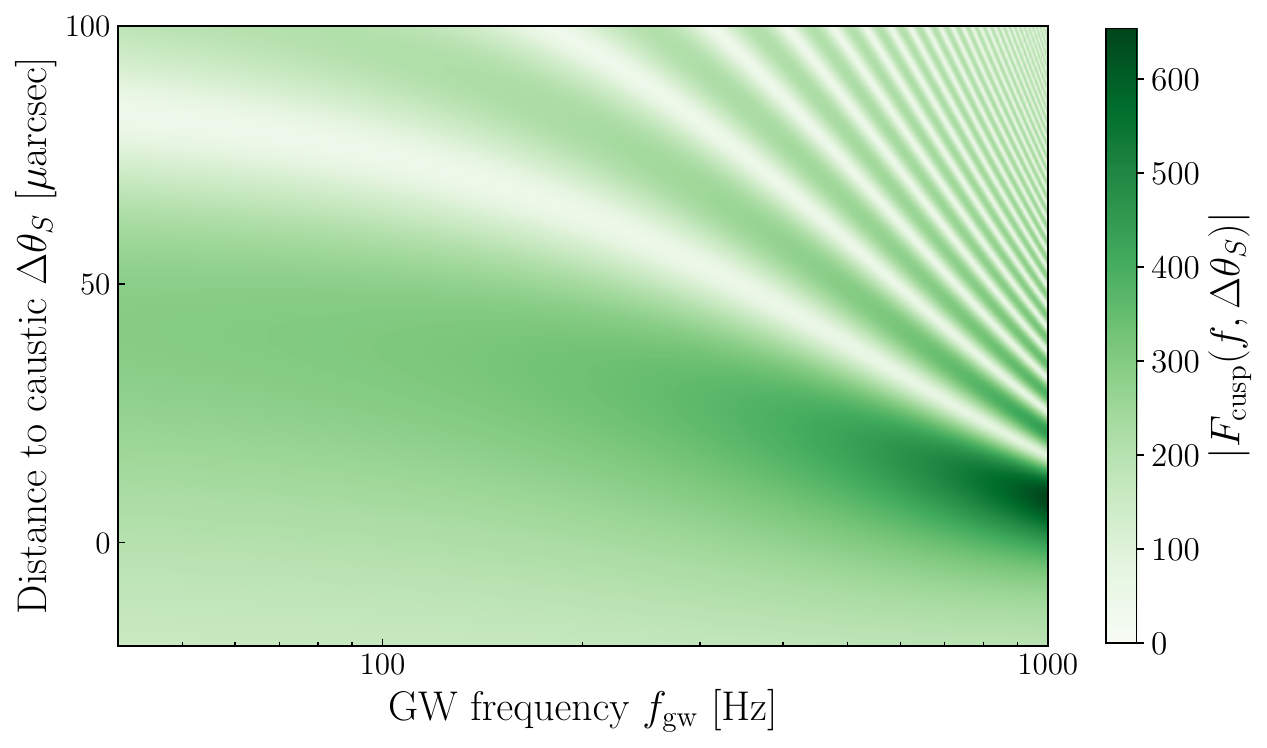}
    \vspace{-20pt}
    \caption{Amplification function around a cusp caustic as a function of the distance to the cusp $\Delta\theta_\mathrm{S}$ along its symmetry axis ($\theta_{S,2}=0$) and the GW frequency $f_\mathrm{gw}$ for a lens at $z_L=0.5$ with velocity dispersion $\sigma=200$km/s and a source at $z_S=2$. 
    Positive $\Delta\theta_\mathrm{S}$ corresponds to source positions in the interior of the caustic leading to three images. }
    \label{fig:F_cusp_fy1}
\end{figure}

The absolute value of the Pearcey integral has a maximum value of $\approx 2.6$ at $\tilde y_1=-2$ and $\tilde y_2=0$. 
Therefore, the magnification, $\mu=|F|^2$, scales with $w^{1/2}$. This is to be compared with the $w^{1/3}$ scaling for the fold caustic. 
Introducing physical units in the problem, we find
\begin{equation} \label{eq:max_mu_cusp}
\begin{split}
    \left(\frac{\mu_\mathrm{max}}{10^5}\right) \simeq &\frac{(1+z_L)^{1/2}}{|\gamma|^{1/2}|T_{11}|}\left(\frac{f}{100\mathrm{Hz}}\right)^{1/2} \\
    &\left(\frac{c\tau_{\mathrm{D}}}{1\mathrm{Gpc}}\right)^{1/2}\left(\frac{\sigma}{200\mathrm{km\,s}^{-1}}\right)^{2}\,,
\end{split}
\end{equation}
where we have introduced the same fiducial isothermal galaxy-scale lens, cf. (\ref{eq:Einstein_angle_sis}), as in the fold case. 
If we plot $|F|$ along the symmetry axis of the cusp as a function of the GW frequency and the physical distance to the cusp ($\Delta \theta_\mathrm{S}=y_1\cdot\theta_*$), we obtain Fig. \ref{fig:F_cusp_fy1}. 
Note that in this example, we are using the same choices for the dimensionless time delay as before, $T_{11}=T_{122}=-T_{2222}=1$, and as a consequence, $\gamma \simeq 1.3$. 
We have checked that for $\sim10\mu$arcsec source distances to a cusp in a realistic galaxy-scale lens, the leading order expansion of the time delay in (\ref{eq:Td_cusp}) is an accurate description. 

We observe that the maximum amplification in Fig. \ref{fig:F_cusp_fy1} matches the estimate in (\ref{eq:max_mu_cusp}). 
The maximum magnification around the cusp will be limited by diffraction. 
Diffraction will be relevant when the time delay between the stationary points is of the order of the period of the wave, $\omega \Delta t_d\sim1$. 
If source approaches the the cusp along its symmetry axis, then the relevant time delay is between the sets of images with different parity. 
In terms of physical quantities, this is
\begin{equation}
\begin{split}
    \omega\Delta t_d \simeq &\frac{(1+z_L)|T_{11}|^2}{|\gamma|}\left(\frac{5\cdot10^4}{|\mu|}\right)^2 \left(\frac{f}{100\mathrm{Hz}}\right) \\
    &\left(\frac{c\tau_{\mathrm{D}}}{1\mathrm{Gpc}}\right)\left(\frac{\sigma}{200\mathrm{km\,s}^{-1}}\right)^4\,.
\end{split}
\end{equation}
This means that within the same galaxy, a point source can be magnified more by a cusp than by a fold. 
However, the probability of encountering a cusp is significantly lower, and as a consequence, the high-magnification tail of the distribution is typically dominated by the fold caustics.

%-----
%SECTION: IMPLICATIONS
%-----
\section{Implications for gravitational wave observations}
\label{sec:implications}

After exploring the diffraction patterns for the point, fold, and cusp caustics, we proceed to quantify the detectability of the distortions induced on GW waveforms. 
Following the standard approach, see e.g.~\cite{Maggiore:1900zz}, we characterize a given GW detector in terms of its noise $n(t)$, which we assume to be Gaussian and stationary. 
Therefore, the noise is fully characterized by its power spectral density in frequency space $S_n(f)$. 
When there is a signal $s$ present, the data $d$ will be a superposition of the signal and noise:
\begin{equation}
    d(t)=s(t) + n(t)\,. 
\end{equation}
If we were to know the template $h$ that characterizes this signal, then its optimal signal-to-noise ratio (SNR) or $\rho$ will be given by
\begin{equation} \label{eq:SNR}
    \rho^2 = (h|h)\,,
\end{equation}
where we have defined the noise-weighted inner product as
\begin{equation}
    (a|b)=4\mathrm{Re}\left[\int_0^\infty df \frac{a\cdot b^*}{S_n(f)}\right]
\end{equation}
In order to look for the best template $h$, one explores its parameter space (typically 15 dimensional for non-eccentric, non-lensed binaries) searching for the combination that returns Gaussian noise, $d-h=n$. 
We can therefore construct a likelihood in the parameter inference by
\begin{equation}
    \mathcal{L}\sim e^{-\frac{1}{2}(d-h|d-h)}\equiv e^{-\chi^2}\,,
\end{equation}
which describes a $\chi^2$ distribution. 
In return, if our objective is to compare the fit of our template $h$ to the true signal $s$, we can compute 
the increase in the $\chi^2$ of the template with respect to the truth, which is given by
\begin{equation}
    \Delta \chi^2 = \chi^2_\mathrm{template}-\chi^2_\mathrm{truth}=(s|s)-2(h|s)+(h|h)\,,
\end{equation}
where the common noise term vanishes. 
$\Delta\chi^2$ has the advantage that, in the frequentist approach, it can be interpreted as the preference for one hypothesis over the other. 
For example, in a single-parameter model, an increase in $\Delta\chi^2=X^2$ directly corresponds to a preference to add that parameter of $X\sigma$. 
Similar relations follow for higher dimensionalities. 

Another metric to quantify the dissimilarity of two waveforms $h_1$ and $h_2$ is the mismatch, which is defined by
\begin{equation}
    \epsilon = 1 - \mathrm{max}\left(\frac{(h_1|h_2)}{\sqrt{(h_1|h_1),(h_2|h_2)}}\right)_{t_0,\phi_0}\,,
\end{equation}
where we maximize the match over global time delays $t_0$ and phases $\phi_0$. 
The mismatch has the benefit that its value is insensitive to the overall amplitude of either the signal or the template. 
Therefore, it is good to capture morphological differences in the waveforms, irrespective of their SNRs. 
It is interesting that in the limit of small mismatches, $(s|s)\approx(h|h)$, we can simply relate the mismatch and $\Delta\chi^2$:
\begin{equation}
    \Delta \chi^2 \approx 2 \epsilon \rho^2\,.
\end{equation}

In the following, we compute the SNR, mismatch, and $\Delta\chi^2$ for a (detector-frame) 30--30 $M_\odot$ binary black hole at $z=2$ lensed by the different caustics. 
We compare how well an unlensed template could fit the lensed signal. 
This gives us information about the overall detectability of the lensing distortions. 
We also compare how waveforms constructed using the stationary phase approximation match the wave optics results. 
This gives us a measure of whether those simpler models could be used in searches. 
In addition, we look closer into some representative examples to understand how search and inference pipelines would be affected when a GW is lensed near a caustic. 
We first compute the SNR time series obtained when filtering lensed signals on simulated noise. 
We then perform parameter estimation to determine both the impact of the lensing distortions of the inference of the parameters when lensing is not accounted for and the possibility to measure the properties of the lens. 
In all the coming results we simulate gravitational wave detectors at current sensitivity.\footnote{Specifically, we set the LIGO detectors to follow ``aligo$\_$O4high.txt'' and Virgo ``avirgo$\_$O4high$\_$NEW.txt'' from the public LIGO Document T2000012-v1 (\href{https://dcc.ligo.org/LIGO-T2000012-v1/public}{https://dcc.ligo.org/LIGO-T2000012-v1/public}). 
Their corresponding power spectral densities are also available in our github repository \cite{githubmodwaveforms}.}

\begin{figure*}
    \vspace{-10pt}
    \centering
    \includegraphics[width=\linewidth]{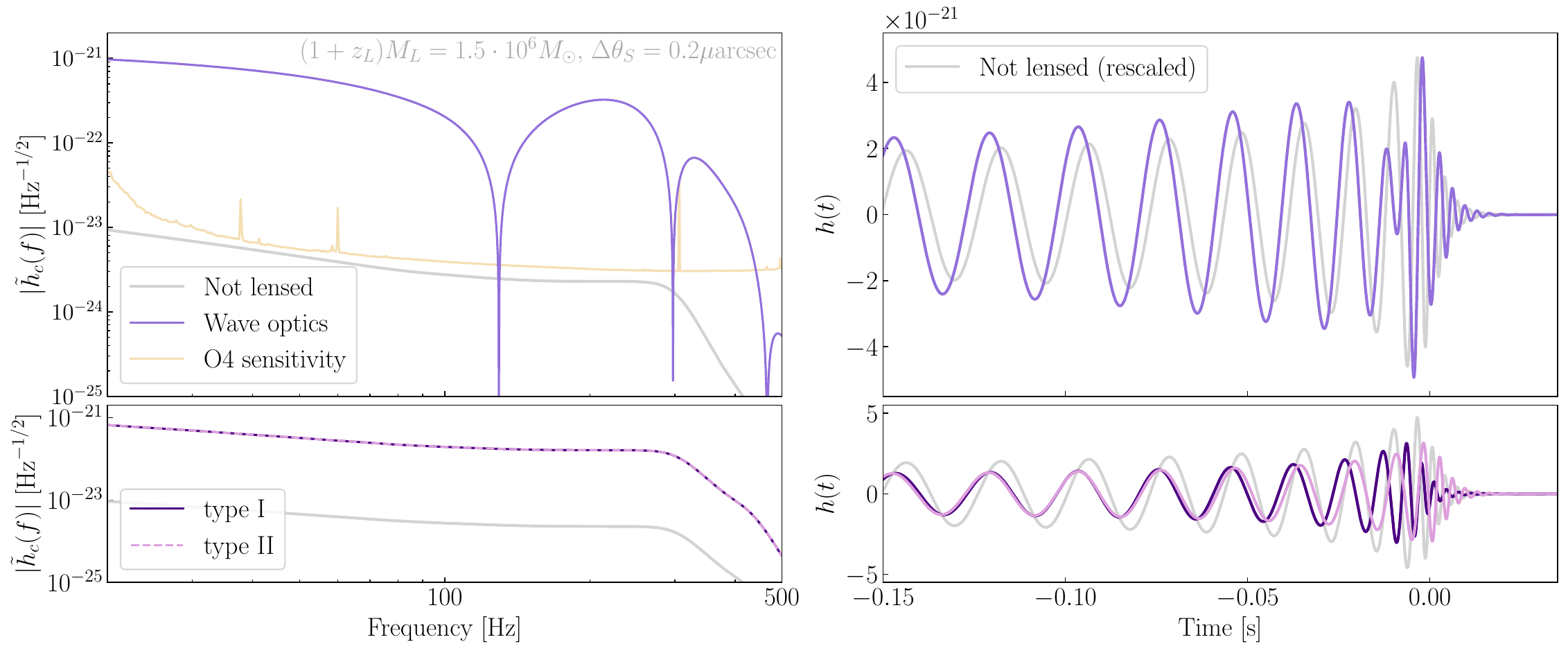}
    \vspace{-20pt}
    \caption{Frequency (left) and time (right) domain waveforms for a 30--30$M_\odot$ binary black hole lensed by a point caustic at a distance of $0.2\mu$arcsec. 
    We plot the characteristic strain $\tilde h_c(f)=\sqrt{f}\tilde h(f)$, which can be directly compared with the amplitude spectral density at O4 sensitivity. 
    The bottom panels show the corresponding geometric optics images.}
    \label{fig:point_caustic_waveform}
\end{figure*}

%-------------
%IMPLICATIONS: POINT CAUSTIC
%-------------
\subsection{Point caustics}
\label{sec:implications_point}

In order to get a sense of the detectability of the wave optics distortions, we first look into the impact of lensing on the detected waveform for a representative example.  
In Fig. \ref{fig:point_caustic_waveform} we plot both the frequency and time domain lensed waveforms. 
The upper panels show the entire signal, while the bottom ones decompose it into its geometric optics images. 
For this case of a point caustic, there are two opposite parity images arriving at different times. 
The fact that their amplitudes are equal becomes obvious in the frequency domain (left) panels, where we are presenting the modulus of the strain. 
There, the large magnification compared to the unlensed case is clear, which is presented in gray for reference. 
This is a case in which the signal could only be detected because of this strong lensing. 
This can be seen when comparing the frequency domain signal with the sensitivity of current ground-based detector, which is also displayed. 
The interference of the two images is seen in frequency domain as the modulation of the amplitude. 
The frequency of this modulation is inversely proportional to the time delay between the images, which in this case corresponds to 5ms. 

Focusing on the time domain waveforms (right panels), we can see the distortion of the signal. 
To facilitate the comparison, we have rescaled the amplitude of the unlensed signal. 
In this time window, the larger distortion corresponds to the interference of the arrival of the peak of the first image with the second images, whose peak arrives shortly after. 
Of course, the particular distortion is subject to the choices of the source-lens configuration. 
In this case, we are choosing a point caustic with a (redshifted) mass of $1.5\cdot 10^6 M_\odot$ and a source located only 0.2$\mu$arsec from its center. 

\begin{figure}[h]
    \centering
    \includegraphics[width=\columnwidth]{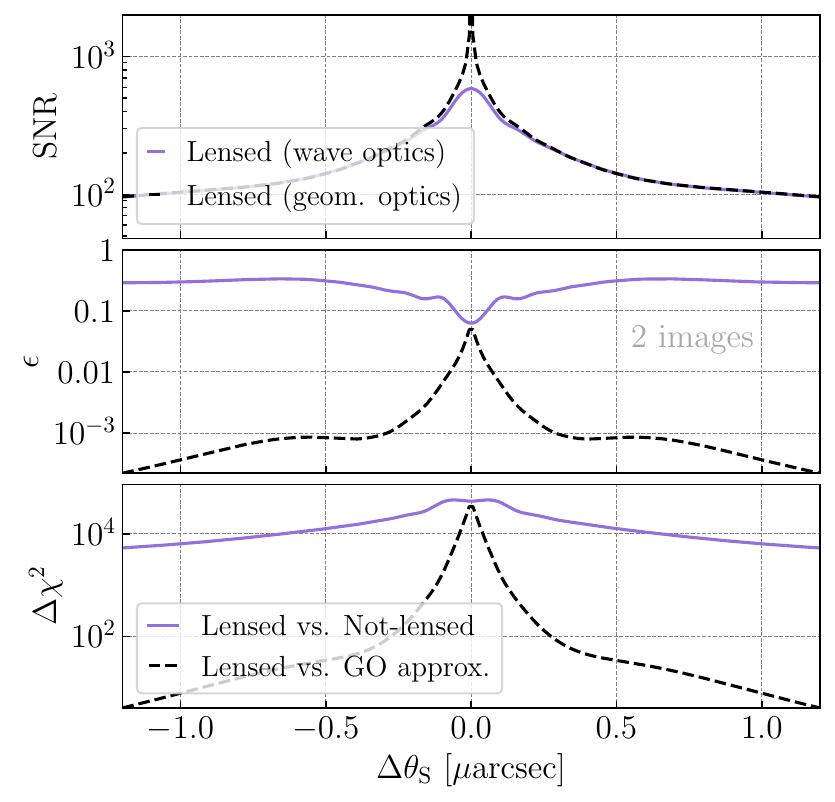}
    \vspace{-20pt}
    \caption{Detectability of diffraction around a point caustic for a 30--30$M_\odot$ binary black hole merger at $z_S=2$ and supermassive black holes lens for different source distances $\Delta\theta_\mathrm{S}$. 
    The top, middle and bottom panels show the signal-to-noise ratio, mismatch $\epsilon$, and goodness-of-fit $\Delta\chi^2$, respectively. 
    We compare the SNR of a lensed signal in wave optics or using the stationary phase approximation, also known as geometric optics (GO), which diverges at the caustic. 
    For the mismatch and $\Delta\chi^2$, the lensed signal is matched against a non-lensed waveform and a GO one. 
    The GO waveform contains always 2 images.}
    \vspace{-20pt}
    \label{fig:point_caustic_detectability}
\end{figure}

After getting an understanding of the physical phenomena taking place, we now look to quantify the potential detectability of these wave optics features for a range of source positions. 
To this end, we present the mismatch $\epsilon$ and $\Delta \chi^2$ in Fig. \ref{fig:point_caustic_detectability} as a function of the distance to the caustic. 
Because of the axi-symmetry of the lens, the results are symmetric around the caustic. 
For reference, we also compute the comparison of the SNR computed using the wave optics formalism (solid line), or using the stationary phase approximation (dashed line). 
This serves to exemplify the regularization of the magnification divergence at the caustic. 
Moreover, because the maximum magnification occurs at the point singularity, both in wave optics and in the stationary phase approximation, we observe the maximum SNR at that point. 

\begin{figure*}
    \vspace{-10pt}
    \centering
    \includegraphics[width=\linewidth]{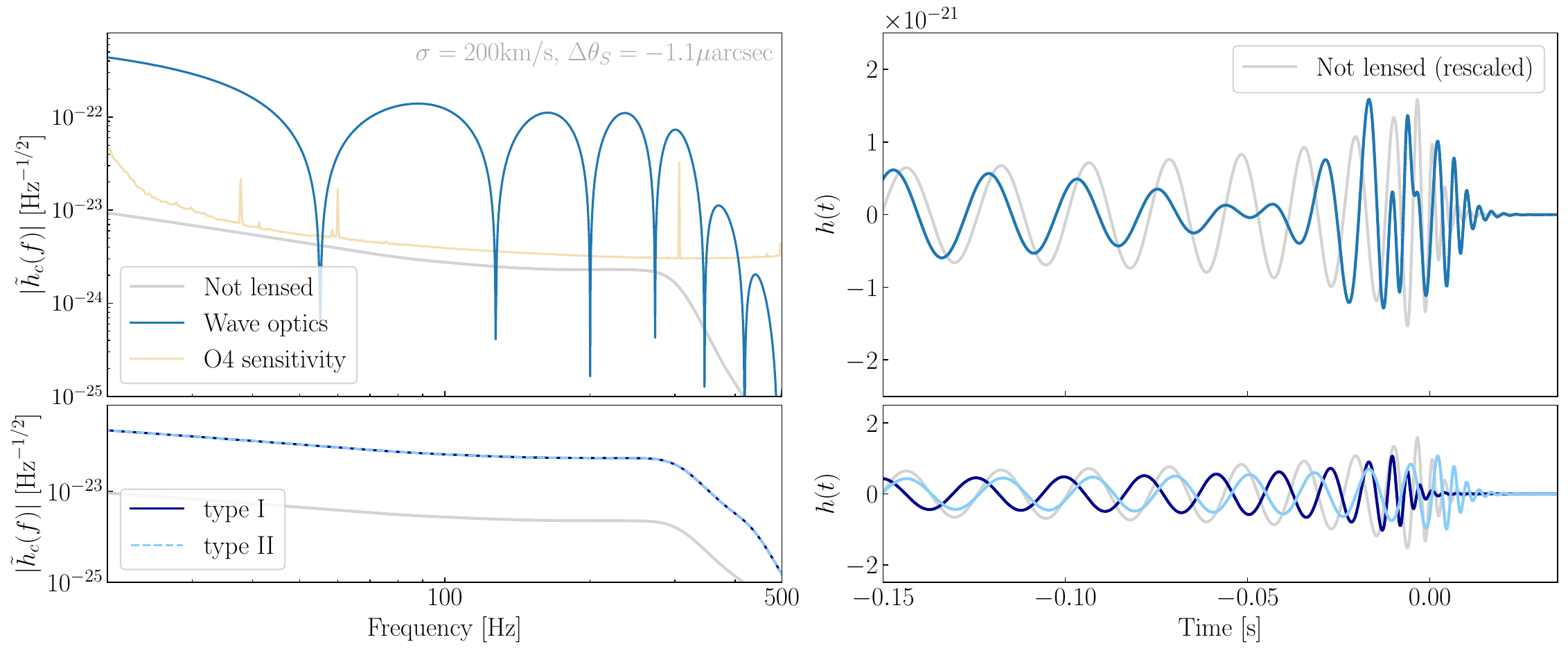}
    \vspace{-20pt}
    \caption{Frequency (left) and time (right) domain waveforms for a 30--30$M_\odot$ binary black hole lensed by a fold caustic at a distance of $-1.1\mu$arcsec. 
    We plot the characteristic strain $\tilde h_c(f)=\sqrt{f}\tilde h(f)$, which can be directly compared with the amplitude spectral density at O4 sensitivity. 
    The bottom panels show the corresponding geometric optics images.}
    \label{fig:fold_caustic_waveform}
    \vspace{-10pt}
\end{figure*}

In terms of the match of the unlensed waveform to the lensed signal, this represents a bad fit, with mismatches larger than 0.1 in most cases. 
This is not surprising, as a single waveform has large difficulties describing two overlapping signals with large interference patterns. 
It is to be noted though, that when computing mismatches, here and throughout the paper, we are fixing the source parameters. 
In other words, we are not exploring the possible degeneracies with other parameters. 
To address this caveat, we perform Bayesian parameter estimation over the entire parameter space later on. 
Nevertheless, we can see that the mismatch is smallest at the caustic, where the images forming the Einstein ring all arrive at the same time. 
The fact that the mismatch is large implies that the effect of lensing is potentially detectable. 
For this particular case, we can quantify this further by computing the $\Delta\chi^2$. 
Because this metric encodes information about the SNR, consequently, we find that the most detectable distortion is at the caustic itself, where the magnification is maximal. 

After quantifying the significance of the distortions compared to the unlensed waveforms, it is interesting to understand how good the geometric optics description is for this system. 
With this goal in mind, we compute the mismatch between the lensed signal against a template that uses the stationary phase approximation, using the analytical results derived in \S\ref{sec:point_caustic}. 
In the middle and bottom panels, this corresponds to the dashed line. 
We find that the geometric optics approximation is indeed an excellent description of the system, with a match that rapidly improves as the source moves away from the perfect alignment. 
The mismatch at the caustic itself (almost) meets for the unlensed and geometric optics cases. This is because at this point, the two geometric optics images have no time delay difference, and only a phase difference. Since the chosen binary has equal masses, the effect of the constant phase shift is negligible~\cite{Ezquiaga:2020gdt}. 
When looking at the goodness-of-fit in the bottom panel, the fact that the difference between wave optics and geometric optics could only be measured very close to the caustic becomes even more evident. 

%-------------
%IMPLICATIONS: FOLD CAUSTIC
%-------------
\subsection{Fold caustics}
\label{sec:implications_fold}

Next, we consider the case of a fold caustic. 
Looking at the particular case of a 30--30$M_\odot$ GW source at $z_S=2$ and 1$\mu$arsec inside the fold caustic of a lens with a velocity dispersion of 200km/s, we present the lensed waveforms in frequency and time domain in Fig. \ref{fig:fold_caustic_waveform}. 
Because the source is within the caustic, there are two stationary points. If the source where to be outside, then no geometric optics images would form. 
For this choice, as in the point caustic, the signal is composed of two equal amplitude but opposite parity images that arrive at different times. 
For this example, the time delay corresponds to $12$ms. 
The interference pattern is visible in both the frequency and time domains. 
Because both images have the same amplitude, at certain frequencies or times, the interference becomes fully destructive. 
In the frequency domain, this is seen as the zeros of the strain amplitude. 
In the time domain, we observe a modulation of the signal. 

\begin{figure}
    \vspace{-20pt}
    \centering
    \includegraphics[width=\columnwidth]{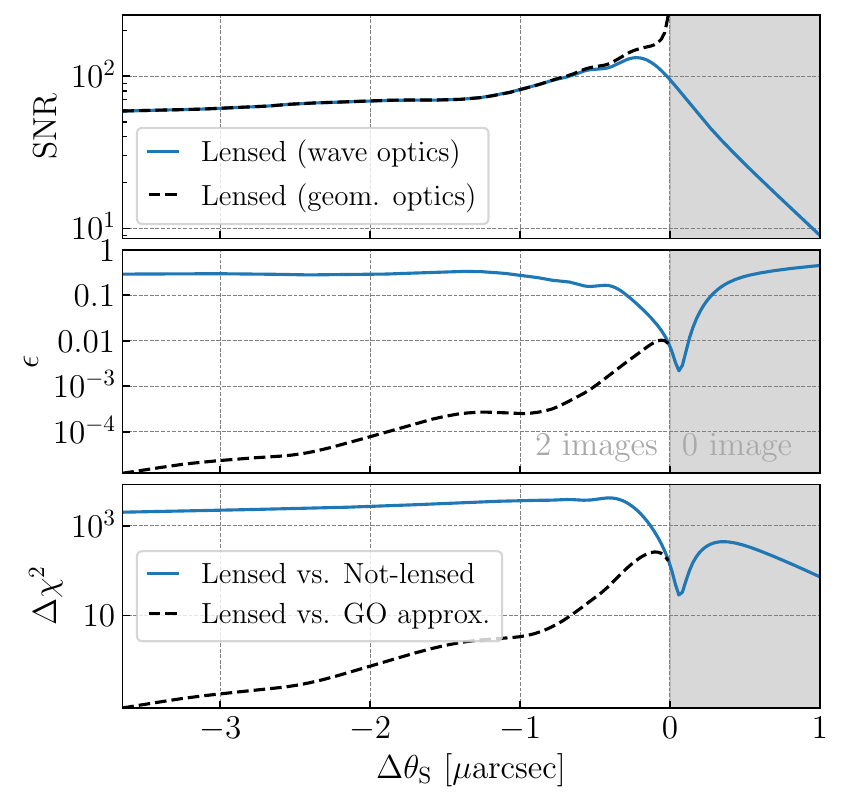}
    \vspace{-20pt}
    \caption{Detectability of diffraction around a fold caustic for a 30--30$M_\odot$ binary black hole merger at $z_S=2$ and galaxy scale lens for different source positions $\Delta\theta_\mathrm{S}$. 
    The top, middle and bottom panels show the signal-to-noise ratio, mismatch $\epsilon$, and goodness-of-fit $\Delta\chi^2$, respectively. 
    We compare the SNR of a lensed signal in wave optics or using the stationary phase approximation, also known as geometric optics (GO), which diverges at the caustic. 
    For the mismatch and $\Delta\chi^2$, the lensed signal is matched against a non-lensed waveform and a GO one. 
    The GO waveform contains 2/0 images for negative/positive $\Delta \theta_\mathrm{S}$.}
    \label{fig:fold_caustic_detectability}
    \vspace{-10pt}
\end{figure}

Then, we examine the similarities of the non-lensed and the geometric optics templates with the wave optics lensed waveform for different distances to the caustic $\Delta \theta_\mathrm{S}$. 
Since the fold is symmetric along its $\theta_{S,1}$ axis, the perpendicular distance to the caustic is equivalent to moving in the $\theta_{S,2}$ direction. 
Moreover, there will be two qualitatively different regimes at each side of the fold. 
The interior will have two images that interfere as in Fig. \ref{fig:fold_caustic_detectability}, while the exterior will have a diffracted image with an exponentially decaying amplitude. 
We present the results in Fig. \ref{fig:fold_caustic_detectability}

Expanding our previous study in \cite{Lo:2024wqm}, we find that the unlensed template only gives a good match to the lensed signal close to the caustic. 
If one moves to the interior (negative values), the interference of the two images introduces dissimilarities, while on the outside (positive values), a single image is diffracted. 
To understand for which source positions the effect will be more measurable, we need to look at $\Delta\chi^2$. 
There, it is evident that because the SNR decays rapidly outside of the fold, those distortions are highly suppressed. 
On the other hand, the modifications to the original waveform remain significant. 
Interestingly, the maximum SNR does not occur at the caustic, $\Delta\theta_\mathrm{S}=0$, as a direct consequence of the maximum of the Airy function not being at zero, which differs from the predictions in geometric optics, cf. discussion around Eq. (\ref{eq:maximum_magnification_fold}). 
This makes the largest detectable distortions occur just inside the caustic. 

\begin{figure*}
    \centering
    \vspace{-20pt}
    \includegraphics[width=\linewidth]{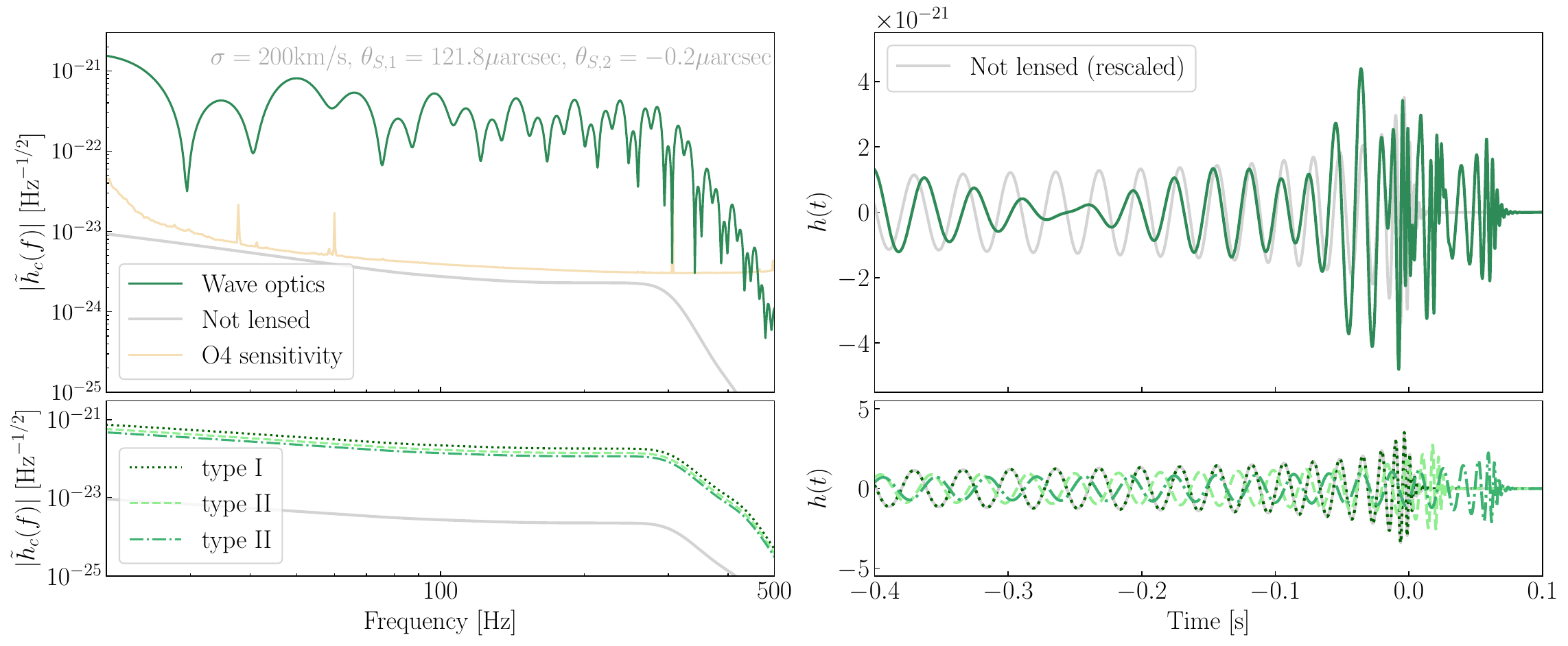}
    \vspace{-20pt}
    \caption{Frequency (left) and time (right) domain waveforms for a 30--30$M_\odot$ binary black hole lensed by a cusp caustic at a distance of $\theta_{S,1}=121\mu$arcsec and $\theta_{S,2}=-0.2\mu$arcsec in the source plane. 
    We plot the characteristic strain $\tilde h_c(f)=\sqrt{f}\tilde h(f)$, which can be directly compared with the amplitude spectral density at O4 sensitivity. 
    The bottom panels show the corresponding geometric optics images, which consist of one type I image arriving first, and two type II images arriving later consecutively.    }
    \label{fig:cusp_caustic_waveform}
    \vspace{-10pt}
\end{figure*}

When comparing with the geometric optics prediction (dashed lines), we need to restrict ourselves to the region of multiple images. 
There, we observe how the approximation provides a very accurate description of the lensed signal, with mismatches below 0.01. 
This means that the simple phenomenological model with a single additional parameter, $\Delta t$, could be used to model and analyze this caustic. 
Because geometric optics predicts no image on the outside of the fold, any observation in that region would be a pure signature of diffraction. 

%-------------
%IMPLICATIONS: CUSP CAUSTIC
%-------------
\subsection{Cusp caustics}
\label{sec:implications_cusp}

\begin{figure}
    \centering
    \includegraphics[width=\columnwidth]{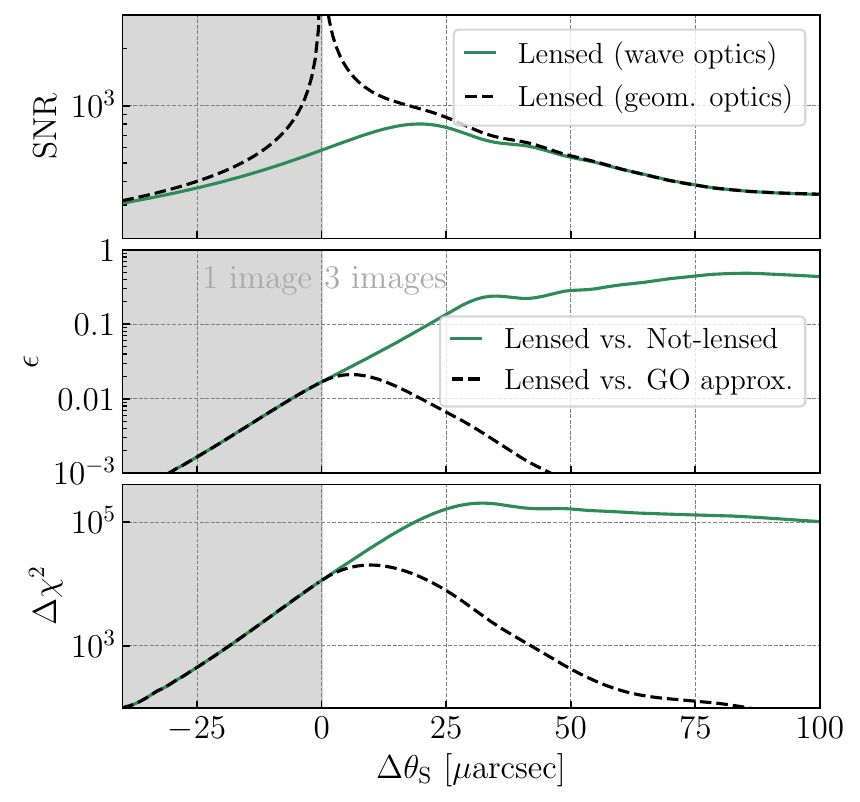}
    \vspace{-20pt}
    \caption{Detectability of diffraction around a cusp caustic for a 30--30$M_\odot$ binary black hole merger at $z_S=2$ and galaxy scale lens for different source distances to the cusp $\Delta\theta_\mathrm{S}$ along its symmetry axis ($\theta_{S,2}=0$). 
    The top, middle and bottom panels show the signal-to-noise ratio (SNR), mismatch $\epsilon$, and goodness-of-fit $\Delta\chi^2$, respectively. 
    We compare the SNR of a lensed signal in wave optics or using the stationary phase approximation, also known as geometric optics (GO), which diverges at the caustic. 
    For the mismatch and $\Delta\chi^2$, the lensed signal is matched against a non-lensed waveform and a GO one. 
    The GO waveform contains 1/3 images for negative/positive $\Delta \theta_\mathrm{S}$. 
    }
    \label{fig:cusp_caustic_detectability}
\end{figure}

Finally, we study the distortions induced by a cusp caustic. 
As discussed in \S\ref{sec:cusp_caustic}, the phenomenology around this type of caustic is richer than in previous cases. 
The three-image region corresponds to the interior of the cusp, while the outside region leads to a single image. 
The multiple-image region could lead to very different time delays and relative magnifications, but there will always be a pair of images inside/outside the critical curve with opposite parity to the one image outside/inside. 
Because the cusp is formed by the junction of two fold caustics, if the source position is far from the cusp and close to the fold, the universal behavior of the fold reappears for two of the images. 
Moreover, if the source is at the symmetry axis of the cusp, the two images with the same parity will arrive at the same time, effectively becoming a two image system.

For our waveform distortion example, we choose a source position such that the three images are at play. Specifically, we position the GW at $\theta_{S,1}=121\mu$arcsec and $\theta_{S,2}=-0.2\mu$arcsec in the source plane. 
We show the lensed signals in the frequency and time domain in Fig. \ref{fig:cusp_caustic_waveform}. 
The overall signal is composed of three images with different amplitudes and arrival times. 
Their magnifications decay with later arrival times. 
The first two images are separated by 0.02 sec, while the last one follows 0.04sec after. 
The second image has a relative magnification of 0.6 with respect to the first one, while the last has a relative magnification of 0.4. 
The first image is type I and the others type II. 
The presence of three images makes the frequency domain pattern more involved. 
As in the previous examples, this is a signal that would only be detected because of the high magnifications.

Focusing on the time domain lensed signal, we also observe a more convoluted pattern. 
The overall modulation of the signal appearing in the inspiral is added to the chaotic distortions of the three mergers at the end of the signal. 
Because of the particular time delays of this example, the merger and ring-down of the last image is mostly unaffected, while the other two images superimpose and distort each other. 
The fact that the three images have similar amplitudes makes the interference pattern more pronounced. 

Subsequently, we investigate the detectability of cusp caustic distortions. 
We begin by considering the case in which the source is located along the symmetry axis of the cusp, $\theta_{S,2}=0$. 
In that case, the distance to the cusp $\Delta\theta_\mathrm{S}$ can be parametrized by the $\theta_{S,1}$ coordinate. 
In the interior of the cusp (positive distances), three images form, while in the exterior (negative distances), only one forms. 
However, for this symmetric case, two images arrive at the same time with the same amplitude and phase, making the geometric optics amplification effectively become a two image model. 
We present our results in Fig. \ref{fig:cusp_caustic_detectability}. 
As one moves through the interior of the cusp, the unlensed template becomes a worse representation of the true lensed signal. 
Outside of the cusp, the unlensed waveform perfectly matches the single lensed image. 
It is to be noted that the maximum SNR occurs in the interior of the cusp, in contrast to the geometric optics prediction of it occurring at the cusp. 
Moreover, even in the single-image region outside of the cusp, the SNR is comparable. 
This is because, unlike the point and fold caustics, the cusp can produce a single highly magnified signal, a fact that has relevant observational implications. 
In terms of the comparison with the geometric optics approximation, we observe a good match, with mismatches around and below 0.01. 

\begin{figure}
    \centering
    \includegraphics[width=\columnwidth]{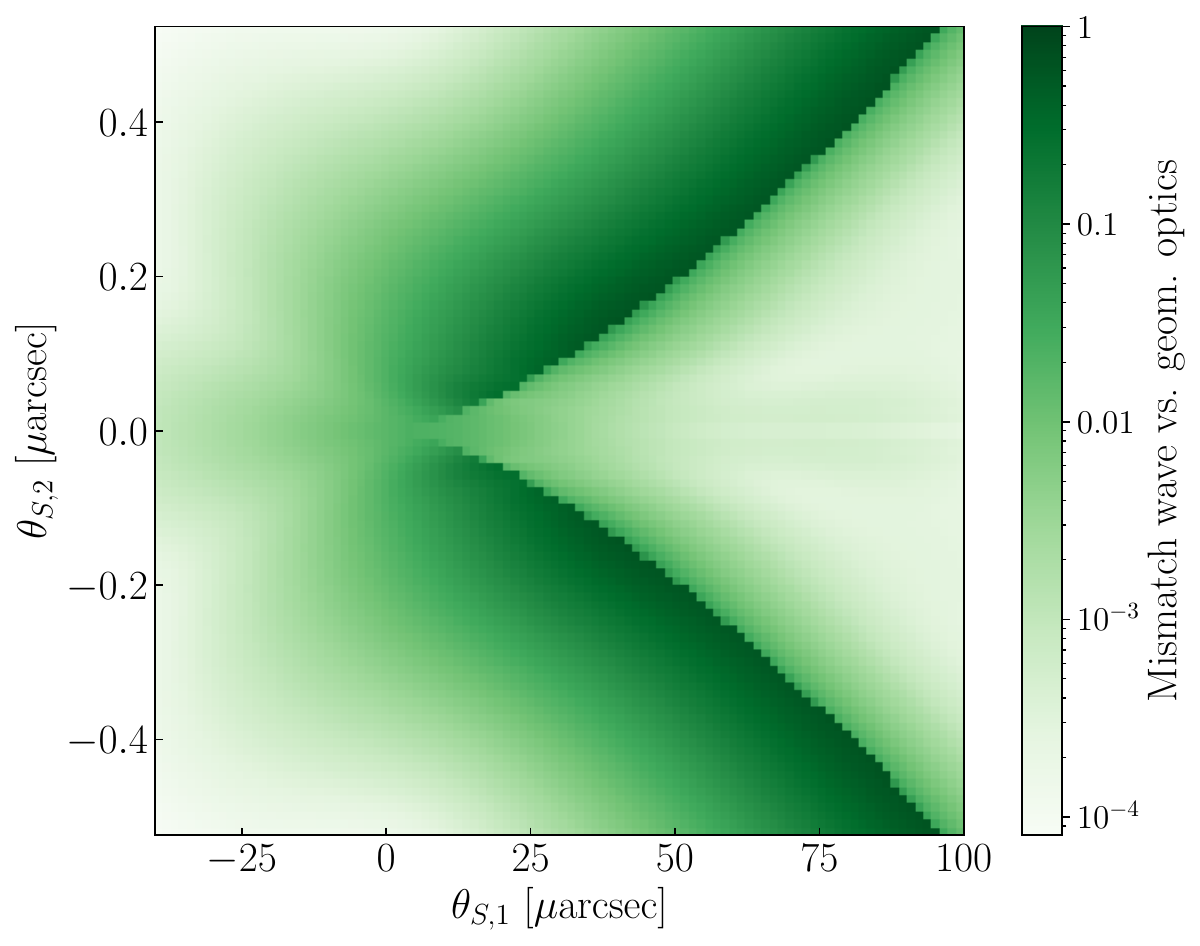}
    \vspace{-20pt}
    \caption{Mismatch between the wave optics lensing around a cusp caustic and the geometric optics approximation as a function of the source position $\vec\theta_\mathrm{S}$. 
    This is an extension of Fig. \ref{fig:cusp_caustic_detectability} which only looked along the line $\theta_{S_2}=0$.
    }
    \label{fig:match_cusp_spa_y1y2}
\end{figure}

However, because the phenomenology of a cusp changes drastically for different source positions, it is important to study the lensing distortions across the source plane. 
We explore this precisely in Fig. \ref{fig:match_cusp_spa_y1y2}, where we compute the mismatch between the lensed signal and its geometric optics approximation. 
We find that this approximation is overall very good, with most of the source plane having mismatches below 0.01. 
It is only very close outside of the caustic that $\epsilon$ can reach up to 0.6.

\begin{figure*}
    \centering
    \vspace{-10pt}
    \includegraphics[width=\linewidth]{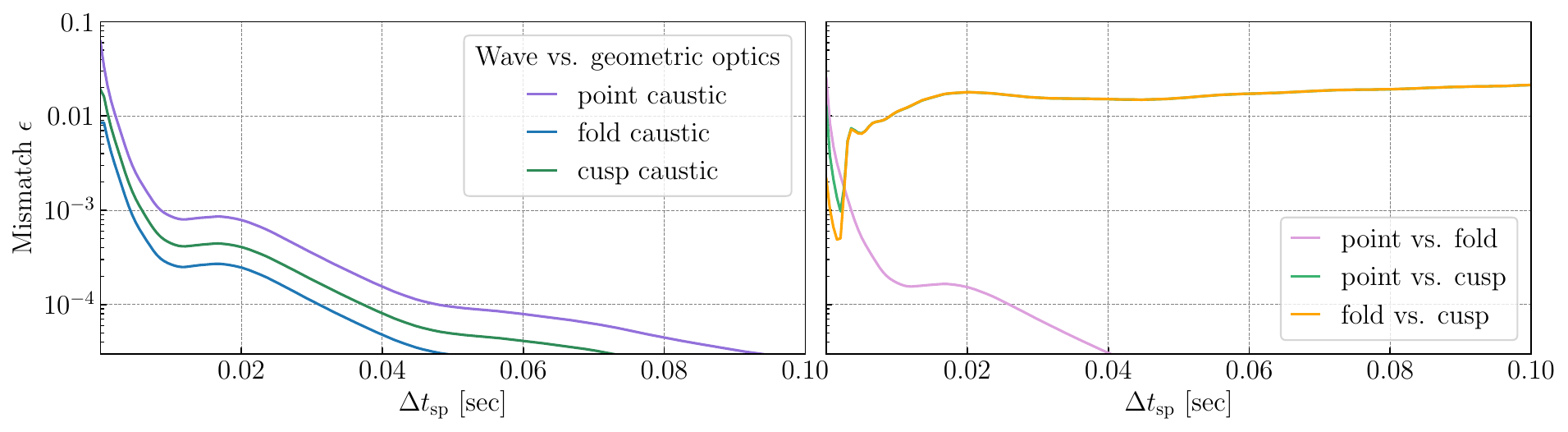}
    \vspace{-20pt}
    \caption{Mismatch as a function of the time delay between the stationary points of the time delay surface $\Delta t_\mathrm{sp}$ for a 30--30$M_\odot$ binary black hole merger at $z_S=2$. 
    On the left, we compare the mismatch between the wave optics result and the geometric optics approximation for each type of caustic. 
    On the right, we compare the mismatch between different caustics in the wave optics lensed signal.
    }
    \label{fig:mismatch_time_delay_comparison}
\end{figure*}

%-------------
%IMPLICATIONS: COMPARISON
%-------------
\subsection{Detectability}
\label{sec:implications_comparison}

After analyzing how well different templates match the lensing distortions for each caustic, it is interesting to explore their similarities, and if those could be disentangled in real data. 
This will serve to quantify the detectability of each feature.
In order to do that, we first quantify the mismatch between the different caustics. 
Then, we simulate different lensed events and explore how they will appear at the detectors. 
Finally, we recover their parameters with different models.

%-------------
%IMPLICATIONS: COMPARISON MISMATCH
%-------------
\subsubsection{Mismatch}

\begin{figure}
    \vspace{-10pt}
    \centering
    \includegraphics[width=\columnwidth]{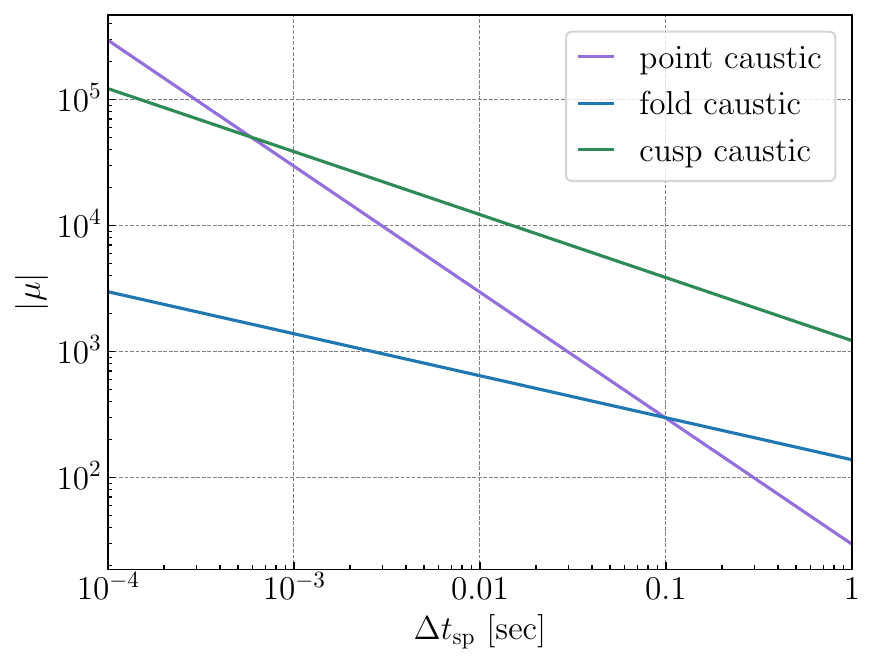}
    \vspace{-20pt}
    \caption{Absolute value of the magnification of the brightest image as a function of the time delay between the stationary points of the time delay surface $\Delta t_\mathrm{sp}$ for the three different caustics. 
    For the cusp caustic this corresponds to the case along its symmetry axis where the pair of equal parity images also have the same magnification and arrival time.  
    }
    \label{fig:mu_vs_DeltaT}
    \vspace{-10pt}
\end{figure}

The comparison between the waveform distortions induced by lensing of each type of caustic is presented in Fig. \ref{fig:mismatch_time_delay_comparison}. 
Since each type of caustic that we have studied corresponds to a different type of lens system with different properties, we choose to parametrize the match between the caustic in terms of common physical observables. 
In particular, the time delay between the stationary points, $\Delta t_\mathrm{sp}$, is the most relevant quantity for both the point and the fold caustic. 
For the cusp, in order to ease the comparison, we restrict ourselves to source positions along its symmetry axis, in which the system effectively becomes two-image.
However, it is important to note that because each caustic has a different Einstein radius, this observed time delay maps to different dimensionless time delays $\Delta T$ for each caustic. 
Because the relative magnification is constant in all of these cases, the most significant quantity to understand the actual signal is how the magnification of the first and brightest image changes as a function of the time delay. 
We plot this in Fig. \ref{fig:mu_vs_DeltaT}. 
The point caustic has the steepest slope and the fold the most shallow. 
For current GW detectors, we are mostly interested in time delays above $\sim0.01$ sec to give sizable distortions.

On the left of Fig. \ref{fig:mismatch_time_delay_comparison}, we compute the mismatch between wave optics and geometric optics for each caustic. 
The results here are the same as those obtained in Figs. \ref{fig:point_caustic_detectability}, \ref{fig:fold_caustic_detectability}, and \ref{fig:cusp_caustic_detectability}, but expressed in terms of the time delay between the stationary points, which is a physical observable common to all. 
We observe that in this parametrization, all caustics have a similar behavior. 
Moreover, their mismatch decays rapidly, reaching values below 1/1000 for images separated more than 0.02sec. 
This means that once again, the geometric optics approximation rapidly becomes a very good approximation. 

On the right panel of Fig. \ref{fig:mismatch_time_delay_comparison}, we show the mismatch between different caustics. 
Noticeably, the point and the fold caustic have an excellent match because, as a function of the time delay between the stationary points, both models predict two equally magnified images with opposite parity in geometric optics. 
On the other hand, the point and fold compared to the cusp have a mismatch that is minimal close to the caustic, $\Delta t_\mathrm{sp}\to0$, but then increases and plateaus. 
This can be explained by the fact that when moving along the symmetry axis of the cusp, there are two of the three images arriving at the same time, but with different magnification compared to the first one. 
Therefore, when the stationary phase approximation is valid, there is an offset in the relative magnification that leads to this mismatch. 
Because the point and the fold give essentially the same interference patterns, in the next analyses, we will only consider the fold and the cusp.

%-------------
%IMPLICATIONS: SNR
%-------------
\subsubsection{SNR time series}
\label{sec:implications_pe}

\begin{figure*}[t!]
    \centering
    \includegraphics[width=2\columnwidth]{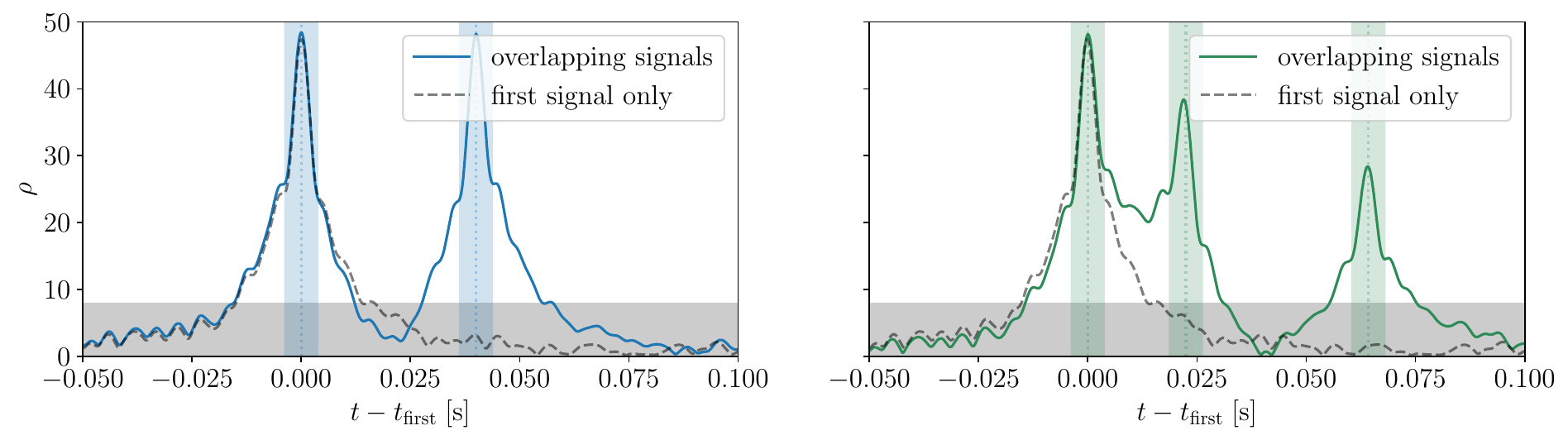}
    \caption{Signal-to-noise ratio time series, $\rho(t)$, for a binary black hole merger lensed by a fold (left) and cusp (right) caustic. See specific parameters in Tab. \ref{table:injections_parameters}. 
    The signals are highly magnified and their peak SNR distinguishable and clearly above the SNR $\sim8$ single-detector detection threshold (gray shaded region).  
    The template’s autocorrelation function is indicated by the vertical bands around the peaks, which correlate well with the SNR peak's width. 
    } 
    \label{fig:snr_time_series}
\end{figure*}

Next, we simulate a set of lensed and unlensed signals in detector noise. 
Similarly to before, we are modeling the noise as Gaussian and stationary with O4 projected sensitivity. 
The concrete parameters chosen for the lensed and unlensed binary black holes are summarized in Table \ref{table:injections_parameters}. 
We consider three situations: an unlensed binary black hole, a binary black hole lensed near a fold caustic, and near a cusp caustic. 
In all cases, we use an inspiral-merger-ringdown phenomenological model with higher modes and precession, \texttt{IMRPhenomXPHM}~\cite{Pratten:2020ceb}, to generate the unlensed signals. 

The first step in GW data analysis is to filter the time-series data at the detectors in order to find suitable triggers that could be GW candidates. 
This is typically achieved using matched filtering, a technique in which the data is cross-checked against a large set of representative waveform models, i.e. a template bank, using some ranking statistic, in its simplest form the SNR (\ref{eq:SNR}). 
Although conceptually simple, matched filtering is a computationally demanding task, since the template bank typically contains millions of waveforms, and the ranking statistic is only meaningful when compared to a statistically significant background. 
A study of the detectability of (simulated) lensed GWs with an actual search pipeline has only recently been completed for the first time in Ref.~\cite{Chan:2024qmb}. 
There, the authors find that precisely the most interesting regions in the parameter space, those of largest waveform distortions, and therefore highest mismatches, are the regions in which the detectability is most affected, dimming highly distorted lensed events as undetected by current matched-filtering-based search pipelines.  
The study was performed using a point lens model, but we expect similar conclusions to extend to the set ups considered here. 

In this work, we do not aim at performing a full fledged detectability study as in~\cite{Chan:2024qmb}, but rather take a more similar approach to that of Ref.~\cite{Lo:2024wqm}. 
That is, we point out that although the time domain waveform is highly distorted, and the matched filtering search pipelines down rank strongly such lensed signals when using an unlensed template bank, in its essence, the signal is just a superposition of different copies of the same signal with some short time delay, amplitude change and (possibly) a phase shift. 
Therefore, when unlensed templates are slid across the data to compute the SNR time series, $\rho(t)$, templates similar to the original unlensed signal should generate peaks in $\rho(t)$ around the merger time of each repeated chirp. 
The ability of a GW detector to resolve different peaks in the SNR time series depends on the autocorrelation time of the template~\cite{Lo:2024wqm}, which for a $\sim30$--30$M_\odot$ binary is $\sim$ms. 
What this means is that GW detectors should be sensitive to lensed signals in which the repeated chirps interfere with each other, but as found in Ref.~\cite{Chan:2024qmb} this requires extensions of current matched-filtering search pipelines. 
 
\begin{table*}
\centering
\begin{tabular}{lcccccccccccccccccc}
\hline
\hline
Injection & $m_1\,[M_\odot]$ & $m_2\,[M_\odot]$ & $a_1$ & $a_2$ & $\phi_1$ & $\phi_2$ & $\phi_{12}$ & $\phi_{JL}$ & $\theta_{JN}$ & ra & dec & $\psi$ & $\phi_\text{ref}$ & $d_L\,$[Mpc] & $t_\text{ref}\,$[ms] & $\Delta t_{10}$ [ms] & $\Delta t_{20}$ [ms] & $\mu_\mathrm{10}$\\
\hline
\hline
unlensed & 35 & 30 & 0.2 & 0.1 & 0.1 & 0.2 & 0.2 & 0.3 & 0.4 & 1.0 & 0.52 & 0.7 & 2.0 & 1000 & 0 & - & - & -\\
fold & 35 & 30 & 0.2 & 0.1 & 0.1 & 0.2 & 0.2 & 0.3 & 0.4 & 1.0 & 0.52 & 0.7 & 2.0 & 1000 & 0 & 41 & - & 1 \\
cusp & 35 & 30 & 0.2 & 0.1 & 0.1 & 0.2 & 0.2 & 0.3 & 0.4 & 1.0 & 0.52 & 0.7 & 2.0 & 1000 & 0 & 22 & 64 & 0.6 \\
\hline
\end{tabular}
\caption{\label{table:injections_parameters}
Summary of the physical parameters for the simulated gravitational waves, aka. injections. All the injections share the following parameters: detector-frame component masses ($m_{1,2}$), dimensionless spin magnitudes ($a_{1,2}$), the azimuthal angle of the spin vectors ($\phi_{1,2}$), the difference between the azimuthal angles of the individual spin vectors ($\phi_{12}$), the difference between total and orbital angular momentum azimuthal angles ($\phi_{JL}$), the angle between the total angular momentum and the line of sight ($\theta_{JN}$), right ascension (ra), declination (dec), polarization angle ($\psi$), phase at reference frequency ($\phi_\text{ref}$), luminosity distance ($d_L$) and reference time ($t_\text{ref}$). The reference frequency is at $20$Hz. 
The lensed injections contain additional parameters: the time delays between the images $\Delta t$ and the relative magnification. 
The fold has two opposite parity images with equal magnification. 
The cusp has three images, the first one with opposite parity to the other two, and all of them with different magnifications. 
The relative magnification is with respect the first two images $\mu_{10}=|\mu_1/\mu_0|$. The last cusp image satisfies $\mu_{20}=1-\mu_{10}$, cf. (\ref{eq:cusp_relative_magnifications}).}
\end{table*}

We demonstrate the ability to distinguish the arrival of each repeated chirp in a signal lensed by a fold and cusp caustic in Fig.~\ref{fig:snr_time_series}. 
The simulated fold lensed signal is composed of two chirps of equal amplitude, opposite phases and a time delay of $\sim40$ ms. 
On the other hand, the simulated cusp lensed signal has three chirps. The first arriving has opposite parity to the other two, which are separated by $\sim20$ and $\sim40$ ms, and with amplitudes $\sim0.6$ and $\sim0.4$ smaller than the first one. 
The apparent luminosity distance (1000Mpc) and detector-frame masses (35--30$M_\odot$) are the same in both cases, which lead to the same large SNR peaks, reaching above $40$. 
The data is filtered with a template that has the same properties as the unlensed waveform. 
We observe that in both cases the SNR peaks are clearly distinguished. 
Moreover, their width also matches the template's autocorrelation time, as determined by the full width at half maximum of the autocorrelation function. 
It should be noted that we are only computing the absolute value of the SNR time series. 
However, it is also possible to obtain the phase, which could serve to, in low-latency, test the hypothesis that the repeated chirps have opposite parity~\cite{Lo:2024wqm}. 

%-------------
%IMPLICATIONS: PARAMETER ESTIMATION
%-------------
\subsubsection{Parameter estimation}
\label{sec:implications_pe}

After exploring the possibility of GW search pipelines identifying lensed signals near caustics, we proceed to study the inference properties of the signals. 
So far we have seen that diffraction around caustics leads to sizable waveform distortions, as measured by the mismatch between the lensed and unlensed waveform. 
Provided that the signal is loud enough, the goodness-of-fit $\Delta\chi^2$ indicates that the lensing distortions could be detectable.
However, because waveform models are built in a high dimension and highly degenerate parameter space, it is crucial to perform Bayesian inference to extract the measurable properties of the signal, their correlations with other parameters, and any possible bias. 
Here we do not wish to conduct a systematic study of many simulated events and their recovery, but rather get some representative examples of the potential measurement uncertainties and biases. 
We perform parameter estimation using nested sampling with \texttt{bilby}~\cite{Ashton:2018jfp}.
Our lensed waveforms are publicly available in the \texttt{modwaveforms} library~\cite{githubmodwaveforms}, and are already integrated to work with \texttt{bilby}. 

\begin{figure}
    \centering
    \includegraphics[width=\columnwidth]{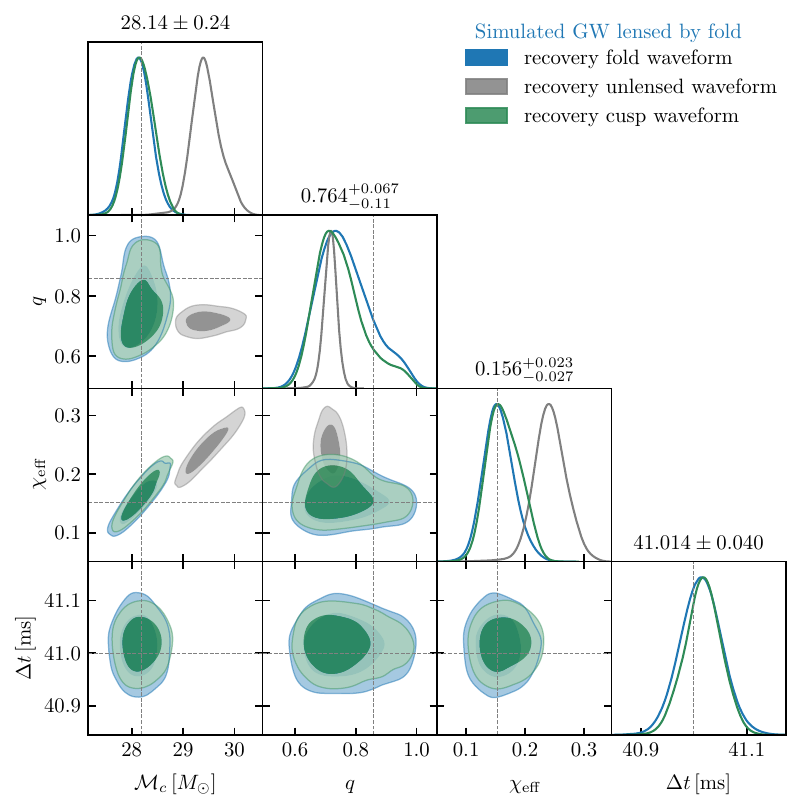}
    \vspace{-20pt}
    \caption{Parameter estimation for a simulated GW lensed by a fold caustic. 
    The recovery is executed with a fold waveform (same model as injection), an unlensed waveform, and a cusp waveform. 
    From the 15D+ inference, we present the posterior samples in some representative parameters, chirp mass $\mathcal{M}_c$, mass ratio $q$, effective spin $\chi_\mathrm{eff}$, and time delay between the images $\Delta t$, to demonstrate the possibility to infer correctly the injected values (dashed lines) when using the right template, while getting large biases in an unlensed reconstruction. 
    For the cusp recovery, $\Delta t$ corresponds to the time delay between two of the images. 
    }
    \label{fig:fold_injection}
\end{figure}
 
We begin by simulating a black hole binary lensed by a fold caustic, cf. Tab. \ref{table:injections_parameters}. 
We perform parameter estimation with three waveform models: unlensed template (15 parameters, gray in upcoming plots), fold template (15+1 parameters, blue in upcoming plots), and cusp template (15+3 parameters, green in upcoming plots). 
It is to be noted that the cusp template contains three images, but reduces to the fold model in the limit of $\mu_{10}\to1$. 
In fact, we find precisely that in the inference when recovering the signal with the cusp model, $\mu_{10}$ peaks narrowly at 1, leaving $\Delta t_{20}$ unconstrained. 
Therefore, we obtain that both the fold and cusp models accurately recover the properties of the simulated fold signal. 
Conversely, the unlensed model is unable to recover the signal properties, exhibiting large biases. 
To exemplify this, in Fig. \ref{fig:fold_injection} we present  a subset of the parameters obtained with the three recoveries. 
Specifically, we plot the chirp mass $\mathcal{M}_c=(m_1m_2)^{1/5}/(m_1+m_2)^{3/5}$, mass ratio $q=m_2/m_1$, effective spin $\chi_\mathrm{eff}=(m_1\chi_{1,z}+m_2\chi_{2,z})/(m_1+m_2)$, and time delay between the images $\Delta t$. 
In this figure, it is evident that the posterior samples obtained with the unlensed waveform do not overlap with the values of the simulation (indicated with dashed lines). 
Moreover, we can quantify how good or bad the fit is to the data by looking at the optimal SNR obtained with each recovery. 
Whilst the fold and cusp have $\rho_\mathrm{fold}=60.1\pm1.5$ and $\rho_\mathrm{cusp}=60.1\pm1.6$ respectively (at $90\%$ credible interval), the unlensed model only has $\rho_\mathrm{unlensed}=45.8\pm1.6$. 
These demonstrate that the fold and cusp models give an equally good fit to the data, while the unlensed waveform significantly struggles. 
Having more intrinsic parameters, the cusp waveform will be disfavored against the fold one in a Bayesian model comparison. 

\begin{figure}
    \centering
    \includegraphics[width=\columnwidth]{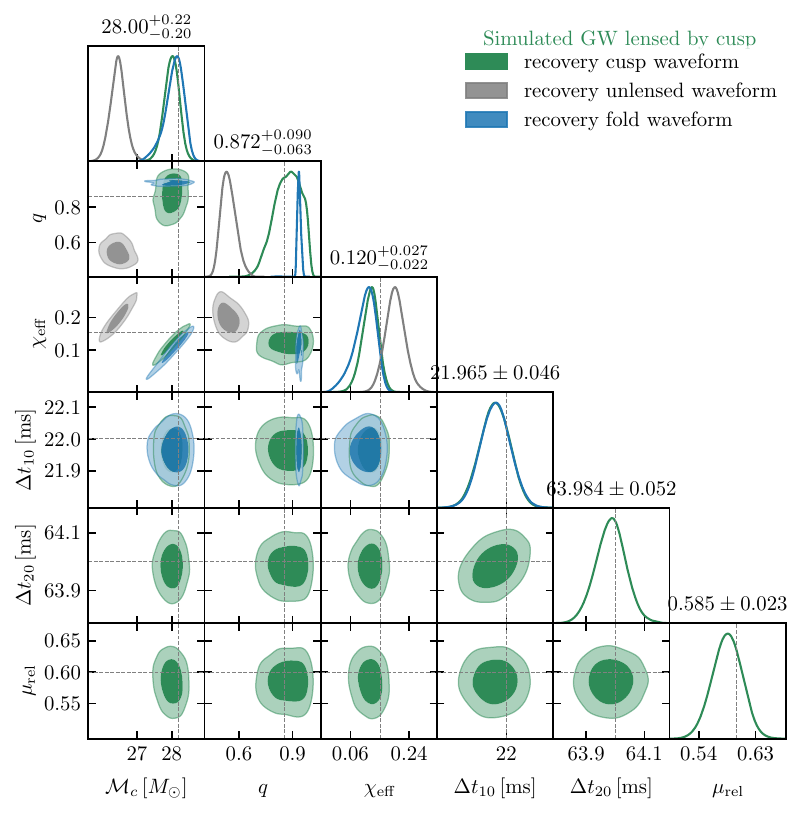}
    \vspace{-20pt}
    \caption{Parameter estimation for a simulated GW lensed by a cusp caustic. 
    The recovery is executed with a cusp waveform (same model as injection), an unlensed waveform, and a fold waveform. 
    From the 15D+ inference, we present the posterior samples in some representative parameters, chirp mass $\mathcal{M}_c$, mass ratio $q$, effective spin $\chi_\mathrm{eff}$, time delay between the first two images $\Delta t_{10}$, the first and third one $\Delta t_{20}$, and their relative magnification $\mu_\mathrm{rel}$. 
    The posteriors contain the injected values (dashed lines) when using the right template, while biased otherwise. 
    For the fold recovery, because it only has two images, there is only one time delay recovery. 
    }
    \label{fig:cusp_injection}
\end{figure}

We repeat the same exercise, but for a simulated GW signal lensed near a cusp. 
Again, we find that the unlensed waveform is unable to adequately fit the data, and the recovered parameters are strongly biased. 
However in this case, the fold and cusp models do not produce the same recoveries, as the injected signals have three component images that cannot be fully reproduced with the two fold images. 
In terms of the recovered SNRs, we obtain $\rho_\mathrm{unlensed}=43.9\pm1.7$, $\rho_\mathrm{fold}=51.0\pm1.6$ and $\rho_\mathrm{cusp}=57.5\pm1.6$ at $90\%$ credible intervals, which corroborates our previous statement.  
The corresponding posterior samples for a reduced set of parameters are presented in Fig. \ref{fig:cusp_injection}. 
We plot the same ones as in Fig. \ref{fig:fold_injection}, and add the additional cusp parameters ($\Delta t_{20}$ and $\mu_\mathrm{rel}$), to demonstrate that they are recovered accurately. 
In fact, it is noticeable that for this high SNR event the time delays and relative magnifications are exquisitely recovered with $\sim$few percent level errors. 
In the figure, it is again clear that the unlensed recovery is highly biased. 
We also observe that the fold recovery converges to match the first pair of first arriving cusp images as can be seen by their similar $\Delta t_{10}$. 
Nonetheless, because of the presence of the third image and the relative magnification between the first two that are not accounted for in the fold model, we find that the chirp mass and the mass ratio recovered by the fold are biased for this example, although to a much lesser degree than the unlensed model.  

Finally, we have also explored the recovery of an unlensed signal with the three models. 
We find that all models are able to reproduce the simulated values and lead to a similar SNR $\sim 41.8\pm 1.6$. 
Moreover, the time delays of the fold and cusp narrowly peak at 0, aligning the peak amplitude of all images. 
For the cusp, the relative magnification also narrowly peaks at 1, to maximize the similarity with a single image. 
Note that there are known degeneracies between precessing systems and lensed signals~\cite{Liu:2023emk}, as both could lead to modulations of the time domain signal. 
However, our simulated event does not exhibit significant precession, so we are not as susceptible to these degeneracies. 

%-----
%SECTION: CONCLUSIONS
%-----
\section{Conclusions}
\label{sec:conclusions}

Gravitational lensing opens the possibility to observe the high-redshift Universe, and look deeper into gravitational interactions and astrophysical environments. 
For certain source-lens configurations, gravitational lensing can significantly boost the source's amplitude, making otherwise unreachable sources detectable. 
In the context of gravitational-wave detectors, because their current sensitivities only allow them to observe relatively nearby compact binary coalescences, $z\lesssim1$, there is a selection bias for the potential lensed signals to have large magnifications~\cite{Oguri:2018muv}.  
Interestingly, when we focus on the high-magnification tail of the distribution,  
current gravitational-wave detectors are already essentially sensitive to any highly magnified lensed signal in the Universe~\cite{Lo:2024wqm}. 

Highly magnified lensed signals occur near caustics. 
Caustics correspond to the locations in the source plane that, in the stationary phase approximation, limit regions of different image multiplicities and where the magnification diverges.
For gravitational waves from compact binary coalescences, this divergence is prevented by the diffraction due to their large wavelengths, which is proportional to their orbital separation during the inspiral. 
Diffraction, therefore, sets the maximum magnification of a gravitational wave near a caustic and also its potential waveform distortions. 
Conveniently, close enough to a caustic, its properties follow universal relations, regardless of the large-scale lens model. 
For galaxy-scale lenses and a population of binary black holes observed with current ground-based detectors, we expect about 1 in every 1000 detectable events to be strongly lensed \cite{Ng:2017yiu,Lai:2018rto,Xu:2021bfn}.
Due to selection effects, such lensed events will have an average magnification larger than 10~\cite{Oguri:2018muv}. 
Moreover, 1 every 100 strongly lensed GW events by galaxies will have $\mu>100$~\cite{Oguri:2018muv}. 
Cluster-scale lenses could increase the rate of strongly lensed GWs, and their substructures can enhance the high magnification tail~\cite{Vujeva:2025kko}. 
The regime of diffraction around caustics increases for longer signals such as binary neutron stars, whose rate of strong lensing will only become significant after a few years with current detectors at their design sensitivity~\cite{Smith:2022vbp}.
 
We have explored diffraction around different types of caustic: point, fold and cusp. 
Point caustics only appear in axi-symmetric lenses and are therefore unstable under any non axi-symmetric perturbation of the lensing potential. 
Folds and cusps are the only stable caustics on two dimensional mappings. 
Higher-order stable caustics are also possible, but they require additional parameters in the lens model. 
For these three caustic models, we have derived analytical expressions for the amplification function within the stationary phase approximation and in the wave optics regime. 
The results in the stationary phase approximation have all been derived long ago, cf. Ref.~\cite{Schneider:1992}, but they are useful to compare with the wave optics amplification. 
On the other hand, the amplitude and phase of the diffraction patterns had only been computed for the point caustic~\cite{NakamuraDeguchi}. 
For the fold and the cusps, only the amplitude was computed, see e.g.~\cite{Schneider:1992}, because this is the relevant quantity for electromagnetic lensing. 
In Ref.~\cite{Lo:2024wqm}, we took this additional step for the fold caustic, which allowed us to compute the waveform distortions in lensed gravitational waves. 
Here, we have provided further details on the derivation of this result for the fold, and also computed it for the cusp. 

With the wave optics amplification functions at hand, 
we investigated the detectability of these features on gravitational waves. 
We found that lensing near caustics leads to significant distortions compared to unlensed waveforms. 
We quantified this both with mismatch calculations and Bayesian parameter estimation on simulated lensed signals. 
These large modifications in the morphology of the signal could hinder its detectability with standard matched-filtering search pipelines~\cite{Chan:2024qmb}. 
However, we point out that the overlapping repeated chirps produced by lensing lead to distinguishable peaks in the signal-to-noise time series. 
This suggests that these signals could be detectable with current facilities, provided that the search pipelines are extended to incorporate this kind of phenomenon. 

While wave optics incorporates both interference and diffraction, we have found that for gravitational waves, even in regions close to the caustic, the waveform distortions are very well described by the interference of the repeated chirps computed in the stationary phase approximation. 
Again, we have tested this with mismatch calculations and Bayesian inference. 
This result has important implications for future data analysis campaigns, as waveforms described in the stationary phase approximation are significantly simpler and more computationally efficient. 
This is relevant for both search and parameter estimation pipelines. 
We have made these lensed waveforms near caustics publicly available. 

Our work demonstrates that the waveform distortions produced by gravitational lensing near caustics could be detectable and used to single out a gravitational wave event as lensed. 
Our analysis of the accuracy and degeneracies in the parameter estimation represents a first step, but 
should be complemented in the future with a comprehensive systematic study. 
For example, in this work we have not explored potential waveform systematics or simulated multiple signals with different source properties. 
Along these lines, we have focused our detectability analysis to ground-based GW detectors. 
For future gravitational-wave detectors in space such as LISA, wave optics effects will also be relevant~\cite{LISACosmologyWorkingGroup:2022jok}, shedding light on the abundance of low-mass dark matter halos~\cite{Savastano:2023spl,Caliskan:2023zqm}.  
Moreover, we have restricted our study to the most common caustics. 
Higher order catastrophes are also possible~\cite{Berry_Upstill,Feldbrugge:2019fjs}. 
Finally, our analysis has only considered isolated caustics. 
Potential substructures could lead to deviations from the universal relations, 
for example in the presence of microlenses~\cite{Kayser_Refsdal_Stabell:1986}. 
We leave these relevant additions for future work. 

Observing diffraction around caustics will probe new aspects of gravity in the wave optics regime. 
Understanding these phenomena is essential to determine the detectability of lensed gravitational-wave signals and their possible connection with electromagnetic signals in multi-messenger lensing. 
Moreover, by setting the maximum magnification of a given source, diffraction is linked to the furthest binaries that a given GW detector could ever observe. 
Our work paves the way to answer these questions.

%-- Acknowledgments
\begin{acknowledgments}
We would like to thank Oleg Bulashenko for his valuable comments on the manuscript, Juno C. L. Chan for his feedback on the detectability of lensed signals, and Miguel Zumalac\'arregui for useful correspondence. 
We are thankful to the Erwin Schr\"odinger International
Institute for Mathematics and Physics for hosting the Workshop ``Lensing and Wave Optics in Strong Gravity", and to the workshop's organizers and participants, who created a vibrant environment where we got useful feedback on the results of this work. 
This project was supported by the research grant no.~VIL37766 and no.~VIL53101 from Villum Fonden, and the DNRF Chair program grant no. DNRF162 by the Danish National Research Foundation.
This project has received funding from the European Union's Horizon 2020 research and innovation programme under the Marie Sklodowska-Curie grant agreement No 101131233.  
JME is also supported by the Marie Sklodowska-Curie grant agreement No.~847523 INTERACTIONS. 
The Center of Gravity is a Center of Excellence funded by the Danish National Research Foundation under grant No. 184. 
The Tycho supercomputer hosted at the SCIENCE HPC center at the University of Copenhagen was used for supporting this work. 
\end{acknowledgments}

\begin{figure*}
    \centering
    \includegraphics[width=\linewidth]{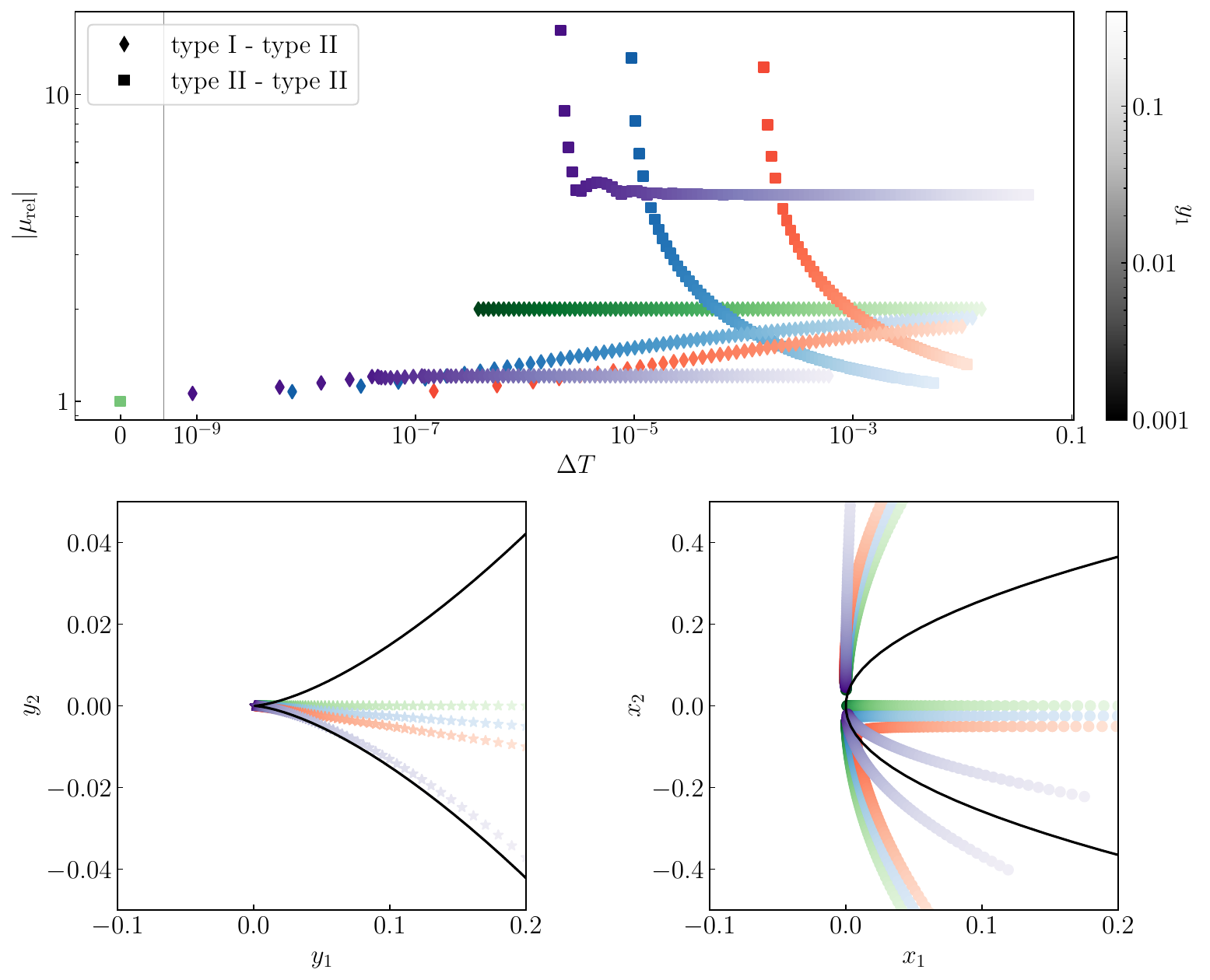}
    \vspace{-20pt}
    \caption{Relative magnifications $\mu_\mathrm{rel}$ and (dimensionless) time delays $\Delta T$ for different source positions within a cusp caustic, i.e. 3 image region. 
    The trajectories are displayed in the source plane in the bottom left panel. 
    Their corresponding image positions are presented in the bottom right panel. 
    We show the properties of $(\Delta T,\mu_\mathrm{rel})$ for the pair of opposite parity images and the pair of images with equal parity. 
    We consider the same cusp properties as in the main text: $T_{11}(x_c)=1$, $T_{122}(x_c)=1$, and $T_{2222}(x_c)=-1$. 
    }
    \label{fig:xaust_angles}
\end{figure*}

\appendix
\section{Approaching the cusp through different trajectories}
\label{app:caustic_angles}

The properties of the three images generated by a cusp caustic vary significantly depending on the source position. 
Limiting cases are \emph{i)} the cusp itself, where the three images merge, \emph{ii)} the cusp symmetry axis, where the pair of equal parity images arrive at the same time, and \emph{iii)} the folds, where two of the images with opposite parity recover the universal behavior of the fold of equal magnification.

To explore the possible sets of images, in the top panel of Fig. \ref{fig:xaust_angles} we plot the relative magnifications and time delays between the three images. 
As shown in the bottom panels, the cusp is the same as in the main text, cf.~ Fig. \ref{fig:source_image_planes_cusp_caustic}. 
The bottom left panel shows the source plane. 
Each color indicates a different trajectory. 
From top to bottom, the first three are straight lines of different slopes, while the fourth line follows the shape of the cusp caustic. 
Each of the points in these trajectories will produce three images that are depicted in the image plane in the lower right panel. 
As the trajectory gets closer to the fold, the two images at $x_2<0$ get closer to each other. 
The transparency of the color of the points is meant to indicate how close they are to the cusp, as measured by the value of $y_1$ which is color-coded in the color-bar. 

The top panel of this figure shows the relation between the magnifications and time delays for the pair of images with equal parity (squares) and one of the pairs with opposite parity (rhombi). 
For the trajectory along the symmetry axis, the equal-parity images arrive at the same time and have the same magnification for any $y_1$ position. Therefore, there is just a single square at $\mu_\mathrm{rel}=1$ and $\Delta T=0$. 
For the other pair in this trajectory, the relative magnification is fixed, while the time delay decreases as one gets closer to the cusp. 
The other trajectories display a wide range of relative magnifications. 

%------------
%------------
\bibliographystyle{apsrev4-1}
\bibliography{gw_lensing}

\begin{thebibliography}{71}%
\makeatletter
\providecommand \@ifxundefined [1]{%
 \@ifx{#1\undefined}
}%
\providecommand \@ifnum [1]{%
 \ifnum #1\expandafter \@firstoftwo
 \else \expandafter \@secondoftwo
 \fi
}%
\providecommand \@ifx [1]{%
 \ifx #1\expandafter \@firstoftwo
 \else \expandafter \@secondoftwo
 \fi
}%
\providecommand \natexlab [1]{#1}%
\providecommand \enquote  [1]{``#1''}%
\providecommand \bibnamefont  [1]{#1}%
\providecommand \bibfnamefont [1]{#1}%
\providecommand \citenamefont [1]{#1}%
\providecommand \href@noop [0]{\@secondoftwo}%
\providecommand \href [0]{\begingroup \@sanitize@url \@href}%
\providecommand \@href[1]{\@@startlink{#1}\@@href}%
\providecommand \@@href[1]{\endgroup#1\@@endlink}%
\providecommand \@sanitize@url [0]{\catcode `\\12\catcode `\$12\catcode
  `\&12\catcode `\#12\catcode `\^12\catcode `\_12\catcode `\%12\relax}%
\providecommand \@@startlink[1]{}%
\providecommand \@@endlink[0]{}%
\providecommand \url  [0]{\begingroup\@sanitize@url \@url }%
\providecommand \@url [1]{\endgroup\@href {#1}{\urlprefix }}%
\providecommand \urlprefix  [0]{URL }%
\providecommand \Eprint [0]{\href }%
\providecommand \doibase [0]{http://dx.doi.org/}%
\providecommand \selectlanguage [0]{\@gobble}%
\providecommand \bibinfo  [0]{\@secondoftwo}%
\providecommand \bibfield  [0]{\@secondoftwo}%
\providecommand \translation [1]{[#1]}%
\providecommand \BibitemOpen [0]{}%
\providecommand \bibitemStop [0]{}%
\providecommand \bibitemNoStop [0]{.\EOS\space}%
\providecommand \EOS [0]{\spacefactor3000\relax}%
\providecommand \BibitemShut  [1]{\csname bibitem#1\endcsname}%
\let\auto@bib@innerbib\@empty
%</preamble>
\bibitem [{\citenamefont {Diego}(2019)}]{Diego:2018fzr}%
  \BibitemOpen
  \bibfield  {author} {\bibinfo {author} {\bibfnamefont {J.~M.}\ \bibnamefont
  {Diego}},\ }\href {\doibase 10.1051/0004-6361/201833670} {\bibfield
  {journal} {\bibinfo  {journal} {Astron. Astrophys.}\ }\textbf {\bibinfo
  {volume} {625}},\ \bibinfo {pages} {A84} (\bibinfo {year} {2019})},\ \Eprint
  {http://arxiv.org/abs/1806.04668} {arXiv:1806.04668 [astro-ph.GA]}
  \BibitemShut {NoStop}%
\bibitem [{\citenamefont {Welch~et al.}(2022)}]{Welch_2022}%
  \BibitemOpen
  \bibfield  {author} {\bibinfo {author} {\bibfnamefont {B.}~\bibnamefont
  {Welch~et al.}},\ }\href {\doibase 10.1038/s41586-022-04449-y} {\bibfield
  {journal} {\bibinfo  {journal} {Nature}\ }\textbf {\bibinfo {volume} {603}},\
  \bibinfo {pages} {815–818} (\bibinfo {year} {2022})}\BibitemShut {NoStop}%
\bibitem [{\citenamefont {{Wang et al.}}(2023)}]{2023ApJ...957L..34W}%
  \BibitemOpen
  \bibfield  {author} {\bibinfo {author} {\bibfnamefont {B.}~\bibnamefont
  {{Wang et al.}}},\ }\href {\doibase 10.3847/2041-8213/acfe07} {\bibfield
  {journal} {\bibinfo  {journal} {Astrophysical Journal, Letters}\ }\textbf
  {\bibinfo {volume} {957}},\ \bibinfo {eid} {L34} (\bibinfo {year} {2023})},\
  \Eprint {http://arxiv.org/abs/2308.03745} {arXiv:2308.03745 [astro-ph.GA]}
  \BibitemShut {NoStop}%
\bibitem [{\citenamefont {{Ohanian}}(1983)}]{Ohanian:1983}%
  \BibitemOpen
  \bibfield  {author} {\bibinfo {author} {\bibfnamefont {H.~C.}\ \bibnamefont
  {{Ohanian}}},\ }\href {\doibase 10.1086/161221} {\bibfield  {journal}
  {\bibinfo  {journal} {Astrophysical Journal}\ }\textbf {\bibinfo {volume}
  {271}},\ \bibinfo {pages} {551} (\bibinfo {year} {1983})}\BibitemShut
  {NoStop}%
\bibitem [{\citenamefont {Schneider}\ \emph {et~al.}(1992)\citenamefont
  {Schneider}, \citenamefont {Ehlers},\ and\ \citenamefont
  {Falco}}]{Schneider:1992}%
  \BibitemOpen
  \bibfield  {author} {\bibinfo {author} {\bibfnamefont {P.}~\bibnamefont
  {Schneider}}, \bibinfo {author} {\bibfnamefont {J.}~\bibnamefont {Ehlers}}, \
  and\ \bibinfo {author} {\bibfnamefont {E.}~\bibnamefont {Falco}},\ }\href
  {\doibase 10.1007/978-3-662-03758-4} {\emph {\bibinfo {title} {{
  Gravitational Lenses}}}}\ (\bibinfo  {publisher} {Springer-Verlag Berlin
  Heidelberg},\ \bibinfo {year} {1992})\BibitemShut {NoStop}%
\bibitem [{\citenamefont {{Miralda-Escude}}(1991)}]{Miralda-Escude:1991}%
  \BibitemOpen
  \bibfield  {author} {\bibinfo {author} {\bibfnamefont {J.}~\bibnamefont
  {{Miralda-Escude}}},\ }\href {\doibase 10.1086/170486} {\bibfield  {journal}
  {\bibinfo  {journal} {Astrophysical Journal}\ }\textbf {\bibinfo {volume}
  {379}},\ \bibinfo {pages} {94} (\bibinfo {year} {1991})}\BibitemShut
  {NoStop}%
\bibitem [{\citenamefont {{Chang}}\ and\ \citenamefont
  {{Refsdal}}(1979)}]{Chang_Refsdal:1979}%
  \BibitemOpen
  \bibfield  {author} {\bibinfo {author} {\bibfnamefont {K.}~\bibnamefont
  {{Chang}}}\ and\ \bibinfo {author} {\bibfnamefont {S.}~\bibnamefont
  {{Refsdal}}},\ }\href {\doibase 10.1038/282561a0} {\bibfield  {journal}
  {\bibinfo  {journal} {Nature}\ }\textbf {\bibinfo {volume} {282}},\ \bibinfo
  {pages} {561} (\bibinfo {year} {1979})}\BibitemShut {NoStop}%
\bibitem [{\citenamefont {{Chang}}\ and\ \citenamefont
  {{Refsdal}}(1984)}]{Chang_Refsdal:1984}%
  \BibitemOpen
  \bibfield  {author} {\bibinfo {author} {\bibfnamefont {K.}~\bibnamefont
  {{Chang}}}\ and\ \bibinfo {author} {\bibfnamefont {S.}~\bibnamefont
  {{Refsdal}}},\ }\href@noop {} {\bibfield  {journal} {\bibinfo  {journal}
  {Astronomy and Astrophysics}\ }\textbf {\bibinfo {volume} {132}},\ \bibinfo
  {pages} {168} (\bibinfo {year} {1984})}\BibitemShut {NoStop}%
\bibitem [{\citenamefont {{Kayser}}\ \emph {et~al.}(1986)\citenamefont
  {{Kayser}}, \citenamefont {{Refsdal}},\ and\ \citenamefont
  {{Stabell}}}]{Kayser_Refsdal_Stabell:1986}%
  \BibitemOpen
  \bibfield  {author} {\bibinfo {author} {\bibfnamefont {R.}~\bibnamefont
  {{Kayser}}}, \bibinfo {author} {\bibfnamefont {S.}~\bibnamefont {{Refsdal}}},
  \ and\ \bibinfo {author} {\bibfnamefont {R.}~\bibnamefont {{Stabell}}},\
  }\href@noop {} {\bibfield  {journal} {\bibinfo  {journal} {Astronomy and
  Astrophysics}\ }\textbf {\bibinfo {volume} {166}},\ \bibinfo {pages} {36}
  (\bibinfo {year} {1986})}\BibitemShut {NoStop}%
\bibitem [{\citenamefont {Blandford}\ and\ \citenamefont
  {Narayan}(1986)}]{Blandford:1986zz}%
  \BibitemOpen
  \bibfield  {author} {\bibinfo {author} {\bibfnamefont {R.}~\bibnamefont
  {Blandford}}\ and\ \bibinfo {author} {\bibfnamefont {R.}~\bibnamefont
  {Narayan}},\ }\href {\doibase 10.1086/164709} {\bibfield  {journal} {\bibinfo
   {journal} {Astrophys. J.}\ }\textbf {\bibinfo {volume} {310}},\ \bibinfo
  {pages} {568} (\bibinfo {year} {1986})}\BibitemShut {NoStop}%
\bibitem [{\citenamefont {{Narayan}}(1992)}]{Narayan1992_pulsar_scintillation}%
  \BibitemOpen
  \bibfield  {author} {\bibinfo {author} {\bibfnamefont {R.}~\bibnamefont
  {{Narayan}}},\ }\href {\doibase 10.1098/rsta.1992.0090} {\bibfield  {journal}
  {\bibinfo  {journal} {Philosophical Transactions of the Royal Society of
  London Series A}\ }\textbf {\bibinfo {volume} {341}},\ \bibinfo {pages} {151}
  (\bibinfo {year} {1992})}\BibitemShut {NoStop}%
\bibitem [{\citenamefont {Jow}\ \emph {et~al.}(2020)\citenamefont {Jow},
  \citenamefont {Foreman}, \citenamefont {Pen},\ and\ \citenamefont
  {Zhu}}]{Jow:2020rcy}%
  \BibitemOpen
  \bibfield  {author} {\bibinfo {author} {\bibfnamefont {D.~L.}\ \bibnamefont
  {Jow}}, \bibinfo {author} {\bibfnamefont {S.}~\bibnamefont {Foreman}},
  \bibinfo {author} {\bibfnamefont {U.-L.}\ \bibnamefont {Pen}}, \ and\
  \bibinfo {author} {\bibfnamefont {W.}~\bibnamefont {Zhu}},\ }\href {\doibase
  10.1093/mnras/staa2230} {\bibfield  {journal} {\bibinfo  {journal} {Mon. Not.
  Roy. Astron. Soc.}\ }\textbf {\bibinfo {volume} {497}},\ \bibinfo {pages}
  {4956} (\bibinfo {year} {2020})},\ \Eprint {http://arxiv.org/abs/2002.01570}
  {arXiv:2002.01570 [astro-ph.HE]} \BibitemShut {NoStop}%
\bibitem [{\citenamefont {Misner}\ \emph {et~al.}(2017)\citenamefont {Misner},
  \citenamefont {Thorne},\ and\ \citenamefont
  {Wheeler}}]{misner2017gravitation}%
  \BibitemOpen
  \bibfield  {author} {\bibinfo {author} {\bibfnamefont {C.~W.}\ \bibnamefont
  {Misner}}, \bibinfo {author} {\bibfnamefont {K.~S.}\ \bibnamefont {Thorne}},
  \ and\ \bibinfo {author} {\bibfnamefont {J.~A.}\ \bibnamefont {Wheeler}},\
  }\href@noop {} {\emph {\bibinfo {title} {Gravitation}}}\ (\bibinfo
  {publisher} {Princeton University Press},\ \bibinfo {year}
  {2017})\BibitemShut {NoStop}%
\bibitem [{\citenamefont {{Bontz}}\ and\ \citenamefont
  {{Haugan}}(1981)}]{Bontz_Haugan:1981}%
  \BibitemOpen
  \bibfield  {author} {\bibinfo {author} {\bibfnamefont {R.~J.}\ \bibnamefont
  {{Bontz}}}\ and\ \bibinfo {author} {\bibfnamefont {M.~P.}\ \bibnamefont
  {{Haugan}}},\ }\href {\doibase 10.1007/BF00654034} {\bibfield  {journal}
  {\bibinfo  {journal} {Astrophysics and Space Science}\ }\textbf {\bibinfo
  {volume} {78}},\ \bibinfo {pages} {199} (\bibinfo {year} {1981})}\BibitemShut
  {NoStop}%
\bibitem [{\citenamefont {Deguchi}\ and\ \citenamefont
  {Watson}(1986)}]{Deguchi:1986zz}%
  \BibitemOpen
  \bibfield  {author} {\bibinfo {author} {\bibfnamefont {S.}~\bibnamefont
  {Deguchi}}\ and\ \bibinfo {author} {\bibfnamefont {W.~D.}\ \bibnamefont
  {Watson}},\ }\href {\doibase 10.1103/PhysRevD.34.1708} {\bibfield  {journal}
  {\bibinfo  {journal} {Phys. Rev. D}\ }\textbf {\bibinfo {volume} {34}},\
  \bibinfo {pages} {1708} (\bibinfo {year} {1986})}\BibitemShut {NoStop}%
\bibitem [{\citenamefont {Nakamura}\ and\ \citenamefont
  {Deguchi}(1999)}]{NakamuraDeguchi}%
  \BibitemOpen
  \bibfield  {author} {\bibinfo {author} {\bibfnamefont {T.~T.}\ \bibnamefont
  {Nakamura}}\ and\ \bibinfo {author} {\bibfnamefont {S.}~\bibnamefont
  {Deguchi}},\ }\href {\doibase 10.1143/PTPS.133.137} {\bibfield  {journal}
  {\bibinfo  {journal} {Progress of Theoretical Physics Supplement}\ }\textbf
  {\bibinfo {volume} {133}},\ \bibinfo {pages} {137} (\bibinfo {year}
  {1999})},\ \Eprint
  {http://arxiv.org/abs/https://academic.oup.com/ptps/article-pdf/doi/10.1143/PTPS.133.137/5283012/133-137.pdf}
  {https://academic.oup.com/ptps/article-pdf/doi/10.1143/PTPS.133.137/5283012/133-137.pdf}
  \BibitemShut {NoStop}%
\bibitem [{\citenamefont {Ezquiaga}\ \emph {et~al.}(2020)\citenamefont
  {Ezquiaga}, \citenamefont {Hu},\ and\ \citenamefont
  {Lagos}}]{Ezquiaga:2020spg}%
  \BibitemOpen
  \bibfield  {author} {\bibinfo {author} {\bibfnamefont {J.~M.}\ \bibnamefont
  {Ezquiaga}}, \bibinfo {author} {\bibfnamefont {W.}~\bibnamefont {Hu}}, \ and\
  \bibinfo {author} {\bibfnamefont {M.}~\bibnamefont {Lagos}},\ }\href
  {\doibase 10.1103/PhysRevD.102.023531} {\bibfield  {journal} {\bibinfo
  {journal} {Phys. Rev. D}\ }\textbf {\bibinfo {volume} {102}},\ \bibinfo
  {pages} {023531} (\bibinfo {year} {2020})},\ \Eprint
  {http://arxiv.org/abs/2005.10702} {arXiv:2005.10702 [astro-ph.CO]}
  \BibitemShut {NoStop}%
\bibitem [{\citenamefont {Cremonese}\ \emph {et~al.}(2021)\citenamefont
  {Cremonese}, \citenamefont {Ezquiaga},\ and\ \citenamefont
  {Salzano}}]{Cremonese:2021puh}%
  \BibitemOpen
  \bibfield  {author} {\bibinfo {author} {\bibfnamefont {P.}~\bibnamefont
  {Cremonese}}, \bibinfo {author} {\bibfnamefont {J.~M.}\ \bibnamefont
  {Ezquiaga}}, \ and\ \bibinfo {author} {\bibfnamefont {V.}~\bibnamefont
  {Salzano}},\ }\href {\doibase 10.1103/PhysRevD.104.023503} {\bibfield
  {journal} {\bibinfo  {journal} {Phys. Rev. D}\ }\textbf {\bibinfo {volume}
  {104}},\ \bibinfo {pages} {023503} (\bibinfo {year} {2021})},\ \Eprint
  {http://arxiv.org/abs/2104.07055} {arXiv:2104.07055 [astro-ph.CO]}
  \BibitemShut {NoStop}%
\bibitem [{\citenamefont {Chen}\ \emph {et~al.}(2024)\citenamefont {Chen},
  \citenamefont {Cremonese}, \citenamefont {Ezquiaga},\ and\ \citenamefont
  {Keitel}}]{Chen:2024xal}%
  \BibitemOpen
  \bibfield  {author} {\bibinfo {author} {\bibfnamefont {A.}~\bibnamefont
  {Chen}}, \bibinfo {author} {\bibfnamefont {P.}~\bibnamefont {Cremonese}},
  \bibinfo {author} {\bibfnamefont {J.~M.}\ \bibnamefont {Ezquiaga}}, \ and\
  \bibinfo {author} {\bibfnamefont {D.}~\bibnamefont {Keitel}},\ }\href
  {\doibase 10.1103/PhysRevD.110.123015} {\bibfield  {journal} {\bibinfo
  {journal} {Phys. Rev. D}\ }\textbf {\bibinfo {volume} {110}},\ \bibinfo
  {pages} {123015} (\bibinfo {year} {2024})},\ \Eprint
  {http://arxiv.org/abs/2408.03856} {arXiv:2408.03856 [astro-ph.CO]}
  \BibitemShut {NoStop}%
\bibitem [{\citenamefont {Misner}\ \emph {et~al.}(1973)\citenamefont {Misner},
  \citenamefont {Thorne},\ and\ \citenamefont {Wheeler}}]{Misner:1974qy}%
  \BibitemOpen
  \bibfield  {author} {\bibinfo {author} {\bibfnamefont {C.~W.}\ \bibnamefont
  {Misner}}, \bibinfo {author} {\bibfnamefont {K.~S.}\ \bibnamefont {Thorne}},
  \ and\ \bibinfo {author} {\bibfnamefont {J.~A.}\ \bibnamefont {Wheeler}},\
  }\href@noop {} {\emph {\bibinfo {title} {{Gravitation}}}}\ (\bibinfo
  {publisher} {W. H. Freeman},\ \bibinfo {address} {San Francisco},\ \bibinfo
  {year} {1973})\BibitemShut {NoStop}%
%%CITATION = INSPIRE-95654;%%
\bibitem [{\citenamefont {{Berry}}\ and\ \citenamefont
  {{Upstill}}(1980)}]{Berry_Upstill}%
  \BibitemOpen
  \bibfield  {author} {\bibinfo {author} {\bibfnamefont {M.~V.}\ \bibnamefont
  {{Berry}}}\ and\ \bibinfo {author} {\bibfnamefont {C.}~\bibnamefont
  {{Upstill}}},\ }\href {\doibase 10.1016/S0079-6638(08)70215-4} {\bibfield
  {journal} {\bibinfo  {journal} {Progess in Optics}\ }\textbf {\bibinfo
  {volume} {18}},\ \bibinfo {pages} {257} (\bibinfo {year} {1980})}\BibitemShut
  {NoStop}%
\bibitem [{\citenamefont {Feldbrugge}\ \emph {et~al.}(2023)\citenamefont
  {Feldbrugge}, \citenamefont {Pen},\ and\ \citenamefont
  {Turok}}]{Feldbrugge:2019fjs}%
  \BibitemOpen
  \bibfield  {author} {\bibinfo {author} {\bibfnamefont {J.}~\bibnamefont
  {Feldbrugge}}, \bibinfo {author} {\bibfnamefont {U.-L.}\ \bibnamefont {Pen}},
  \ and\ \bibinfo {author} {\bibfnamefont {N.}~\bibnamefont {Turok}},\ }\href
  {\doibase 10.1016/j.aop.2023.169255} {\bibfield  {journal} {\bibinfo
  {journal} {Annals Phys.}\ }\textbf {\bibinfo {volume} {451}},\ \bibinfo
  {pages} {169255} (\bibinfo {year} {2023})},\ \Eprint
  {http://arxiv.org/abs/1909.04632} {arXiv:1909.04632 [astro-ph.HE]}
  \BibitemShut {NoStop}%
\bibitem [{\citenamefont {Ulmer}\ and\ \citenamefont
  {Goodman}(1995)}]{Ulmer:1994ij}%
  \BibitemOpen
  \bibfield  {author} {\bibinfo {author} {\bibfnamefont {A.}~\bibnamefont
  {Ulmer}}\ and\ \bibinfo {author} {\bibfnamefont {J.}~\bibnamefont
  {Goodman}},\ }\href {\doibase 10.1086/175422} {\bibfield  {journal} {\bibinfo
   {journal} {Astrophys. J.}\ }\textbf {\bibinfo {volume} {442}},\ \bibinfo
  {pages} {67} (\bibinfo {year} {1995})},\ \Eprint
  {http://arxiv.org/abs/astro-ph/9406042} {arXiv:astro-ph/9406042} \BibitemShut
  {NoStop}%
\bibitem [{\citenamefont {Villarrubia-Rojo}\ \emph {et~al.}(2024)\citenamefont
  {Villarrubia-Rojo}, \citenamefont {Savastano}, \citenamefont
  {Zumalac\'arregui}, \citenamefont {Choi}, \citenamefont {Goyal},
  \citenamefont {Dai},\ and\ \citenamefont
  {Tambalo}}]{Villarrubia-Rojo:2024xcj}%
  \BibitemOpen
  \bibfield  {author} {\bibinfo {author} {\bibfnamefont {H.}~\bibnamefont
  {Villarrubia-Rojo}}, \bibinfo {author} {\bibfnamefont {S.}~\bibnamefont
  {Savastano}}, \bibinfo {author} {\bibfnamefont {M.}~\bibnamefont
  {Zumalac\'arregui}}, \bibinfo {author} {\bibfnamefont {L.}~\bibnamefont
  {Choi}}, \bibinfo {author} {\bibfnamefont {S.}~\bibnamefont {Goyal}},
  \bibinfo {author} {\bibfnamefont {L.}~\bibnamefont {Dai}}, \ and\ \bibinfo
  {author} {\bibfnamefont {G.}~\bibnamefont {Tambalo}},\ }\href@noop {} {\
  (\bibinfo {year} {2024})},\ \Eprint {http://arxiv.org/abs/2409.04606}
  {arXiv:2409.04606 [gr-qc]} \BibitemShut {NoStop}%
\bibitem [{\citenamefont {Lo}\ \emph {et~al.}(2025)\citenamefont {Lo},
  \citenamefont {Vujeva}, \citenamefont {Ezquiaga},\ and\ \citenamefont
  {Chan}}]{Lo:2024wqm}%
  \BibitemOpen
  \bibfield  {author} {\bibinfo {author} {\bibfnamefont {R.~K.~L.}\
  \bibnamefont {Lo}}, \bibinfo {author} {\bibfnamefont {L.}~\bibnamefont
  {Vujeva}}, \bibinfo {author} {\bibfnamefont {J.~M.}\ \bibnamefont
  {Ezquiaga}}, \ and\ \bibinfo {author} {\bibfnamefont {J.~C.~L.}\ \bibnamefont
  {Chan}},\ }\href@noop {} {\bibfield  {journal} {\bibinfo  {journal} {to
  appear in Physical Review Letters}\ } (\bibinfo {year} {2025})},\ \Eprint
  {http://arxiv.org/abs/2407.17547} {arXiv:2407.17547 [gr-qc]} \BibitemShut
  {NoStop}%
\bibitem [{\citenamefont {Aasi}\ \emph {et~al.}(2015)\citenamefont {Aasi} \emph
  {et~al.}}]{LIGOScientific:2014pky}%
  \BibitemOpen
  \bibfield  {author} {\bibinfo {author} {\bibfnamefont {J.}~\bibnamefont
  {Aasi}} \emph {et~al.} (\bibinfo {collaboration} {LIGO Scientific}),\ }\href
  {\doibase 10.1088/0264-9381/32/7/074001} {\bibfield  {journal} {\bibinfo
  {journal} {Class. Quant. Grav.}\ }\textbf {\bibinfo {volume} {32}},\ \bibinfo
  {pages} {074001} (\bibinfo {year} {2015})},\ \Eprint
  {http://arxiv.org/abs/1411.4547} {arXiv:1411.4547 [gr-qc]} \BibitemShut
  {NoStop}%
\bibitem [{\citenamefont {Acernese}\ \emph {et~al.}(2015)\citenamefont
  {Acernese} \emph {et~al.}}]{VIRGO:2014yos}%
  \BibitemOpen
  \bibfield  {author} {\bibinfo {author} {\bibfnamefont {F.}~\bibnamefont
  {Acernese}} \emph {et~al.} (\bibinfo {collaboration} {VIRGO}),\ }\href
  {\doibase 10.1088/0264-9381/32/2/024001} {\bibfield  {journal} {\bibinfo
  {journal} {Class. Quant. Grav.}\ }\textbf {\bibinfo {volume} {32}},\ \bibinfo
  {pages} {024001} (\bibinfo {year} {2015})},\ \Eprint
  {http://arxiv.org/abs/1408.3978} {arXiv:1408.3978 [gr-qc]} \BibitemShut
  {NoStop}%
\bibitem [{\citenamefont {Akutsu}\ \emph {et~al.}(2021)\citenamefont {Akutsu}
  \emph {et~al.}}]{KAGRA:2020tym}%
  \BibitemOpen
  \bibfield  {author} {\bibinfo {author} {\bibfnamefont {T.}~\bibnamefont
  {Akutsu}} \emph {et~al.} (\bibinfo {collaboration} {KAGRA}),\ }\href
  {\doibase 10.1093/ptep/ptaa125} {\bibfield  {journal} {\bibinfo  {journal}
  {PTEP}\ }\textbf {\bibinfo {volume} {2021}},\ \bibinfo {pages} {05A101}
  (\bibinfo {year} {2021})},\ \Eprint {http://arxiv.org/abs/2005.05574}
  {arXiv:2005.05574 [physics.ins-det]} \BibitemShut {NoStop}%
\bibitem [{\citenamefont {Hannuksela}\ \emph {et~al.}(2019)\citenamefont
  {Hannuksela}, \citenamefont {Haris}, \citenamefont {Ng}, \citenamefont
  {Kumar}, \citenamefont {Mehta}, \citenamefont {Keitel}, \citenamefont {Li},\
  and\ \citenamefont {Ajith}}]{Hannuksela:2019kle}%
  \BibitemOpen
  \bibfield  {author} {\bibinfo {author} {\bibfnamefont {O.}~\bibnamefont
  {Hannuksela}}, \bibinfo {author} {\bibfnamefont {K.}~\bibnamefont {Haris}},
  \bibinfo {author} {\bibfnamefont {K.}~\bibnamefont {Ng}}, \bibinfo {author}
  {\bibfnamefont {S.}~\bibnamefont {Kumar}}, \bibinfo {author} {\bibfnamefont
  {A.}~\bibnamefont {Mehta}}, \bibinfo {author} {\bibfnamefont
  {D.}~\bibnamefont {Keitel}}, \bibinfo {author} {\bibfnamefont
  {T.}~\bibnamefont {Li}}, \ and\ \bibinfo {author} {\bibfnamefont
  {P.}~\bibnamefont {Ajith}},\ }\href {\doibase 10.3847/2041-8213/ab0c0f}
  {\bibfield  {journal} {\bibinfo  {journal} {Astrophys. J. Lett.}\ }\textbf
  {\bibinfo {volume} {874}},\ \bibinfo {pages} {L2} (\bibinfo {year} {2019})},\
  \Eprint {http://arxiv.org/abs/1901.02674} {arXiv:1901.02674 [gr-qc]}
  \BibitemShut {NoStop}%
\bibitem [{\citenamefont {Abbott}\ \emph {et~al.}(2021)\citenamefont {Abbott}
  \emph {et~al.}}]{LIGOScientific:2021izm}%
  \BibitemOpen
  \bibfield  {author} {\bibinfo {author} {\bibfnamefont {R.}~\bibnamefont
  {Abbott}} \emph {et~al.} (\bibinfo {collaboration} {LIGO Scientific,
  VIRGO}),\ }\href {\doibase 10.3847/1538-4357/ac23db} {\bibfield  {journal}
  {\bibinfo  {journal} {Astrophys. J.}\ }\textbf {\bibinfo {volume} {923}},\
  \bibinfo {pages} {14} (\bibinfo {year} {2021})},\ \Eprint
  {http://arxiv.org/abs/2105.06384} {arXiv:2105.06384 [gr-qc]} \BibitemShut
  {NoStop}%
\bibitem [{\citenamefont {Janquart}\ \emph {et~al.}(2023)\citenamefont
  {Janquart} \emph {et~al.}}]{Janquart:2023mvf}%
  \BibitemOpen
  \bibfield  {author} {\bibinfo {author} {\bibfnamefont {J.}~\bibnamefont
  {Janquart}} \emph {et~al.},\ }\href@noop {} {\  (\bibinfo {year} {2023})},\
  \Eprint {http://arxiv.org/abs/2306.03827} {arXiv:2306.03827 [gr-qc]}
  \BibitemShut {NoStop}%
\bibitem [{\citenamefont {Abbott}\ \emph {et~al.}(2023)\citenamefont {Abbott}
  \emph {et~al.}}]{LIGOScientific:2023bwz}%
  \BibitemOpen
  \bibfield  {author} {\bibinfo {author} {\bibfnamefont {R.}~\bibnamefont
  {Abbott}} \emph {et~al.} (\bibinfo {collaboration} {LIGO Scientific, VIRGO,
  KAGRA}),\ }\href@noop {} {\  (\bibinfo {year} {2023})},\ \Eprint
  {http://arxiv.org/abs/2304.08393} {arXiv:2304.08393 [gr-qc]} \BibitemShut
  {NoStop}%
\bibitem [{\citenamefont {Oguri}(2019)}]{Oguri:2019fix}%
  \BibitemOpen
  \bibfield  {author} {\bibinfo {author} {\bibfnamefont {M.}~\bibnamefont
  {Oguri}},\ }\href {\doibase 10.1088/1361-6633/ab4fc5} {\bibfield  {journal}
  {\bibinfo  {journal} {Rept. Prog. Phys.}\ }\textbf {\bibinfo {volume} {82}},\
  \bibinfo {pages} {126901} (\bibinfo {year} {2019})},\ \Eprint
  {http://arxiv.org/abs/1907.06830} {arXiv:1907.06830 [astro-ph.CO]}
  \BibitemShut {NoStop}%
\bibitem [{\citenamefont {Xu}\ \emph {et~al.}(2022)\citenamefont {Xu},
  \citenamefont {Ezquiaga},\ and\ \citenamefont {Holz}}]{Xu:2021bfn}%
  \BibitemOpen
  \bibfield  {author} {\bibinfo {author} {\bibfnamefont {F.}~\bibnamefont
  {Xu}}, \bibinfo {author} {\bibfnamefont {J.~M.}\ \bibnamefont {Ezquiaga}}, \
  and\ \bibinfo {author} {\bibfnamefont {D.~E.}\ \bibnamefont {Holz}},\ }\href
  {\doibase 10.3847/1538-4357/ac58f8} {\bibfield  {journal} {\bibinfo
  {journal} {Astrophys. J.}\ }\textbf {\bibinfo {volume} {929}},\ \bibinfo
  {pages} {9} (\bibinfo {year} {2022})},\ \Eprint
  {http://arxiv.org/abs/2105.14390} {arXiv:2105.14390 [astro-ph.CO]}
  \BibitemShut {NoStop}%
\bibitem [{\citenamefont {Vujeva}\ \emph {et~al.}(2025)\citenamefont {Vujeva},
  \citenamefont {Ezquiaga}, \citenamefont {Lo},\ and\ \citenamefont
  {Chan}}]{Vujeva:2025kko}%
  \BibitemOpen
  \bibfield  {author} {\bibinfo {author} {\bibfnamefont {L.}~\bibnamefont
  {Vujeva}}, \bibinfo {author} {\bibfnamefont {J.~M.}\ \bibnamefont
  {Ezquiaga}}, \bibinfo {author} {\bibfnamefont {R.~K.~L.}\ \bibnamefont {Lo}},
  \ and\ \bibinfo {author} {\bibfnamefont {J.~C.~L.}\ \bibnamefont {Chan}},\
  }\href@noop {} {\  (\bibinfo {year} {2025})},\ \Eprint
  {http://arxiv.org/abs/2501.02096} {arXiv:2501.02096 [astro-ph.CO]}
  \BibitemShut {NoStop}%
\bibitem [{\citenamefont {Takahashi}\ and\ \citenamefont
  {Nakamura}(2003)}]{Takahashi:2003ix}%
  \BibitemOpen
  \bibfield  {author} {\bibinfo {author} {\bibfnamefont {R.}~\bibnamefont
  {Takahashi}}\ and\ \bibinfo {author} {\bibfnamefont {T.}~\bibnamefont
  {Nakamura}},\ }\href {\doibase 10.1086/377430} {\bibfield  {journal}
  {\bibinfo  {journal} {Astrophys. J.}\ }\textbf {\bibinfo {volume} {595}},\
  \bibinfo {pages} {1039} (\bibinfo {year} {2003})},\ \Eprint
  {http://arxiv.org/abs/astro-ph/0305055} {arXiv:astro-ph/0305055} \BibitemShut
  {NoStop}%
\bibitem [{\citenamefont {Dai}\ and\ \citenamefont
  {Venumadhav}(2017)}]{Dai:2017huk}%
  \BibitemOpen
  \bibfield  {author} {\bibinfo {author} {\bibfnamefont {L.}~\bibnamefont
  {Dai}}\ and\ \bibinfo {author} {\bibfnamefont {T.}~\bibnamefont
  {Venumadhav}},\ }\href@noop {} {\  (\bibinfo {year} {2017})},\ \Eprint
  {http://arxiv.org/abs/1702.04724} {arXiv:1702.04724 [gr-qc]} \BibitemShut
  {NoStop}%
\bibitem [{\citenamefont {Ezquiaga}\ \emph {et~al.}(2021)\citenamefont
  {Ezquiaga}, \citenamefont {Holz}, \citenamefont {Hu}, \citenamefont {Lagos},\
  and\ \citenamefont {Wald}}]{Ezquiaga:2020gdt}%
  \BibitemOpen
  \bibfield  {author} {\bibinfo {author} {\bibfnamefont {J.~M.}\ \bibnamefont
  {Ezquiaga}}, \bibinfo {author} {\bibfnamefont {D.~E.}\ \bibnamefont {Holz}},
  \bibinfo {author} {\bibfnamefont {W.}~\bibnamefont {Hu}}, \bibinfo {author}
  {\bibfnamefont {M.}~\bibnamefont {Lagos}}, \ and\ \bibinfo {author}
  {\bibfnamefont {R.~M.}\ \bibnamefont {Wald}},\ }\href {\doibase
  10.1103/PhysRevD.103.064047} {\bibfield  {journal} {\bibinfo  {journal}
  {Phys. Rev. D}\ }\textbf {\bibinfo {volume} {103}},\ \bibinfo {pages}
  {064047} (\bibinfo {year} {2021})},\ \Eprint
  {http://arxiv.org/abs/2008.12814} {arXiv:2008.12814 [gr-qc]} \BibitemShut
  {NoStop}%
\bibitem [{\citenamefont {Wright}\ and\ \citenamefont
  {Hendry}(2021)}]{Wright:2021cbn}%
  \BibitemOpen
  \bibfield  {author} {\bibinfo {author} {\bibfnamefont {M.}~\bibnamefont
  {Wright}}\ and\ \bibinfo {author} {\bibfnamefont {M.}~\bibnamefont
  {Hendry}},\ }\href {\doibase 10.3847/1538-4357/ac7ec2} {\  (\bibinfo {year}
  {2021}),\ 10.3847/1538-4357/ac7ec2},\ \Eprint
  {http://arxiv.org/abs/2112.07012} {arXiv:2112.07012 [astro-ph.HE]}
  \BibitemShut {NoStop}%
\bibitem [{\citenamefont {Liu}\ \emph {et~al.}(2023)\citenamefont {Liu},
  \citenamefont {Wong}, \citenamefont {Leong}, \citenamefont {More},
  \citenamefont {Hannuksela},\ and\ \citenamefont {Li}}]{Liu:2023ikc}%
  \BibitemOpen
  \bibfield  {author} {\bibinfo {author} {\bibfnamefont {A.}~\bibnamefont
  {Liu}}, \bibinfo {author} {\bibfnamefont {I.~C.~F.}\ \bibnamefont {Wong}},
  \bibinfo {author} {\bibfnamefont {S.~H.~W.}\ \bibnamefont {Leong}}, \bibinfo
  {author} {\bibfnamefont {A.}~\bibnamefont {More}}, \bibinfo {author}
  {\bibfnamefont {O.~A.}\ \bibnamefont {Hannuksela}}, \ and\ \bibinfo {author}
  {\bibfnamefont {T.~G.~F.}\ \bibnamefont {Li}},\ }\href {\doibase
  10.1093/mnras/stad1302} {\bibfield  {journal} {\bibinfo  {journal} {Mon. Not.
  Roy. Astron. Soc.}\ }\textbf {\bibinfo {volume} {525}},\ \bibinfo {pages}
  {4149} (\bibinfo {year} {2023})},\ \Eprint {http://arxiv.org/abs/2302.09870}
  {arXiv:2302.09870 [gr-qc]} \BibitemShut {NoStop}%
\bibitem [{\citenamefont {{Burke}}(1981)}]{Burke:1981}%
  \BibitemOpen
  \bibfield  {author} {\bibinfo {author} {\bibfnamefont {W.~L.}\ \bibnamefont
  {{Burke}}},\ }\href {\doibase 10.1086/183466} {\bibfield  {journal} {\bibinfo
   {journal} {Astrophysical Journal, Letters}\ }\textbf {\bibinfo {volume}
  {244}},\ \bibinfo {pages} {L1} (\bibinfo {year} {1981})}\BibitemShut
  {NoStop}%
\bibitem [{\citenamefont {{Schneider}}(1984)}]{Schneider:1984}%
  \BibitemOpen
  \bibfield  {author} {\bibinfo {author} {\bibfnamefont {P.}~\bibnamefont
  {{Schneider}}},\ }\href@noop {} {\bibfield  {journal} {\bibinfo  {journal}
  {Astronomy and Astrophysics}\ }\textbf {\bibinfo {volume} {140}},\ \bibinfo
  {pages} {119} (\bibinfo {year} {1984})}\BibitemShut {NoStop}%
\bibitem [{\citenamefont {Serra}\ and\ \citenamefont
  {Bulashenko}(2025)}]{Serra:2025kbw}%
  \BibitemOpen
  \bibfield  {author} {\bibinfo {author} {\bibfnamefont {A.~M.}\ \bibnamefont
  {Serra}}\ and\ \bibinfo {author} {\bibfnamefont {O.}~\bibnamefont
  {Bulashenko}},\ }\href@noop {} {\  (\bibinfo {year} {2025})},\ \Eprint
  {http://arxiv.org/abs/2504.10128} {arXiv:2504.10128 [gr-qc]} \BibitemShut
  {NoStop}%
\bibitem [{\citenamefont {Bulashenko}\ and\ \citenamefont
  {Ubach}(2022)}]{Bulashenko:2021fes}%
  \BibitemOpen
  \bibfield  {author} {\bibinfo {author} {\bibfnamefont {O.}~\bibnamefont
  {Bulashenko}}\ and\ \bibinfo {author} {\bibfnamefont {H.}~\bibnamefont
  {Ubach}},\ }\href {\doibase 10.1088/1475-7516/2022/07/022} {\bibfield
  {journal} {\bibinfo  {journal} {JCAP}\ }\textbf {\bibinfo {volume} {07}},\
  \bibinfo {pages} {022} (\bibinfo {year} {2022})},\ \Eprint
  {http://arxiv.org/abs/2112.10773} {arXiv:2112.10773 [gr-qc]} \BibitemShut
  {NoStop}%
\bibitem [{\citenamefont {Tambalo}\ \emph {et~al.}(2023)\citenamefont
  {Tambalo}, \citenamefont {Zumalac\'arregui}, \citenamefont {Dai},\ and\
  \citenamefont {Cheung}}]{Tambalo:2022plm}%
  \BibitemOpen
  \bibfield  {author} {\bibinfo {author} {\bibfnamefont {G.}~\bibnamefont
  {Tambalo}}, \bibinfo {author} {\bibfnamefont {M.}~\bibnamefont
  {Zumalac\'arregui}}, \bibinfo {author} {\bibfnamefont {L.}~\bibnamefont
  {Dai}}, \ and\ \bibinfo {author} {\bibfnamefont {M.~H.-Y.}\ \bibnamefont
  {Cheung}},\ }\href {\doibase 10.1103/PhysRevD.108.043527} {\bibfield
  {journal} {\bibinfo  {journal} {Phys. Rev. D}\ }\textbf {\bibinfo {volume}
  {108}},\ \bibinfo {pages} {043527} (\bibinfo {year} {2023})},\ \Eprint
  {http://arxiv.org/abs/2210.05658} {arXiv:2210.05658 [gr-qc]} \BibitemShut
  {NoStop}%
\bibitem [{\citenamefont {Feldbrugge}\ and\ \citenamefont
  {Turok}(2020)}]{Feldbrugge:2020ycp}%
  \BibitemOpen
  \bibfield  {author} {\bibinfo {author} {\bibfnamefont {J.}~\bibnamefont
  {Feldbrugge}}\ and\ \bibinfo {author} {\bibfnamefont {N.}~\bibnamefont
  {Turok}},\ }\href@noop {} {\  (\bibinfo {year} {2020})},\ \Eprint
  {http://arxiv.org/abs/2008.01154} {arXiv:2008.01154 [gr-qc]} \BibitemShut
  {NoStop}%
\bibitem [{\citenamefont {Mishra}\ \emph {et~al.}(2021)\citenamefont {Mishra},
  \citenamefont {Meena}, \citenamefont {More}, \citenamefont {Bose},\ and\
  \citenamefont {Bagla}}]{Mishra:2021xzz}%
  \BibitemOpen
  \bibfield  {author} {\bibinfo {author} {\bibfnamefont {A.}~\bibnamefont
  {Mishra}}, \bibinfo {author} {\bibfnamefont {A.~K.}\ \bibnamefont {Meena}},
  \bibinfo {author} {\bibfnamefont {A.}~\bibnamefont {More}}, \bibinfo {author}
  {\bibfnamefont {S.}~\bibnamefont {Bose}}, \ and\ \bibinfo {author}
  {\bibfnamefont {J.~S.}\ \bibnamefont {Bagla}},\ }\href {\doibase
  10.1093/mnras/stab2875} {\bibfield  {journal} {\bibinfo  {journal} {Mon. Not.
  Roy. Astron. Soc.}\ }\textbf {\bibinfo {volume} {508}},\ \bibinfo {pages}
  {4869} (\bibinfo {year} {2021})},\ \Eprint {http://arxiv.org/abs/2102.03946}
  {arXiv:2102.03946 [astro-ph.CO]} \BibitemShut {NoStop}%
\bibitem [{\citenamefont {Yeung}\ \emph {et~al.}(2023)\citenamefont {Yeung},
  \citenamefont {Cheung}, \citenamefont {Seo}, \citenamefont {Gais},
  \citenamefont {Hannuksela},\ and\ \citenamefont {Li}}]{Yeung:2021chy}%
  \BibitemOpen
  \bibfield  {author} {\bibinfo {author} {\bibfnamefont {S.~M.~C.}\
  \bibnamefont {Yeung}}, \bibinfo {author} {\bibfnamefont {M.~H.~Y.}\
  \bibnamefont {Cheung}}, \bibinfo {author} {\bibfnamefont {E.}~\bibnamefont
  {Seo}}, \bibinfo {author} {\bibfnamefont {J.~A.~J.}\ \bibnamefont {Gais}},
  \bibinfo {author} {\bibfnamefont {O.~A.}\ \bibnamefont {Hannuksela}}, \ and\
  \bibinfo {author} {\bibfnamefont {T.~G.~F.}\ \bibnamefont {Li}},\ }\href
  {\doibase 10.1093/mnras/stad2772} {\bibfield  {journal} {\bibinfo  {journal}
  {Mon. Not. Roy. Astron. Soc.}\ }\textbf {\bibinfo {volume} {526}},\ \bibinfo
  {pages} {2230} (\bibinfo {year} {2023})},\ \Eprint
  {http://arxiv.org/abs/2112.07635} {arXiv:2112.07635 [gr-qc]} \BibitemShut
  {NoStop}%
\bibitem [{\citenamefont {Meena}(2025)}]{Meena:2025gry}%
  \BibitemOpen
  \bibfield  {author} {\bibinfo {author} {\bibfnamefont {A.~K.}\ \bibnamefont
  {Meena}},\ }\href@noop {} {\  (\bibinfo {year} {2025})},\ \Eprint
  {http://arxiv.org/abs/2502.11488} {arXiv:2502.11488 [gr-qc]} \BibitemShut
  {NoStop}%
\bibitem [{\citenamefont {Abramowitz}\ and\ \citenamefont
  {Stegun}(1964)}]{abramowitz+stegun}%
  \BibitemOpen
  \bibfield  {author} {\bibinfo {author} {\bibfnamefont {M.}~\bibnamefont
  {Abramowitz}}\ and\ \bibinfo {author} {\bibfnamefont {I.~A.}\ \bibnamefont
  {Stegun}},\ }\href@noop {} {\emph {\bibinfo {title} {Handbook of Mathematical
  Functions with Formulas, Graphs, and Mathematical Tables}}},\ \bibinfo
  {edition} {ninth dover printing, tenth gpo printing}\ ed.\ (\bibinfo
  {publisher} {Dover},\ \bibinfo {address} {New York},\ \bibinfo {year}
  {1964})\BibitemShut {NoStop}%
\bibitem [{\citenamefont {Whitney}(1992)}]{Whitney1992}%
  \BibitemOpen
  \bibfield  {author} {\bibinfo {author} {\bibfnamefont {H.}~\bibnamefont
  {Whitney}},\ }\enquote {\bibinfo {title} {On singularities of mappings of
  euclidean spaces. i. mappings of the plane into the plane},}\ in\ \href
  {\doibase 10.1007/978-1-4612-2972-8_27} {\emph {\bibinfo {booktitle} {Hassler
  Whitney Collected Papers}}},\ \bibinfo {editor} {edited by\ \bibinfo {editor}
  {\bibfnamefont {J.}~\bibnamefont {Eells}}\ and\ \bibinfo {editor}
  {\bibfnamefont {D.}~\bibnamefont {Toledo}}}\ (\bibinfo  {publisher}
  {Birkh{\"a}user Boston},\ \bibinfo {address} {Boston, MA},\ \bibinfo {year}
  {1992})\ pp.\ \bibinfo {pages} {370--406}\BibitemShut {NoStop}%
\bibitem [{\citenamefont {Gaudi}\ and\ \citenamefont
  {Petters}(2002)}]{Gaudi:2001fp}%
  \BibitemOpen
  \bibfield  {author} {\bibinfo {author} {\bibfnamefont {B.~S.}\ \bibnamefont
  {Gaudi}}\ and\ \bibinfo {author} {\bibfnamefont {A.~O.}\ \bibnamefont
  {Petters}},\ }\href {\doibase 10.1086/341063} {\bibfield  {journal} {\bibinfo
   {journal} {Astrophys. J.}\ }\textbf {\bibinfo {volume} {574}},\ \bibinfo
  {pages} {970} (\bibinfo {year} {2002})},\ \Eprint
  {http://arxiv.org/abs/astro-ph/0112531} {arXiv:astro-ph/0112531} \BibitemShut
  {NoStop}%
\bibitem [{\citenamefont {Gaudi}(2002)}]{Gaudi:2002hu}%
  \BibitemOpen
  \bibfield  {author} {\bibinfo {author} {\bibfnamefont {B.~S.}\ \bibnamefont
  {Gaudi}},\ }\href {\doibase 10.1086/343114} {\bibfield  {journal} {\bibinfo
  {journal} {Astrophys. J.}\ }\textbf {\bibinfo {volume} {580}},\ \bibinfo
  {pages} {468} (\bibinfo {year} {2002})},\ \Eprint
  {http://arxiv.org/abs/astro-ph/0206162} {arXiv:astro-ph/0206162} \BibitemShut
  {NoStop}%
\bibitem [{\citenamefont {Congdon}\ \emph {et~al.}(2008)\citenamefont
  {Congdon}, \citenamefont {Keeton},\ and\ \citenamefont
  {Nordgren}}]{Congdon:2008pm}%
  \BibitemOpen
  \bibfield  {author} {\bibinfo {author} {\bibfnamefont {A.~B.}\ \bibnamefont
  {Congdon}}, \bibinfo {author} {\bibfnamefont {C.~R.}\ \bibnamefont {Keeton}},
  \ and\ \bibinfo {author} {\bibfnamefont {C.~E.}\ \bibnamefont {Nordgren}},\
  }\href {\doibase 10.1111/j.1365-2966.2008.13604.x} {\bibfield  {journal}
  {\bibinfo  {journal} {Mon. Not. Roy. Astron. Soc.}\ }\textbf {\bibinfo
  {volume} {389}},\ \bibinfo {pages} {398} (\bibinfo {year} {2008})},\ \Eprint
  {http://arxiv.org/abs/0806.2841} {arXiv:0806.2841 [astro-ph]} \BibitemShut
  {NoStop}%
\bibitem [{\citenamefont {{Schneider}}\ and\ \citenamefont
  {{Weiss}}(1992)}]{Schneider_cusp_1992}%
  \BibitemOpen
  \bibfield  {author} {\bibinfo {author} {\bibfnamefont {P.}~\bibnamefont
  {{Schneider}}}\ and\ \bibinfo {author} {\bibfnamefont {A.}~\bibnamefont
  {{Weiss}}},\ }\href@noop {} {\bibfield  {journal} {\bibinfo  {journal}
  {Astronomy and Astrophysics}\ }\textbf {\bibinfo {volume} {260}},\ \bibinfo
  {pages} {1} (\bibinfo {year} {1992})}\BibitemShut {NoStop}%
\bibitem [{\citenamefont {Pearcey}(1946)}]{Pearcey01051946}%
  \BibitemOpen
  \bibfield  {author} {\bibinfo {author} {\bibfnamefont {T.}~\bibnamefont
  {Pearcey}},\ }\href {\doibase 10.1080/14786444608561335} {\bibfield
  {journal} {\bibinfo  {journal} {The London, Edinburgh, and Dublin
  Philosophical Magazine and Journal of Science}\ }\textbf {\bibinfo {volume}
  {37}},\ \bibinfo {pages} {311} (\bibinfo {year} {1946})},\ \Eprint
  {http://arxiv.org/abs/https://doi.org/10.1080/14786444608561335}
  {https://doi.org/10.1080/14786444608561335} \BibitemShut {NoStop}%
\bibitem [{git(2025{\natexlab{a}})}]{githubpearcey}%
  \BibitemOpen
  \href {https://github.com/ricokaloklo/pearcey} {\emph {\bibinfo {title}
  {https://github.com/ricokaloklo/pearcey}}} (\bibinfo {year}
  {2025}{\natexlab{a}})\BibitemShut {NoStop}%
\bibitem [{\citenamefont {{Nye}}\ and\ \citenamefont
  {{Berry}}(1974)}]{Nye_Berry:1974}%
  \BibitemOpen
  \bibfield  {author} {\bibinfo {author} {\bibfnamefont {J.~F.}\ \bibnamefont
  {{Nye}}}\ and\ \bibinfo {author} {\bibfnamefont {M.~V.}\ \bibnamefont
  {{Berry}}},\ }\href {\doibase 10.1098/rspa.1974.0012} {\bibfield  {journal}
  {\bibinfo  {journal} {Proceedings of the Royal Society of London Series A}\
  }\textbf {\bibinfo {volume} {336}},\ \bibinfo {pages} {165} (\bibinfo {year}
  {1974})}\BibitemShut {NoStop}%
\bibitem [{\citenamefont {Maggiore}(2007)}]{Maggiore:1900zz}%
  \BibitemOpen
  \bibfield  {author} {\bibinfo {author} {\bibfnamefont {M.}~\bibnamefont
  {Maggiore}},\ }\href@noop {} {\emph {\bibinfo {title} {{Gravitational Waves.
  Vol. 1: Theory and Experiments}}}},\ Oxford Master Series in Physics\
  (\bibinfo  {publisher} {Oxford University Press},\ \bibinfo {year}
  {2007})\BibitemShut {NoStop}%
\bibitem [{git(2025{\natexlab{b}})}]{githubmodwaveforms}%
  \BibitemOpen
  \href {https://github.com/ezquiaga/modwaveforms} {\emph {\bibinfo {title}
  {https://github.com/ezquiaga/modwaveforms}}} (\bibinfo {year}
  {2025}{\natexlab{b}})\BibitemShut {NoStop}%
\bibitem [{\citenamefont {Pratten}\ \emph {et~al.}(2021)\citenamefont {Pratten}
  \emph {et~al.}}]{Pratten:2020ceb}%
  \BibitemOpen
  \bibfield  {author} {\bibinfo {author} {\bibfnamefont {G.}~\bibnamefont
  {Pratten}} \emph {et~al.},\ }\href {\doibase 10.1103/PhysRevD.103.104056}
  {\bibfield  {journal} {\bibinfo  {journal} {Phys. Rev. D}\ }\textbf {\bibinfo
  {volume} {103}},\ \bibinfo {pages} {104056} (\bibinfo {year} {2021})},\
  \Eprint {http://arxiv.org/abs/2004.06503} {arXiv:2004.06503 [gr-qc]}
  \BibitemShut {NoStop}%
\bibitem [{\citenamefont {Chan}\ \emph {et~al.}(2024)\citenamefont {Chan},
  \citenamefont {Seo}, \citenamefont {Li}, \citenamefont {Fong},\ and\
  \citenamefont {Ezquiaga}}]{Chan:2024qmb}%
  \BibitemOpen
  \bibfield  {author} {\bibinfo {author} {\bibfnamefont {J.~C.~L.}\
  \bibnamefont {Chan}}, \bibinfo {author} {\bibfnamefont {E.}~\bibnamefont
  {Seo}}, \bibinfo {author} {\bibfnamefont {A.~K.~Y.}\ \bibnamefont {Li}},
  \bibinfo {author} {\bibfnamefont {H.}~\bibnamefont {Fong}}, \ and\ \bibinfo
  {author} {\bibfnamefont {J.~M.}\ \bibnamefont {Ezquiaga}},\ }\href@noop {} {\
   (\bibinfo {year} {2024})},\ \Eprint {http://arxiv.org/abs/2411.13058}
  {arXiv:2411.13058 [gr-qc]} \BibitemShut {NoStop}%
\bibitem [{\citenamefont {Ashton}\ \emph {et~al.}(2019)\citenamefont {Ashton}
  \emph {et~al.}}]{Ashton:2018jfp}%
  \BibitemOpen
  \bibfield  {author} {\bibinfo {author} {\bibfnamefont {G.}~\bibnamefont
  {Ashton}} \emph {et~al.},\ }\href {\doibase 10.3847/1538-4365/ab06fc}
  {\bibfield  {journal} {\bibinfo  {journal} {Astrophys. J. Suppl.}\ }\textbf
  {\bibinfo {volume} {241}},\ \bibinfo {pages} {27} (\bibinfo {year} {2019})},\
  \Eprint {http://arxiv.org/abs/1811.02042} {arXiv:1811.02042 [astro-ph.IM]}
  \BibitemShut {NoStop}%
\bibitem [{\citenamefont {Liu}\ and\ \citenamefont {Kim}(2024)}]{Liu:2023emk}%
  \BibitemOpen
  \bibfield  {author} {\bibinfo {author} {\bibfnamefont {A.}~\bibnamefont
  {Liu}}\ and\ \bibinfo {author} {\bibfnamefont {K.}~\bibnamefont {Kim}},\
  }\href {\doibase 10.1103/PhysRevD.110.123008} {\bibfield  {journal} {\bibinfo
   {journal} {Phys. Rev. D}\ }\textbf {\bibinfo {volume} {110}},\ \bibinfo
  {pages} {123008} (\bibinfo {year} {2024})},\ \Eprint
  {http://arxiv.org/abs/2301.07253} {arXiv:2301.07253 [gr-qc]} \BibitemShut
  {NoStop}%
\bibitem [{\citenamefont {Oguri}(2018)}]{Oguri:2018muv}%
  \BibitemOpen
  \bibfield  {author} {\bibinfo {author} {\bibfnamefont {M.}~\bibnamefont
  {Oguri}},\ }\href {\doibase 10.1093/mnras/sty2145} {\bibfield  {journal}
  {\bibinfo  {journal} {Mon. Not. Roy. Astron. Soc.}\ }\textbf {\bibinfo
  {volume} {480}},\ \bibinfo {pages} {3842} (\bibinfo {year} {2018})},\ \Eprint
  {http://arxiv.org/abs/1807.02584} {arXiv:1807.02584 [astro-ph.CO]}
  \BibitemShut {NoStop}%
\bibitem [{\citenamefont {Ng}\ \emph {et~al.}(2018)\citenamefont {Ng},
  \citenamefont {Wong}, \citenamefont {Broadhurst},\ and\ \citenamefont
  {Li}}]{Ng:2017yiu}%
  \BibitemOpen
  \bibfield  {author} {\bibinfo {author} {\bibfnamefont {K.~K.~Y.}\
  \bibnamefont {Ng}}, \bibinfo {author} {\bibfnamefont {K.~W.~K.}\ \bibnamefont
  {Wong}}, \bibinfo {author} {\bibfnamefont {T.}~\bibnamefont {Broadhurst}}, \
  and\ \bibinfo {author} {\bibfnamefont {T.~G.~F.}\ \bibnamefont {Li}},\ }\href
  {\doibase 10.1103/PhysRevD.97.023012} {\bibfield  {journal} {\bibinfo
  {journal} {Phys. Rev. D}\ }\textbf {\bibinfo {volume} {97}},\ \bibinfo
  {pages} {023012} (\bibinfo {year} {2018})},\ \Eprint
  {http://arxiv.org/abs/1703.06319} {arXiv:1703.06319 [astro-ph.CO]}
  \BibitemShut {NoStop}%
\bibitem [{\citenamefont {Lai}\ \emph {et~al.}(2018)\citenamefont {Lai},
  \citenamefont {Hannuksela}, \citenamefont {Herrera-Mart{\'\i}n},
  \citenamefont {Diego}, \citenamefont {Broadhurst},\ and\ \citenamefont
  {Li}}]{Lai:2018rto}%
  \BibitemOpen
  \bibfield  {author} {\bibinfo {author} {\bibfnamefont {K.-H.}\ \bibnamefont
  {Lai}}, \bibinfo {author} {\bibfnamefont {O.~A.}\ \bibnamefont {Hannuksela}},
  \bibinfo {author} {\bibfnamefont {A.}~\bibnamefont {Herrera-Mart{\'\i}n}},
  \bibinfo {author} {\bibfnamefont {J.~M.}\ \bibnamefont {Diego}}, \bibinfo
  {author} {\bibfnamefont {T.}~\bibnamefont {Broadhurst}}, \ and\ \bibinfo
  {author} {\bibfnamefont {T.~G.~F.}\ \bibnamefont {Li}},\ }\href {\doibase
  10.1103/PhysRevD.98.083005} {\bibfield  {journal} {\bibinfo  {journal} {Phys.
  Rev. D}\ }\textbf {\bibinfo {volume} {98}},\ \bibinfo {pages} {083005}
  (\bibinfo {year} {2018})},\ \Eprint {http://arxiv.org/abs/1801.07840}
  {arXiv:1801.07840 [gr-qc]} \BibitemShut {NoStop}%
\bibitem [{\citenamefont {Smith}\ \emph {et~al.}(2023)\citenamefont {Smith},
  \citenamefont {Robertson}, \citenamefont {Mahler}, \citenamefont {Nicholl},
  \citenamefont {Ryczanowski}, \citenamefont {Bianconi}, \citenamefont
  {Sharon}, \citenamefont {Massey}, \citenamefont {Richard},\ and\
  \citenamefont {Jauzac}}]{Smith:2022vbp}%
  \BibitemOpen
  \bibfield  {author} {\bibinfo {author} {\bibfnamefont {G.~P.}\ \bibnamefont
  {Smith}}, \bibinfo {author} {\bibfnamefont {A.}~\bibnamefont {Robertson}},
  \bibinfo {author} {\bibfnamefont {G.}~\bibnamefont {Mahler}}, \bibinfo
  {author} {\bibfnamefont {M.}~\bibnamefont {Nicholl}}, \bibinfo {author}
  {\bibfnamefont {D.}~\bibnamefont {Ryczanowski}}, \bibinfo {author}
  {\bibfnamefont {M.}~\bibnamefont {Bianconi}}, \bibinfo {author}
  {\bibfnamefont {K.}~\bibnamefont {Sharon}}, \bibinfo {author} {\bibfnamefont
  {R.}~\bibnamefont {Massey}}, \bibinfo {author} {\bibfnamefont
  {J.}~\bibnamefont {Richard}}, \ and\ \bibinfo {author} {\bibfnamefont
  {M.}~\bibnamefont {Jauzac}},\ }\href {\doibase 10.1093/mnras/stad140}
  {\bibfield  {journal} {\bibinfo  {journal} {Mon. Not. Roy. Astron. Soc.}\
  }\textbf {\bibinfo {volume} {520}},\ \bibinfo {pages} {702} (\bibinfo {year}
  {2023})},\ \Eprint {http://arxiv.org/abs/2204.12977} {arXiv:2204.12977
  [astro-ph.HE]} \BibitemShut {NoStop}%
\bibitem [{\citenamefont {Auclair}\ \emph {et~al.}(2023)\citenamefont {Auclair}
  \emph {et~al.}}]{LISACosmologyWorkingGroup:2022jok}%
  \BibitemOpen
  \bibfield  {author} {\bibinfo {author} {\bibfnamefont {P.}~\bibnamefont
  {Auclair}} \emph {et~al.} (\bibinfo {collaboration} {LISA Cosmology Working
  Group}),\ }\href {\doibase 10.1007/s41114-023-00045-2} {\bibfield  {journal}
  {\bibinfo  {journal} {Living Rev. Rel.}\ }\textbf {\bibinfo {volume} {26}},\
  \bibinfo {pages} {5} (\bibinfo {year} {2023})},\ \Eprint
  {http://arxiv.org/abs/2204.05434} {arXiv:2204.05434 [astro-ph.CO]}
  \BibitemShut {NoStop}%
\bibitem [{\citenamefont {Savastano}\ \emph {et~al.}(2023)\citenamefont
  {Savastano}, \citenamefont {Tambalo}, \citenamefont {Villarrubia-Rojo},\ and\
  \citenamefont {Zumalacarregui}}]{Savastano:2023spl}%
  \BibitemOpen
  \bibfield  {author} {\bibinfo {author} {\bibfnamefont {S.}~\bibnamefont
  {Savastano}}, \bibinfo {author} {\bibfnamefont {G.}~\bibnamefont {Tambalo}},
  \bibinfo {author} {\bibfnamefont {H.}~\bibnamefont {Villarrubia-Rojo}}, \
  and\ \bibinfo {author} {\bibfnamefont {M.}~\bibnamefont {Zumalacarregui}},\
  }\href {\doibase 10.1103/PhysRevD.108.103532} {\bibfield  {journal} {\bibinfo
   {journal} {Phys. Rev. D}\ }\textbf {\bibinfo {volume} {108}},\ \bibinfo
  {pages} {103532} (\bibinfo {year} {2023})},\ \Eprint
  {http://arxiv.org/abs/2306.05282} {arXiv:2306.05282 [gr-qc]} \BibitemShut
  {NoStop}%
\bibitem [{\citenamefont {\c{C}al\i{}\c{s}kan}\ \emph
  {et~al.}(2023)\citenamefont {\c{C}al\i{}\c{s}kan}, \citenamefont
  {Anil~Kumar}, \citenamefont {Ji}, \citenamefont {Ezquiaga}, \citenamefont
  {Cotesta}, \citenamefont {Berti},\ and\ \citenamefont
  {Kamionkowski}}]{Caliskan:2023zqm}%
  \BibitemOpen
  \bibfield  {author} {\bibinfo {author} {\bibfnamefont {M.}~\bibnamefont
  {\c{C}al\i{}\c{s}kan}}, \bibinfo {author} {\bibfnamefont {N.}~\bibnamefont
  {Anil~Kumar}}, \bibinfo {author} {\bibfnamefont {L.}~\bibnamefont {Ji}},
  \bibinfo {author} {\bibfnamefont {J.~M.}\ \bibnamefont {Ezquiaga}}, \bibinfo
  {author} {\bibfnamefont {R.}~\bibnamefont {Cotesta}}, \bibinfo {author}
  {\bibfnamefont {E.}~\bibnamefont {Berti}}, \ and\ \bibinfo {author}
  {\bibfnamefont {M.}~\bibnamefont {Kamionkowski}},\ }\href {\doibase
  10.1103/PhysRevD.108.123543} {\bibfield  {journal} {\bibinfo  {journal}
  {Phys. Rev. D}\ }\textbf {\bibinfo {volume} {108}},\ \bibinfo {pages}
  {123543} (\bibinfo {year} {2023})},\ \Eprint
  {http://arxiv.org/abs/2307.06990} {arXiv:2307.06990 [astro-ph.CO]}
  \BibitemShut {NoStop}%
\end{thebibliography}%
%------------
%------------
\end{document}